\documentclass[11pt,a4paper]{article}
\usepackage{jheppub}
\usepackage{amsmath,graphicx,amsthm,amssymb,amsfonts,hyperref,srcltx,color,float}
\usepackage[FIGTOPCAP]{subfigure}

\newcommand{\Tbar}{\bar{T}}
\newcommand{\kbar}{\bar{k}}
\newcommand{\wbar}{\bar{\omega}}
\newcommand{\Bbar}{\bar{B}}
\newcommand{\N}{{\mathcal{N}}}
\newcommand{\jt}{\langle J^t \rangle}
\newcommand{\Om}{\mathcal{O}_m}
\newcommand{\Omv}{\langle \mathcal{O}_m \rangle}

\newcommand{\beq}{\begin{equation}}
\newcommand{\eeq}{\end{equation}}
\newcommand{\bea}{\begin{eqnarray}}
\newcommand{\eea}{\end{eqnarray}}

\renewcommand{\Im}{\operatorname{Im}}
\renewcommand{\Re}{\operatorname{Re}}

\title{Collective Excitations of Holographic Quantum Liquids in a Magnetic Field}

\author[a,1]{Daniel K.~Brattan,\note{E-mail address: d.k.brattan@durham.ac.uk}}
\author[b,2]{Richard A.~Davison,\note{E-mail address: r.davison1@physics.ox.ac.uk}}
\author[a,3]{Simon A.~Gentle,\note{E-mail address: s.a.gentle@durham.ac.uk}}
\author[c,4]{and Andy~O'Bannon\note{E-mail address: A.OBannon@damtp.cam.ac.uk}}
\affiliation[a]{Centre for Particle Theory \& Department of Mathematical Sciences, University of Durham,\\ Science Laboratories, South Road, Durham DH1 3LE United Kingdom}
\affiliation[b]{Rudolf Peierls Centre for Theoretical Physics, University of Oxford,\\ 1 Keble Road, Oxford OX1 3NP United Kingdom}
\affiliation[c]{Department of Applied Mathematics and Theoretical Physics, University of Cambridge, \\ Wilberforce Road, Cambridge CB3 0WA United Kingdom}

\abstract{We use holography to study $\N=4$ supersymmetric $SU(N_c)$ Yang-Mills theory in the large-$N_c$ and large-coupling limits coupled to a number $N_f \ll N_c$ of $(n+1)$-dimensional massless supersymmetric hypermultiplets in the fundamental representation of $SU(N_c)$, with $n=2,3$. We introduce a temperature $T$, a baryon number chemical potential $\mu$, and a baryon number magnetic field $B$, and work in a regime with $\mu\gg T,\sqrt{B}$. We study the collective excitations of these holographic quantum liquids by computing the poles in the retarded Green's function of the baryon number charge density operator and the associated peaks in the spectral function. We focus on the evolution of the collective excitations as we increase the frequency relative to $T$, \textit{i.e.}\ the hydrodynamic/collisionless crossover. We find that for all $B$, at low frequencies the tallest peak in the spectral function is associated with hydrodynamic charge diffusion. At high frequencies the tallest peak is associated with  a sound mode similar to the zero sound mode in the collisionless regime of a Landau Fermi liquid. The sound mode has a gap proportional to $B$, and as a result for intermediate frequencies and for $B$ sufficiently large compared to $T$ the spectral weight is strongly suppressed. We find that the hydrodynamic/collisionless crossover occurs at a frequency that is approximately $B$-independent.}

\begin{document}
\begin{flushright}DAMTP-2012-61 \\ DCPT-12/35\\ OUTP-12-18P \end{flushright}
\maketitle
\flushbottom

\section{Introduction}
\label{intro}

Consider a system with a global $U(1)$ symmetry, in a state with a net charge density of that $U(1)$. Such a state is compressible if the charge density varies smoothly as a function of the chemical potential. The best-understood examples of compressible states are solids, in which translational symmetry is spontaneously broken to a discrete subgroup, superfluids, in which the $U(1)$ is spontaneously broken, and Landau Fermi Liquids (LFLs), in which neither translational symmetry nor the $U(1)$ are broken.

The degrees of freedom in a LFL are long-lived fermionic quasi-particle excitations about a Fermi surface. Two characteristic features of a LFL are that at temperatures $T$ low compared to the Fermi energy, the heat capacity $c_V \propto T$ and the electrical resistivity $\rho \propto T^2$. The spectrum of a LFL includes not only particle-hole excitations about the Fermi surface but also collective charge density excitations, which produce poles in the retarded Green's function of the charge density operator. At sufficiently low $T$, in the so-called ``collisionless'' regime in which quantum effects are more important than thermal effects, the spectrum of collective excitations includes ``zero sound," a fluctuation of the shape of the Fermi surface that has the dispersion relation of a sound wave~\cite{landau1,zerosoundreview,pinesnozieres}. At sufficiently high $T$, in the hydrodynamic regime in which thermal effects are more important than quantum effects, the spectrum of collective excitations includes the expected hydrodynamic sound and charge diffusion modes~\cite{pinesnozieres}. The transition between these two regimes as $T$ increases is called the collisionless/hydrodynamic crossover.

Some real materials are compressible, break neither translational symmetry nor the $U(1)$, and exhibit Fermi surfaces, but the low-energy degrees of freedom are \textit{not} those of a LFL. These materials are generically called ``non-Fermi liquids'' (NFLs). A subset of these are the so-called ``strange metals,'' including the normal (non-superconducting) phase of high-$T_c$ superconductors. The most salient characteristic feature of a strange metal is a linear resistivity, $\rho \propto T$. Experimental evidence indicates that, in many cases, the strange metal phase is related to quantum criticality, \textit{i.e.} a continuous phase transition at zero temperature~\cite{Sachdev:2011cs}. In particular, the degrees of freedom at such a quantum critical point, when heated up to non-zero $T$, may give rise to a quantum critical region in which $\rho \propto T$. Demonstrating this from first principles is difficult because generically quantum critical degrees of freedom are strongly interacting. The existence of strange metals and other NFLs raises a crucial question: can we classify compressible ground states?

The Anti-de Sitter/Conformal Field Theory (AdS/CFT) correspondence~\cite{Maldacena:1997re,Witten:1998qj,Gubser:1998bc}, and more generally holography, may help answer this question. In AdS/CFT the conserved current of a global $U(1)$ symmetry of the CFT is dual to a $U(1)$ gauge field in AdS, and compressible states can be described by asymptotically AdS spacetimes with non-zero electric flux at the boundary. Holography provides many examples of compressible states that are not solids, superfluids, or LFLs, the canonical example being the planar Reissner-Nordstr\"om-AdS solution of Einstein-Maxwell theory~\cite{Chamblin:1999tk}. Holography has been used to describe both quantum criticality~\cite{Evans:2010iy,Jensen:2010vd,Jensen:2010ga,Evans:2010hi,Jensen:2010vx,Evans:2010np,Iqbal:2011aj,D'Hoker:2012ih} and compressible states with $\rho \propto T$~\cite{Hartnoll:2009ns,Faulkner:2010zz,Donos:2012} and, more generally, has the potential to reveal some guiding principles for classifying compressible states~\cite{Huijse:2011hp,Sachdev:2011wg}.

With these motivations in mind, we will study two holographic systems describing compressible states that are not solids, superfluids, LFLs, or NFLs, and which exhibit quantum phase transitions for sufficiently large magnetic field $B$. Our main goal is to study the effect of non-zero $B$ on the collisionless/hydrodynamic crossover in these systems.

We will study the field theories arising from the $(n+1)$-dimensional intersection of $N_c$ D3-branes with $N_f$ D$p$-branes in type IIB string theory, with $p=2n+1=5,7$ so that $n=2,3$~\cite{Karch:2002sh,Karch:2000gx,DeWolfe:2001pq,Erdmenger:2002ex}. We will call these the D3/D$p$ systems or D3/D$p$ theories. These theories are $(3+1)$-dimensional $\N=4$ supersymmetric $SU(N_c)$ Yang-Mills (SYM) theory coupled to a number $N_f$ of $(n+1)$-dimensional hypermultiplets in the fundamental representation of $SU(N_c)$, \textit{i.e.} flavor fields. We consider only massless flavor fields. We will work in the 't Hooft large-$N_c$ limit, with large 't Hooft coupling, and in the probe limit $N_f \ll N_c$.

Each D3/D$p$ theory enjoys a global $U(1)_b$ baryon number symmetry, with conserved current $J^{\nu}$, where $\nu=t,x,y$ and for $n=3$ also $\nu=z$. To study compressible states of these theories, we will introduce a baryon number charge density $\jt$, or equivalently a baryon number chemical potential $\mu$. Recall that a hypermultiplet contains fermions and scalars, and that both are charged under $U(1)_b$. States with non-zero $\jt$ are dual holographically to probe D$p$-branes in AdS with non-zero worldvolume electric flux~\cite{Karch:2002sh,Karch:2000gx,Kobayashi:2006sb}.

For temperatures $T \ll \mu$, these compressible states have been studied extensively, using holography, in refs.~\cite{Karch:2007br,Karch:2008fa,Kulaxizi:2008kv,Karch:2009eb,Nickel:2010pr,Ammon:2011hz,Davison:2011ek,Ammon:2012je,Chang:2012ek,Ammon:2012mu}. Neither translational symmetry nor $U(1)_b$ are broken in these states, and to date no evidence of a Fermi surface in these systems has been found. In short, the evidence accumulated so far suggests that these states are not solids, superfluids, LFLs, or NFLs. Indeed, these states exhibit various unusual properties, including an extensive $T=0$ ground state degeneracy~\cite{Karch:2008fa}. We will call these low-temperature compressible states ``holographic quantum liquids.''

The spectrum of excitations in these holographic quantum liquids at $T=0$ includes sound modes, \textit{i.e.}\ two modes producing poles in $J^t$'s retarded Green's function, $G^{tt}_R(\omega,k)$, that, when expressed as complex frequencies $\omega$, have real parts proportional to momentum $k$ and imaginary parts $\propto k^2$~\cite{Karch:2008fa}. Due to the similarity with the zero sound mode in a LFL, we will call these modes ``holographic zero sound.''

As $T$ increases, these holographic quantum liquids exhibit a collisionless/hydrodynamic crossover analogous to that in a LFL. For the D3/D7 system the crossover was studied in ref.~\cite{Davison:2011ek}, and a similar crossover occurs in the D3/D5 system, as we will show. The crossover is simplest to define from the behavior of poles of $G^{tt}_R(\omega,k)$. We can identify poles as holographic zero sound or charge diffusion by computing their dispersion relations. When $k \gg T^2/\mu$, we can identify the two poles of $G^{tt}_R(\omega,k)$ closest to the origin of the complex $\omega$ plane as holographic zero sound. As $T$ increases, these poles move down into the complex $\omega$ plane and towards the imaginary axis. When $k\simeq T^2/\mu$, the two poles collide on the imaginary axis. As $T$ continues to increase the poles split, one pole moving up and one moving down the imaginary axis (see figure~\ref{fig:D3D5hydrocrossover-ZeroB}). When $k\ll T^2/\mu$, we can identify the pole moving up the imaginary axis as that of charge diffusion. We define the crossover as the point where the poles collide on the imaginary axis. The crossover is also apparent in $J^t$'s spectral function, which as a function of $\omega$ exhibits a single peak due to holographic zero sound when $k\gg T^2/\mu$ and a single peak due to charge diffusion when $k \ll T^2/\mu$.

Now let us introduce a non-zero $U(1)_b$ magnetic field $B$, which appears in the bulk as magnetic flux on the D$p$-brane worldvolume~\cite{Filev:2007gb}. When $T=0$ and $\jt$ is non-zero, in each of the D3/D$p$ systems a sufficiently large $B$ triggers a quantum phase transition in which a global symmetry is spontaneously broken~\cite{Evans:2010iy,Jensen:2010vd,Jensen:2010ga,Evans:2010hi}. For $n=3$ the transition is at least second order~\cite{Jensen:2010vd}, while for $n=2$ the transition is of BKT-type~\cite{Jensen:2010ga,Evans:2010hi}. When $T$ is non-zero, each quantum critical point gives rise to a quantum critical region.

We want to determine how $B$ affects the collisionless/hydrodynamic crossover in these holographic quantum liquids. Five scales appear in our problem: $\mu$, $T$, $B$, which characterize the state of the system, and $\omega$, $k$, which characterize perturbations of the state. We will work in the quantum liquid regime, where $\mu \gg T,\sqrt{B}$. Furthermore, we will always work with values of $B$ below any known phase transition, where we already find a rich story. We choose the momentum to point in the $x$ direction. For $n=3$ we choose the magnetic field to point in the $z$ direction, orthogonal to the momentum. We work with $\omega,k\ll \mu$. Our results are qualitatively similar for both of the D3/D$p$ theories. 

In our systems, a non-zero $B$ has two major effects on the spectrum of collective excitations at low temperatures. First, a non-zero $B$ breaks parity, and so operators can mix that could not mix when $B=0$. In particular, we find that when $B$ is non-zero the poles in $G^{tt}_R(\omega,k)$ mix with those of $J^y$'s retarded Green's function, $G^{yy}_R(\omega,k)$. Due to the mixing of poles, for small $T$ the holographic zero sound poles are no longer the only poles near the origin of the complex $\omega$ plane: a purely imaginary pole also appears. As a result, the spectral function of $J^t$, as a function of $\omega$, can exhibit \textit{two} large peaks, one from holographic zero sound and one from the purely imaginary mode.

Second, when $T=0$, a non-zero $B$ produces a gap in the holographic zero sound dispersion relation $\propto B/\mu$~\cite{Goykhman:2012vy,Gorsky:2012gi}. Such a gap is consistent with Kohn's theorem~\cite{Kohn:1961zz}, which states that for non-relativistic particles with pair-wise interactions of any strength, in an external magnetic field the center-of-mass motion decouples from the relative motion of the particles and behaves as a free particle in a magnetic field, which then guarantees a $k=0$ resonance at the cyclotron frequency, which is $\propto B$.

For sufficiently small $B$, meaning $B \lesssim T^2$, and for fixed $k$, we find that as $T$ increases a collision of poles on the imaginary axis still takes place. Due to the mixing of poles, the details of the collision differ from the $B=0$ case, nevertheless we can still use the collision of poles to define the location of the collisionless/hydrodynamic crossover. For sufficiently large $B$, meaning $B \gtrsim T^2$, however, no such collision of poles occurs, so at first glance we seem to be unable to identify a crossover at a specific value of $T$.

After a thorough search, however, we have found a simple alternative definition that allows us to identify a precise location for the crossover, for all values of $B$ that we study. If we express the poles in $G^{tt}_R(\omega,k)$ in terms of real $\omega$ and complex $k$, then in the complex $k$ plane we always find a \textit{single} pole near the origin, for all values of $B$ that we study. For low $T$, meaning $\omega \gg T^2/\mu$, and for $\omega$ larger than any other scale except $\mu$, we can identify the pole as that of holographic zero sound. As $T$ increases, the pole becomes that of charge diffusion when $\omega \approx 0.45 \, T^2/\mu$ for the $n=3$ case and when $\omega \approx 0.30 \, T^2/\mu$ for $n=2$. Most importantly, these values of $T$ are \textit{independent of $B$}, and indeed are exactly the same values as those at $B=0$. We thus define the location of the collisionless/hydrodynamic crossover as the point where the pole in the \textit{complex $k$ plane} becomes that of charge diffusion.

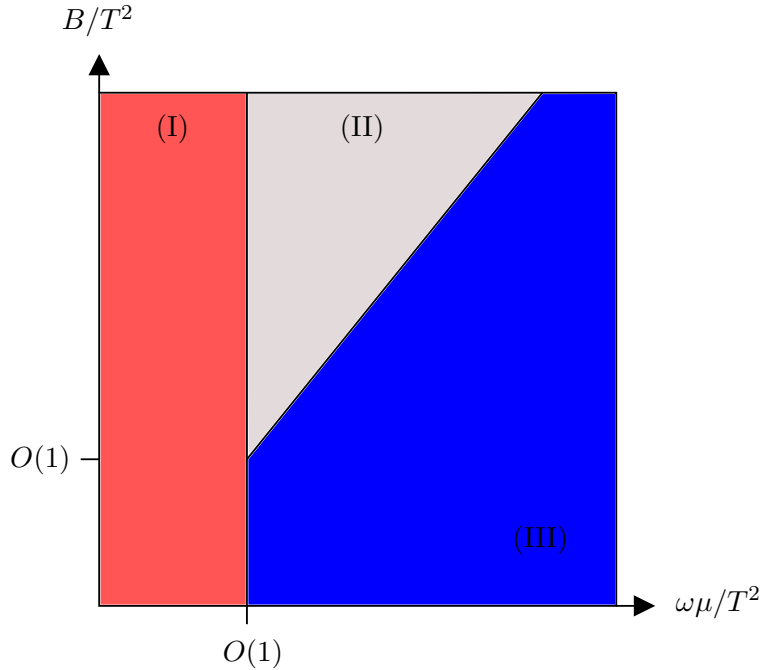
\begin{figure}[!htb]
\centering
\vskip1em
\includegraphics[width=0.5\textwidth]{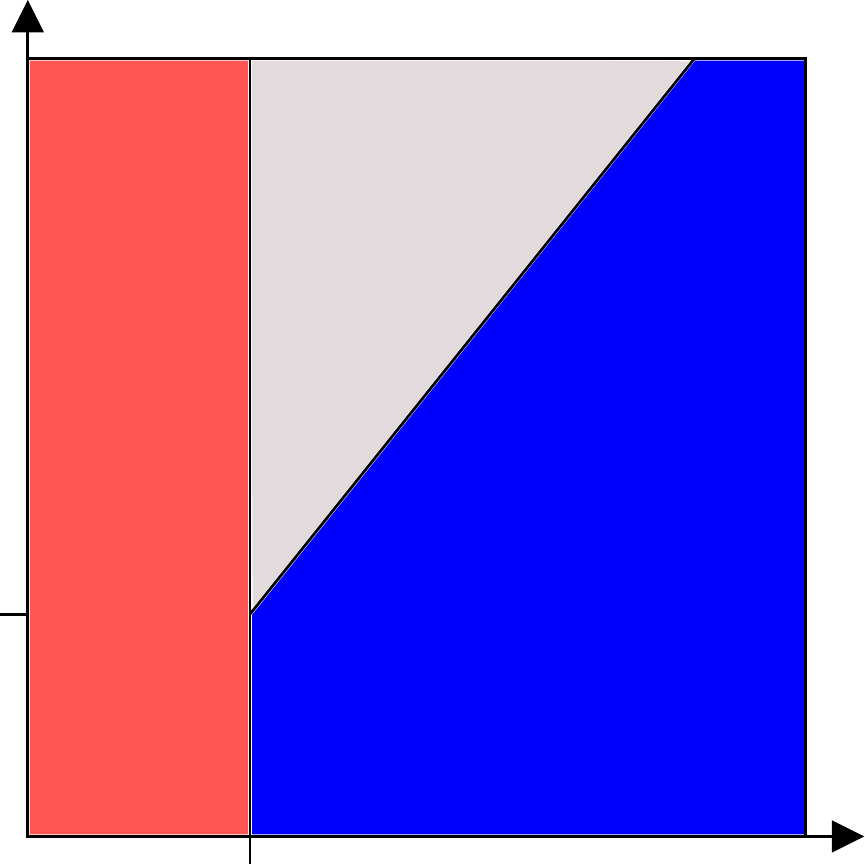}
\begin{center}
\setlength{\unitlength}{0.1\columnwidth}
\begin{picture}(0.1,0.25)(0,0)
\thicklines
\put(-2.3,5.85){\makebox(0,0){$B/T^2$}}
\put(3.1,0.75){\makebox(0,0){$\omega\mu/T^2$}}
\put(-0.95,0.3){\makebox(0,0){$O(1)$}}
\put(-2.8,2.0){\makebox(0,0){$O(1)$}}
\put(-1.65,4.9){\makebox(0,0){(I)}}
\put(0.0,4.9){\makebox(0,0){(II)}}
\put(1.55,1.3){\makebox(0,0){(III)}}
\end{picture}
\vskip-1em
\caption{A schematic summary of our main results. We depict the plane of $B/T^2$ versus $\omega \mu /T^2$, with colors indicating the nature of the peak in $J^t$'s spectral function, at fixed $\omega$, as a function of $k$. In region (I), shaded red, the peak is due to charge diffusion. In region (III), shaded blue, the peak is due to holographic zero sound. In region (II), shaded grey, the peak is not associated with either charge diffusion or holographic zero sound, and in fact in region (II) the peak is suppressed by several orders of magnitude compared to the peaks in regions (I) and (III). The suppression is due to the gap $\propto B/\mu$ in the holographic zero sound's dispersion relation, which defines the diagonal line between regions (II) and (III). We define the collisionless/hydrodynamic crossover as the point where, as we increase $T$, the pole in $J^t$'s retarded Green's function becomes that of charge diffusion. The crossover so defined is independent of $B$, and appears at the vertical line at 
the right-hand boundary of region (I).}
\label{fig:cartoon2}
\end{center}
\vskip-1.5em
\end{figure}

What happens to the pole in the complex $k$ plane between $\omega \approx T^2/\mu$ and $\omega \gg T^2/\mu$ depends on the value of $B$, and is simplest to understand by studying the spectral function of $J^t$. For any $B$, we find that the spectral function, as a function of $k$ at a fixed $\omega$, exhibits only a single peak. For any $B$, when $\omega \lesssim T^2/\mu$ the peak is due to charge diffusion. For small $B$, as we increase $\omega$, the peak changes to that of holographic zero sound at $\omega \approx T^2/\mu$. At large $B$, however, when $\omega \approx T^2/\mu$ the peak is suppressed by orders of magnitude, and only when $\omega \approx B/\mu$ does the peak grow again, becoming that of holographic zero sound. The suppression is simple to understand: once $B \gtrsim T^2$ the holographic zero sound dispersion relation acquires a gap $\propto B/\mu$, so if we perturb the system with a frequency in the collisionless regime, $\omega \gtrsim T^2/\mu$, but with $\omega \lesssim B/\mu$, then the frequency is not high enough to overcome the gap and excite the holographic zero sound. We thus find an intermediate region $T^2/\mu \lesssim \omega < B/\mu$, with relatively low spectral weight. We summarize these results schematically in fig.~\ref{fig:cartoon2} (see also fig.~\ref{fig:Andywedge2}).

We have several reasons to believe that results qualitatively similar to ours will appear for any compressible state with a holographic description in terms of probe D-branes with worldvolume electric flux. The only ingredient in our calculation is a probe D-brane in AdS space. We see qualitatively similar results for both of our D3/D$p$ systems. A gap appears in the holographic zero sound dispersion relation in several different holographic probe D-brane systems~\cite{Jokela:2012vn,Goykhman:2012vy,Gorsky:2012gi}. Indeed, Kohn's theorem is very general, and suggests that at high frequencies and high magnetic fields, compared to $T$, we should always find a region of suppressed spectral weight, like region (II) in fig.~\ref{fig:cartoon2}. In short, our fig.~\ref{fig:cartoon2} may be characteristic of holographic quantum liquids described by probe D-branes with worldvolume electric flux, and hence may help in classifying them.

This paper is organized as follows. In section~\ref{review} we review the thermodynamics of our holographic quantum liquids and in section~\ref{eoms} we review how to compute poles in retarded Green's functions, and spectral functions, using holography. In section~\ref{zerob} we discuss the collisionless/hydrodynamic crossover in our systems when $B=0$, which for the D3/D7 system is a review but for the D3/D5 system includes novel results. In section \ref{finiteb} we discuss the crossover in our systems at non-zero $B$. We conclude in section~\ref{conclusions} with some suggestions for future research in these (and similar) systems.

\section{Review of Holographic Quantum Liquids}
\label{review}

In type IIB string theory, we will study the $(n+1)$-dimensional intersection of $N_c$ D3-branes with $N_f$ D$p$-branes, with $p=2n+1=5,7$, which we call the D3/D$p$ systems~\cite{Karch:2002sh,Karch:2000gx,DeWolfe:2001pq,Erdmenger:2002ex}. We summarize both systems in the following array:
\beq
\begin{array}{c|cccccccccc}
   & t & x & y & z & X^1 & X^2 & X^3 & X^4 & X^5 & X^6\\ \hline
N_c \,\,\, \mbox{D3} & \times & \times & \times & \times & & &  &  & & \\
N_f \,\,\,\mbox{D5} & \times & \times & \times & & \times & \times & \times & & & \\
N_f \,\,\, \mbox{D7} & \times & \times & \times & \times & \times  & \times & \times & \times &  &   \\
\end{array}
\eeq
In these intersections the low-energy theory on the D3-brane worldvolume is $(3+1)$-dimensional $\N=4$ SYM theory with gauge group $SU(N_c)$ and Yang-Mills coupling squared $g_{YM}^2=4\pi g_s$, with string coupling $g_s$, coupled to $N_f$ $(n+1)$-dimensional massless hypermultiplets in the $N_c$ representation of $SU(N_c)$, \textit{i.e.} flavor fields, preserving eight Poincar\'e supercharges. In the $n=2$ case the flavor fields propagate along a codimension-one defect, which we take to be the plane $z=0$. We will take the 't Hooft limit $N_c \to \infty$ with $g_{YM}^2 \to 0$, with the 't Hooft coupling $\lambda \equiv g_{YM}^2 N_c$ fixed, followed by the large-coupling limit $\lambda \gg 1$. We will always work in the probe limit: we keep $N_f$ fixed as $N_c \to \infty$, expand all observables in the small parameter $N_f/N_c$, and retain terms only up to order $N_f N_c$.

The R-symmetry of $\N=4$ SYM is $SO(6)$, corresponding to rotations in the $(X^1,\dots,X^6)$ directions. The $(n+1)$-dimensional flavor fields break the $SO(6)$ to $SO(n+1) \times SO(5-n)$, corresponding to rotations in the $(X^1,\dots,X^{n+1})$ and $(X^{n+2},\dots,X^6)$ directions, respectively. In the $n=2$ case the $SO(3)\times SO(3)$ is the R-symmetry of the remaining $(2+1)$-dimensional $\N=4$ supersymmetry. In the $n=3$ case an $SU(2) \times U(1)$ subgroup of $SO(4) \times U(1)$ is the R-symmetry of the remaining $\N=2$ supersymmetry.

We want to study thermal equilibrium states of these theories at temperature $T$. A thermal state of large-$N_c$, strongly-coupled $\N=4$ SYM is dual to type IIB supergravity in the near-horizon geometry of non-extremal D3-branes, $AdS_5$-Schwarzschild times an $S^5$, whose metric takes the form
\bea
\label{eq:backgroundmetric}
ds^2 & = & g_{tt}(r) dt^2  + g_{xx}(r) \left( dx^2+dy^2+dz^2\right) + g_{rr}(r) dr^2 + R^2 ds^2_{S^5} \nonumber \\
& = & -\frac{r^2}{R^2}f(r)dt^2+\frac{r^2}{R^2}\left(dx^2+dy^2+dz^2\right)+\frac{R^2}{r^2}\frac{dr^2}{f(r)}+R^2ds^2_{S^5},
\eea
where
\beq
\label{eq:metricfdef}
f(r)=1-\frac{r_H^4}{r^4}.
\eeq
In eqs.~\eqref{eq:backgroundmetric} and~\eqref{eq:metricfdef}, $r$ is the $AdS_5$ radial coordinate, with the $AdS_5$ boundary at $r \to \infty$ and the planar Schwarzschild horizon at $r=r_H$, $R$ is the $AdS_5$ radius of curvature, which is fixed to be $R^4 = 4 \pi g_s N_c \alpha'^2$, with $\alpha'$ the string length squared, and $ds_{S^5}^2$ is the metric of a unit-radius $S^5$. The full type IIB supergravity solution includes $N_c$ units of Ramond-Ramond (RR) five-form flux on the $S^5$. The Hawking temperature $T$ of this planar black hole is given by $r_H=\pi T R^2$. We also identify $T$ as the temperature of the dual field theory state~\cite{Witten:1998zw}. Starting now, we use units in which $R\equiv1$. In these units we can convert between supergravity and field theory quantities using $\alpha'=1/\sqrt{\lambda}$.

The massless probe flavors are dual to probe D$p$-branes extended along $AdS_{n+2} \times S^n$ inside the $AdS_5$-Schwarzschild times $S^5$, where the $AdS_{n+2}$ is spanned by $\left(t,x,y,r\right)$ and for $n=3$ also $z$, and the $S^n \in S^5$~\cite{Karch:2002sh,Karch:2000gx}. The $S^5$ has an $SO(6)$ isometry, dual to the $SO(6)$ R-symmetry of $\N=4$ SYM. The D$p$-brane breaks that to $SO(n+1)\times SO(5-n)$, dual to the corresponding global symmetries in the field theory. The $SO(n+1)$ factor is the isometry of the $S^n$.

The non-Abelian D$p$-brane action is not known in full generality, however in what follows we will only need the Abelian D$p$-brane action, which consists of two types of terms, a Dirac-Born-Infeld (DBI) term and Wess-Zumino (WZ) terms describing the coupling to background RR fields. In what follows the WZ terms will play no role, so we will omit their explicit forms. The action for our D$p$-branes, $S_{\textrm{D}p}$, is thus the DBI term alone,
\beq
\label{eq:probebraneaction}
S_{\mathrm{D}p}=-N_fT_{\mathrm{D}p}\int d^{p+1}\xi\sqrt{-\det\left(g_{ab}^{\mathrm{D}p}+F_{ab}\right)},
\eeq
where $T_{\mathrm{D}p} = (2\pi)^{-p}g_s^{-1} \alpha'^{-(p+1)/2}$ is the D$p$-brane tension, $\xi^a$ with $a,b=0,1,\ldots, p$ are the D$p$-brane worldvolume coordinates, $g_{ab}^{\mathrm{D}p}$ is the induced metric of the D$p$-brane, and $F_{ab}$ is the field strength of the $U(1)$ gauge field $A_a$ on the D$p$-brane.\footnote{Notice that with respect to the usual conventions (such as those of ref.~\cite{Polchinski:1998rq}) we have absorbed a factor of $(2\pi\alpha')$ into the field strength.}

The (9-$p$) scalars propagating on the worldvolume of the D$p$-brane, which describe the embedding of the D$p$-brane in the background, enter $S_{\mathrm{D}p}$ via the induced metric $g_{ab}^{\mathrm{D}p}$. In our cases, of the (9-$p$) scalars, the $5-n$ describing the position of the D$p$-brane on the $S^5$ form a vector of $SO(5-n)$, one of which is holographically dual to the supersymmetric mass operator $\Om$. We will always work with trivial worldvolume scalars,\footnote{The one and only place where we discuss solutions with non-trivial worldvolume scalars is at the end of this subsection, when we briefly review the solutions of refs.~\cite{Evans:2010iy,Jensen:2010vd,Jensen:2010ga,Evans:2010hi}.} and as a gauge choice we will identify the worldvolume coordinates with the background coordinates along the $AdS_{n+2}\times S^n$ spanned by the D$p$-brane. In that case, $g_{ab}^{\mathrm{D}p}$ coincides with the background metric in the $AdS_{n+2}\times S^n$ subspace. 

The conserved $U(1)_b$ current $J^{\nu}$ is dual to $A_{\nu}$. We want to study states with non-zero $\jt$ and $B$. We will demand that these states preserve various symmetries: time translations, $n$-dimensional translations, $SO(n+1)\times SO(5-n)$, and the reflection symmetry $z \rightarrow -z$. To describe non-zero $\jt$ and preserve these symmetries, we introduce $A_t(r)$. Employing $A_r=0$ gauge, we thus have a worldvolume electric flux $F_{rt}(r) = A_t'(r)$, where prime denotes $\partial_r$. To describe non-zero $B$ and preserve these symmetries, we introduce the magnetic flux $F_{xy}=B$, which must be constant to satisfy the Bianchi identity. With these fluxes and with trivial worldvolume scalars, the D$p$-brane action reduces to
\beq
\label{eq:dpactionansatz1}
S_{\mathrm{D}p} = - \N_p \, V_{\mathbb{R}^{(n,1)}} \int dr \, g_{xx}^{(n-2)/2} \sqrt{\left(g_{xx}^2 + B^2 \right)\left(g_{rr}|g_{tt}| - A_t'^2\right)},
\eeq
where $\N_p\equiv N_f T_{\textrm{D}p} V_{S^n}$, with $V_{S^n}$ the volume of a unit-radius $S^n$. Explicitly,
\beq
\label{eq:npdef}
\N_p \equiv N_f T_{\mathrm{D}p} V_{S^n} = \left\{ 
  \begin{array}{l l}
    \frac{4\sqrt{\lambda}}{(2\pi)^3} N_f N_c, & \quad p=5,\\
   \frac{\lambda}{(2\pi)^4} N_f N_c & \quad p=7,\\
  \end{array} \right.
\eeq
where in the second equality we have converted to field theory quantities. In eq.~\eqref{eq:dpactionansatz1}, $V_{\mathbb{R}^{(n,1)}}$ is the (infinite) volume of $\mathbb{R}^{(n,1)}$. Starting now, we will divide both sides of eq.~\eqref{eq:dpactionansatz1} by $V_{\mathbb{R}^{(n,1)}}$, and in an abuse of notation we will use $S_{\mathrm{D}p}$ to denote the resulting action density, $S_{\mathrm{D}p}/V_{\mathbb{R}^{(n,1)}} \to S_{\mathrm{D}p}$, which we will henceforth refer to as the D$p$-brane action.

Since $S_{\mathrm{D}p}$ depends only on the derivative of $A_t(r)$, we obtain a constant of integration, $\frac{\delta S_{\mathrm{D}p}}{\delta A_t'}$, which as shown in refs.~\cite{Kobayashi:2006sb,Karch:2007br} determines the charge density in the field theory:
\beq
\jt = (2\pi\alpha')\frac{\delta S_{\mathrm{D}p}}{\delta A_t'} = \N_p \, g_{xx}^{(n-2)/2} \sqrt{g_{xx}^2 + B^2} \frac{(2\pi\alpha')A_t'}{\sqrt{g_{rr}|g_{tt}| - A_t'^2}}.
\eeq
Solving for $A_t'$, we find
\beq
\label{eq:atsol}
A_t'(r) = \frac{d \sqrt{g_{rr}|g_{tt}|}}{\sqrt{g_{xx}^{n-2} \left(g_{xx}^2 + B^2\right)+d^2}},
\eeq
where
\beq
d \equiv \frac{\jt}{(2\pi\alpha')\N_p}.
\eeq
We will use ``charge density'' to refer to either $\jt$ or $d$.

We will now review various properties that characterize the compressible states described by the solution in eq.~\eqref{eq:atsol}, as determined from holographic calculations. We begin with low but non-zero temperatures, $T \ll d^{1/n}$, and with $B=0$. In this regime, in these compressible states neither $n$-dimensional translations nor $U(1)_b$ are broken. In the canonical ensemble, the chemical potential is~\cite{Karch:2008fa}
\beq
\label{eq:muzeroT}
\mu(T,d) = \frac{\Gamma\left(\frac{1}{2}-\frac{1}{2n}\right)\Gamma\left(1+\frac{1}{2n}\right)}{\Gamma(1/2)} \, d^{1/n} - (\pi T) \left[ 1 + O\left(T^{2n}/d^2\right)\right].
\eeq
From eq.~\eqref{eq:muzeroT}, we see that when $T=0$ the density $d^{1/n}\propto\mu$, so the density varies smoothly as a function of $\mu$, \textit{i.e.}\ these states are indeed compressible, and a large density implies a large chemical potential. For convenience, we will denote the value of the chemical potential at $T=0$ as $\mu_0$,
\beq
\label{eq:mu0def}
\mu_0 \equiv  \frac{\Gamma\left(\frac{1}{2}-\frac{1}{2n}\right)\Gamma\left(1+\frac{1}{2n}\right)}{\Gamma(1/2)} \, d^{1/n}.
\eeq
At low temperatures, the leading density-dependent contribution to the heat capacity is $c_V \propto T^{2n}/d$~\cite{Karch:2008fa}, which is markedly different from that of a LFL, where $c_V \propto T$, or a gas of free bosons, where $c_V \propto T^n$, while the leading contribution to the $U(1)_b$ resistivity is $\rho \propto T^2$~\cite{Karch:2007pd}, the same as in a LFL.

Next let us consider $T=0$, still with $B=0$.  The system actually has a non-zero $T=0$ entropy density $s \propto \jt/\sqrt{\lambda}$~\cite{Karch:2008fa}, indicating some degeneracy of states, in stark contrast to solids, superfluids, and LFLs, which are all unique ground states. Degeneracy suggests instability, since generically any perturbation will break the degeneracy and drive the system to a new, presumably unique, ground state. These states are known to be thermodynamically stable, in the sense that they are at least local minima of the free energy~\cite{Benincasa:2009be}. As shown for the $n=3$ case in ref.~\cite{Ammon:2011hz}, these states are also stable against dynamical (non-zero $\omega$ and $k$) fluctuations. To date, no Fermi surfaces have been detected in these states~\cite{Jensen:2010vd,Ammon:2011hz,Goykhman:2012vy}. Remarkably, these states are in fact merely isolated points in an entire moduli space of compressible states parameterized by the expectation values of certain scalar operators, as 
discussed in detail in refs.~\cite{Chang:2012ek,Ammon:2012mu}. The existence of the moduli space may be related to the non-zero $T=0$ entropy density~\cite{Ammon:2012mu}.

All of the above facts together indicate that these states are not solids, superfluids, LFLs, or NFLs. Indeed, as proposed in ref.~\cite{Karch:2008fa}, these states may be examples of some new kind(s) of compressible matter. Some evidence even suggests that the low-energy theory describing excitations about these states may be a (0+1)-dimensional CFT~\cite{Jensen:2010ga,Nickel:2010pr,Ammon:2011hz,Goykhman:2012vy}.

Now let us turn to the thermodynamics of these states with non-zero $B$. For the $n=3$ case, a non-zero $B$ explicitly breaks parity and charge conjugation symmetries, and breaks rotational symmetry from $SO(3)$ down to $SO(2)$. Moreover, with $T=0$ and non-zero $\jt$, a sufficiently large $B$ triggers a quantum phase transition~\cite{Evans:2010iy,Jensen:2010vd}. When $B$ is below a critical value, $B<B_c$, in the ground state $\jt$ is non-zero but $\Omv$ is zero. When $B>B_c$, the ground state changes such that $\jt$ and $\Omv$ are both nonzero. In the bulk, the worldvolume scalar dual to $\Om$ becomes nontrivial when $B\geq B_c$. The non-zero $\Omv$ signals spontaneous breaking of the $SU(2)\times U(1)$ symmetry down to $SU(2)$. The system thus exhibits a quantum phase transition. Notice that $U(1)_b$ is unbroken for all $B$. A numerical analysis indicates that the critical value of $B$ is $B_c \approx 2.138 \, d^{2/3}$ and that the transition is not first order, \textit{i.e.} is at least second order~\cite{Jensen:2010vd}. At non-zero $T$, the quantum critical point gives rise to a quantum critical region~\cite{Evans:2010iy,Jensen:2010vd}.

For the $n=2$ case, a non-zero $B$ breaks parity and charge conjugation symmetries. Here again, with $T=0$ and non-zero $\jt$, a sufficiently large $B$ triggers a quantum phase transition~\cite{Jensen:2010ga,Evans:2010hi}: for $B<B_c$ $\jt$ is non-zero and $\Omv=0$, while for $B>B_c$ both $\jt$ and $\Omv$ are non-zero, and the non-zero $\Omv$ spontaneously breaks the $SO(3) \times SO(3)$ symmetry down to $SO(3) \times U(1)$. In the bulk, the worldvolume scalar dual to $\Om$ becomes nontrivial when $B\geq B_c$. The critical value of $B$ is known exactly, $B_c = d/\sqrt{7} \approx 0.378 \, d$, and the transition is of BKT-type: schematically, as $B$ increases past $B_c$, $\Omv \propto e^{-\sqrt{d/B_c-d/B}}$~\cite{Jensen:2010ga}. At any non-zero $T$ the transition becomes second order, and the quantum critical point gives rise to a quantum critical region~\cite{Jensen:2010ga,Evans:2010hi}.

We hasten to add that we have only reviewed the \textit{currently-known} phase diagrams of these systems. For given values of $T$, $B$, and $d$, other states may exist with lower free energy than any of the states mentioned above. Indeed, in the $n=3$ case, when $B$ is non-zero an instability has been detected at a single point in the grand canonical phase diagram, $(T,\mu) = (\frac{\sqrt{2}}{4\pi}\sqrt{B},\sqrt{B})$~\cite{Kharzeev:2011rw}, where the known ground state has non-zero $\jt$ and $\Omv=0$. The instability has non-zero $k$, suggesting the existence of an \textit{inhomogeneous} state with lower free energy than the known ground state \cite{Kharzeev:2011rw}.  In what follows, we always work in regions of the phase diagram where the known ground state has non-zero $\jt$ and $\Omv=0$.  In bulk terms, we exclusively consider values of $T$, $B$ and $d$ for which the only known solutions for the D$p$-brane worldvolume scalars are the trivial solutions.

\section{Spectral Functions from Holography}
\label{eoms}

To probe a many-body system experimentally, we can perturb the system by a source for some operator. The simplest perturbations to study have either a fixed frequency or a fixed momentum. Within linear response theory the change in the expectation value of an operator, such as our $U(1)_b$ current $J^{\mu}$, is given in terms of the small-amplitude source $\alpha_{\nu}(\omega,k)$ by
\beq
\label{Eq: Greensratio}
\delta\left\langle J^{\mu}(\omega,k) \right\rangle = G_{R}^{\mu \nu}(\omega,k) \, \alpha_{\nu}(\omega,k),
\eeq
where $G_{R}^{\mu \nu}(\omega,k)$ is the retarded Green's function. In general, $G_{R}^{\mu \nu}(\omega,k)$ is a complex function of the two real variables, $\omega$ and $k$. If we na\"ively complexify either $\omega$ or $k$, then often poles appear in $G_{R}^{\mu \nu}(\omega,k)$, indicating a large response to an infinitesimal source. The position of each pole can be expressed as a function of the complex variable in terms of the real variable (as well as $T$, $B$, etc.); this complex function is called the dispersion relation of the excitation corresponding to the pole. Henceforth we will abuse notation and write the dispersion relations simply as $\omega\left(k\right)$ or $k\left(\omega\right)$. Using the Fourier decomposition, we can interpret $\textrm{Re} \,\omega(k)$ as the propagating frequency and $-\textrm{Im} \,\omega(k)$ as the decay rate of the excitation, or alternatively $\textrm{Re} \,k(\omega)$ as the propagating momentum and $\textrm{Im} \,k(\omega)$ as the attenuation, or decay length, of the excitation.

The poles described above are not directly observable: in a real experiment, neither $\omega$ nor $k$ can be complex. A common experimental technique for studying excitations of a many-body system is to measure the spectral function
\beq
\label{eq:spectralfuncdef}
\chi^{\mu \nu}(\omega,k) \equiv i \left[ G_R^{\mu \nu}(\omega,k)-G_R^{\mu \nu}(\omega,k)^{\dagger}\right],
\eeq
which is a real-valued function of real $\omega$ and $k$, and hence is observable.  Physically, $\chi^{\mu \nu}(\omega,k)$ determines the rate of work done on the system by the small external source $\alpha_{\nu}(\omega,k)$~\cite{Hartnoll:2009sz}. A pole in $G_R^{\mu \nu}(\omega,k)$ with sufficiently large residue will produce a large peak in $\chi^{\mu \nu}(\omega,k)$. We can thus identify excitations of the system from peaks in $\chi^{\mu \nu}(\omega,k)$. In general, the precise relation between the dispersion relation of a pole in $G_R^{\mu \nu}(\omega,k)$ and the location of the corresponding peak in $\chi^{\mu \nu}(\omega,k)$ is complicated, but typically a pole with dispersion relation $\omega\left(k\right)$ produces a peak in $\chi^{\mu\nu}\left(\omega\right)$ centred near $\textrm{Re} \,\omega(k)$, with width proportional to $-\textrm{Im} \,\omega(k)$. Notice that given $\chi^{\mu\nu}\left(\omega,k\right)$ we can reconstruct $\textrm{Re} \, G_R^{\mu \nu}(\omega,k)$ using the Kramers-Kronig relation, provided the large-$\omega$ and large-$k$ asymptotics have been suitably regulated~\cite{Hartnoll:2009sz}.

In holography the on-shell bulk action is the field theory generating functional~\cite{Witten:1998qj,Gubser:1998bc}, hence to compute a two-point function such as $G^{\mu\nu}_R(\omega,k)$, we need to compute the bulk on-shell action and take two functional derivatives, which in turn means we need to solve the linearized equation of motion for the bulk fields. We thus introduce fluctuations of all the D$p$-brane worldvolume fields, including all components of $A_a$ and all (9-$p$) scalars. The fluctuations form representations of the symmetry preserved by the background solution. At the linearized level different representations cannot couple. In our case the background solution eq.~\eqref{eq:atsol} preserves $SO(n+1)\times SO(5-n)$ and the reflection symmetry $z\to-z$. Under $SO(n+1)$, the components of $A_a$ on the $S^n$ are in the $n+1$ while all other fluctuations are singlets. Under $SO(5-n)$, the worldvolume scalars describing the position of the D$p$-brane on the $S^5$ are in the $5-n$ and all other fluctuations are singlets. We are interested only in $J^{\nu}$, which is a singlet under $SO(n+1)\times SO(5-n)$, and so is dual to a fluctuation of $A_{\nu}$ that is also a singlet, and thus has no dependence on the $S^n$ directions and will not couple to fluctuations of $A_a$ on the $S^n$ or to the worldvolume scalars describing the position of the D$p$-brane on the $S^5$. In the $n=2$ case, under the $z\to-z$ symmetry the fluctuation of the worldvolume scalar describing the position of the D5-brane in the $z$ direction is clearly odd, and hence does not couple to the $a_{\nu}$, which are even. We thus only need to consider the fluctuations $a_{\nu}$ of $A_{\nu}$,
\beq
\label{eq:aflucdef}
A_{\nu}(r,x^{\mu}) = A_t(r)\,\delta^t_{~\nu} + B x \,\delta^y_{~\nu} +a_{\nu}(r,x^{\mu}).
\eeq
On the right-hand-side of eq.~\eqref{eq:aflucdef} the first two terms produce the electric and magnetic fluxes $F_{rt}(r)$ and $F_{xy}=B$ of the background solution. We will consider fluctuations $a_{\nu}$ that depend on $r$, $t$, and $x$ only. The background solution preserves time translations and $n$-dimensional spatial translations, allowing us to perform a Fourier decomposition,
\beq
a_{\nu}(r,t,x) = \int \frac{d\omega dk}{(2\pi)^2} \, a_{\nu}(r,\omega,k) \, e^{ - i \omega t + i k x }.
\eeq
To preserve $A_r=0$ gauge, we take $a_r=0$.

The background solution preserves $SO(2)$ rotational symmetry in the $(x,y)$ plane. Under this $SO(2)$, $a_t$ is a singlet while $a_x$ and $a_y$ together form a doublet. Our choice of momentum only in $x$ breaks this $SO(2)$. In the D3/D5 system we make that choice with no loss of generality, but in the D3/D7 system we are restricting to a special subset of fluctuations with momentum perpendicular to the magnetic field. In the D3/D7 system, our choice of momentum preserves the $z\to-z$ symmetry, under which $a_z$ is odd and hence will not couple to $a_t$, $a_x$, or $a_y$, which are even.  When $B=0$, with our choice of momentum $a_t$ and $a_x$ can couple to one another, but parity forbids these from coupling to $a_y$. When $B$ is non-zero and parity is broken, all three fluctuations may couple to one another, and indeed do, as we will see shortly.

If different components of $a_{\nu}$ in the bulk couple, then the dual operators will mix, and in particular the poles in different components of $G_{R}^{\mu \nu}(\omega,k)$ will mix. These poles will be shared by all the components of $G_{R}^{\mu \nu}(\omega,k)$ that mix, but the spectral functions can still be different since the \textit{residues} of the poles may be different for each component.

The quadratic action of the fluctuations, $S_{a^2}$, and the resulting equations of motion are straightforward to obtain but unilluminating, so we will omit them. We will record the equation for $a_r$, which in $a_r=0$ gauge is a constraint on the other fluctuations,
\beq
\label{eq:areom}
\omega \, a_t' + u(r)^2 k \, a_x' =0,
\eeq
where
\beq
u(r)^2 \equiv \frac{|g_{tt}|g_{rr} - A_t'^2}{\frac{g_{rr}}{g_{xx}} \left(g_{xx}^2 + B^2\right)} = \frac{|g_{tt}|g_{xx}^{n-1}}{g_{xx}^{n-2}\left(g_{xx}^2 + B^2\right)+d^2}.
\eeq
With our choice of momentum, the gauge-invariant fluctuations are $a_y$ itself and also the electric field (in Fourier space)~\cite{Kovtun:2005ev}
\beq
E(r,\omega,k) \equiv k \, a_t(r,\omega,k) + \omega \, a_x(r,\omega,k),
\eeq
which is dual to the operator
\beq
J^E \equiv k \, J^t + \omega \, J^x.
\eeq
In terms of these gauge-invariant fluctuations, using eq.~\eqref{eq:areom} and performing an integration-by-parts, we can write $S_{a^2}$ as
\bea
\label{eq:saquared}
S_{a^2} &=& \frac{\N_{p}}{2} \int dr \frac{d\omega dk}{(2 \pi)^2} \frac{|g_{tt}| g_{xx}^{(n+1)/2}}{u(r) g_{rr}^{1/2} \left(g_{xx}^2 + B^2\right)} \times \\
& & \left[ \frac{1}{\omega^2 - u(r)^2 k^2} |E'|^2 - \frac{g_{rr}}{|g_{tt}|} |E|^2 - |a_y'|^2 + \frac{g_{rr}}{|g_{tt}|} \left( \omega^2 - u(r)^2 k^2 \right) |a_y|^2 \right . \nonumber \\
& & \left . + i B \frac{u(r) g_{rr}^{1/2} \left(g_{xx}^2 + B^2\right)}{|g_{tt}| g_{xx}^{(n+1)/2}} \partial_{r} \left(\frac{g_{xx}^{(n-1)/2} g_{rr}^{-1/2} A_t'}{u(r)\left(g_{xx}^2 + B^2\right)}\right) \left( E \, a_y^* + a_y \, E^* \right) \right ], \nonumber
\eea
where $E^*$ and $a_y^*$ denote complex conjugates. For the D7-brane, in eq.~\eqref{eq:saquared} we have omitted a boundary term produced by the integration-by-parts since this boundary term vanishes on-shell. The equations of motion that follow from $S_{a^2}$ are
\begin{subequations}
\bea
\label{eq:Eeom}
E'' &+& \left [ \partial_r \log \left( \frac{g_{xx}^{(n+1)/2}|g_{tt}| g_{rr}^{-1/2}}{\left(\omega^2 - u(r)^2 k^2\right)u(r)\left(g_{xx}^2 + B^2\right)} \right)\right] E' + \frac{g_{rr}}{|g_{tt}|}\left(\omega^2 - u(r)^2 k^2\right) E \\
&=& + i B \left[\frac{u(r)\left(g_{xx}^2 + B^2\right)g_{rr}^{1/2}}{|g_{tt}|g_{xx}^{(n+1)/2}}\right] \left [ \partial_r \left(\frac{g_{xx}^{(n-1)/2} g_{rr}^{-1/2} A_t'}{u(r)\left(g_{xx}^2 + B^2\right)}\right)\right]  \left(\omega^2 - u(r)^2 k^2\right) a_y, \nonumber \\
& & \nonumber \\
\label{eq:ayeom}
a_y'' &+& \left [ \partial_r \log \left(\frac{g_{xx}^{(n+1)/2}|g_{tt}|g_{rr}^{-1/2}}{u(r)\left(g_{xx}^2 + B^2\right)}\right) \right] a_y'  + \frac{g_{rr}}{|g_{tt}|} \left(\omega^2 - u(r)^2 k^2\right)a_y \nonumber \\
&=& - i B \left[ 	\frac{u(r) \left(g_{xx}^2 + B^2\right) g_{rr}^{1/2}}{|g_{tt}|g_{xx}^{(n+1)/2}}\right]\left[\partial_r \left(\frac{g_{xx}^{(n-1)/2}g_{rr}^{-1/2}A_t'}{u(r)\left(g_{xx}^2 + B^2\right)}\right)\right]E.
\eea
\end{subequations}
When $B=0$, these reduce to the equations of motion in refs.~\cite{Karch:2008fa,Myers:2008me,Davison:2011ek}, and when $B$ is non-zero, these agree with the equations of motion in refs.~\cite{Wapler:2009tr,Pal:2012gr,Goykhman:2012vy}. Notice in particular that non-zero $B$ leads to couplings between $E$ and $a_y$, as advertised.

All of our results will follow from solutions of eqs.~\eqref{eq:Eeom} and~\eqref{eq:ayeom}, with appropriate boundary conditions. Eqs.~\eqref{eq:Eeom} and~\eqref{eq:ayeom} are second-order, hence for each field, $E$ and $a_y$, we need two boundary conditions to specify a solution completely. At the future horizon, a solution for $E$ or $a_y$ looks like a linear combination of in-going and out-going waves, with some normalizations. To obtain the \textit{retarded} Green's function, we require that the fields be purely in-going~\cite{Son:2002sd,Policastro:2002se,Skenderis:2008dh,vanRees:2009rw}. For each field, the normalization then provides us with a second boundary condition. Inserting in-going solutions into the action $S_{a^2}$, we obtain $G^{\mu\nu}_R(\omega,k)$ via functional differentiation,\footnote{Generically in AdS/CFT the on-shell action is divergent due to integration all the way to the AdS boundary. To compute renormalized correlation functions from the on-shell action, we first regulate the divergence by introducing a cutoff in $r$, introduce covariant counterterms localized on the cutoff surface, and then remove the regulator~\cite{Skenderis:2002wp}. In our cases, the only near-boundary divergence that appears is $\propto F_{\mu\nu}F^{\mu\nu} \log r$ in the $S_{a^2}$ of the D7-brane, which we cancel with a counterterm. As discussed in ref.~\cite{Jensen:2010vd}, a finite counterterm $\propto F_{\mu\nu}F^{\mu\nu}$ is also present, with unfixed coefficient. As in ref.~\cite{Jensen:2010vd} we set the coefficient of the finite counterterm to zero. Our $G^{\mu\nu}_R(\omega,k)$ in eq.~\eqref{eq:greensholodef} is then the renormalized Green's function.}
\beq
\label{eq:greensholodef}
G^{\mu\nu}_R(\omega,k) = \lim_{r \to \infty} \frac{\delta^2 S_{a^2}}{\delta a_{\mu} \delta a_{\nu}^*}.
\eeq
Notice that
\beq
G_{R}^{tt}(\omega,k) = \lim_{r \to \infty} \frac{\delta^2 S_{a^2}}{\delta a_t \delta a_t^*} = \lim_{r \to \infty}  \left|\frac{\delta E}{\delta a_t}\right|^2 \frac{\delta^2 S_{a^2}}{\delta E \,\delta E^*}  = k^2 G_{R}^{EE}(\omega,k),
\eeq
so any pole in $G_{R}^{EE}(\omega,k)$ not of the form $1/k^2$ will produce a pole in $G_{R}^{tt}(\omega,k)$. Given $G^{\mu\nu}_R(\omega,k)$, we obtain the spectral function $\chi^{\mu\nu}(\omega,k)$ via its definition, eq.~\eqref{eq:spectralfuncdef}.

Most of our results will be numerical. We will use the numerical method of ref.~\cite{Kaminski:2009dh} to extract the locations of poles in $G^{tt}_R(\omega,k)$ and the value of $\chi^{tt}(\omega,k)$, which we will now briefly review. Let us define a vector of fields
\beq
\vec{V}(r,\omega,k)\equiv\begin{pmatrix} E(r,\omega,k) \\ a_y(r,\omega,k) \end{pmatrix}.
\eeq
We need four boundary conditions to specify a solution for $\vec{V}$ completely. As stated above, at the future horizon we choose solutions to be purely in-going waves. We next need to fix the normalizations of these in-going waves. Explicitly, the vector of near-horizon normalization factors, $\vec{V}_{\mathrm{nh}}$, is, when $T$ is non-zero,\footnote{We use non-zero $T$ in all of our numerical analysis. In what follows, we consider $T=0$ only in section~\ref{hzsnonzerob}, where we compute the dispersion relation of holographic zero sound at $T=0$. In that case the near-horizon behavior of $\vec{V}(r,\omega,k)$ is different from that at non-zero $T$: see eq.~\eqref{Eq: NHExpansion}.}
\beq
\label{eq:vnhdef}
\vec{V}_{\mathrm{nh}} \equiv \lim_{r \rightarrow r_{H}} \, \exp \left( i \omega \int dr \sqrt{g_{rr}/|g_{tt}|} \right) \vec{V}(r,\omega,k).
\eeq
Notice that $\vec{V}_{\mathrm{nh}}$ is constant, independent of $r$, $\omega$, and $k$. On the right-hand-side of eq.~\eqref{eq:vnhdef}, the exponential factor is designed to \textit{cancel} the exponential factor that represents an in-going wave at the future horizon.

In the method of ref.~\cite{Kaminski:2009dh}, we first pick any convenient value of $\vec{V}_{\mathrm{nh}}$ and solve the equations, then we pick another $\vec{V}_{\mathrm{nh}}$, linearly independent of the first, and solve the equations, thus constructing a basis of solutions. We can then write any solution in that basis. For example, suppose we solve the equations twice, with two normalizations $\vec{V}^{(1)}_{\mathrm{nh}}=(1,0)^T$ and $\vec{V}^{(2)}_{\mathrm{nh}}=(0,1)^T$. Let the corresponding solutions be $\vec{V}^{(1)}$ and $\vec{V}^{(2)}$. Using these solutions, we define a $2\times2$ matrix $P(r,\omega,k)$ as
\beq
\label{eq:pmatrixdef}
P(r,\omega,k) \equiv \left(\vec{V}^{(1)}(r,\omega,k),\vec{V}^{(2)}(r,\omega,k)\right),
\eeq
allowing us to write any solution as
\beq
\label{Eq:Diffproblem}
\vec{V}(r,\omega,k) = P(r,\omega,k) \, \vec{V}_{\mathrm{nh}},
\eeq
for some choice of $\vec{V}_{\mathrm{nh}}$. Inserting a solution for $\vec{V}(r,\omega,k)$ of the form in eq.~\eqref{Eq:Diffproblem} into $S_{a^2}$ and using eq.~\eqref{eq:greensholodef}, we can express $G_{R}^{\mu\nu}(\omega,k)$ in terms of $P(r,\omega,k)$,
\beq
\label{eq:holoresultgr}
G_{R}^{\mu\nu}(\omega,k) = - \N_{p} \lim_{r \rightarrow \infty} r^n\left[ \left( \begin{array}{cc} \left(-\omega^2 + k^2\right)^{-1} & 0 \\ 0 & 1 \end{array} \right)P'(r,\omega,k) P^{-1}(r,\omega,k)\right]^{\mu\nu}.
\eeq

Generically, a pole will occur in $G_R^{\mu\nu}(\omega,k)$ when $\lim_{r\to\infty} \det P^{-1}(r,\omega,k)=\infty$, or equivalently when $\lim_{r\to\infty} \det P(r,\omega,k)=0$. We can thus locate poles in $G_R^{\mu\nu}(\omega,k)$ simply by computing $\lim_{r\to\infty} \det P(r,\omega,k)$, rather than by computing the entire retarded Green's function, eq.~\eqref{eq:holoresultgr}. A pole in $G_R^{\mu\nu}(\omega,k)$ corresponds to a quasi-normal mode in the bulk, that is, a non-trivial solution for a gauge-invariant fluctuation with the in-going boundary condition at the horizon and a Dirichlet boundary condition at the AdS boundary~\cite{Kovtun:2005ev}. Explicitly, imposing the Dirichlet condition at the AdS boundary means
\beq
\label{eq:qnmdef}
\lim_{r \to \infty} \vec{V}(r,\omega,k) = \lim_{r \to \infty} P(r,\omega,k) \vec{V}_{\mathrm{nh}} = 0.
\eeq
If the solution $\vec{V}(r,\omega,k)$, and hence $\vec{V}_{\mathrm{nh}}$, is non-trivial, then eq.~\eqref{eq:qnmdef} is satisfied only when $\lim_{r\to\infty} \det P(r,\omega,k)=0$, demonstrating that a pole in $G_R^{\mu\nu}(\omega,k)$ is dual to a quasi-normal mode in the bulk.

Let us summarize our numerical procedure. First, we choose values of $T$, $d$, $B$, $\omega$, and $k$. Next we solve the equations of motion, eqs.~\eqref{eq:Eeom} and ~\eqref{eq:ayeom}, twice, with two linearly-independent values of $\vec{V}_{\mathrm{nh}}$. With the two resulting solutions we construct the matrix $P(r,\omega,k)$ as in eq.~\eqref{eq:pmatrixdef}. At that point we can test for the presence of a pole in $G_R^{\mu\nu}(\omega,k)$ by testing whether $\lim_{r\to\infty} \det P(r,\omega,k)=0$. We obtain the spectral function $\chi^{\mu\nu}(\omega,k)$ via eqs.~\eqref{eq:holoresultgr} and~\eqref{eq:spectralfuncdef}. Repeating this process for various values of  $T$, $d$, $B$, $\omega$, and $k$, we obtained all of the numerical results that follow.

\section{Crossover at Zero Magnetic Field}
\label{zerob}

In this section we study the collisionless/hydrodynamic crossover in the D3/D5 system, in the limit $B=0$. The results are qualitatively similar to those in the D3/D7 theory~\cite{Davison:2011ek}. We begin by reviewing the crossover in a LFL, which provides a useful reference point for studying the collective density excitations of our systems.

\subsection{Collective Modes in a Landau Fermi Liquid}
\label{collectivelfl}

\begin{figure}[!htb]
  \centering
  \includegraphics[width=0.9\textwidth]{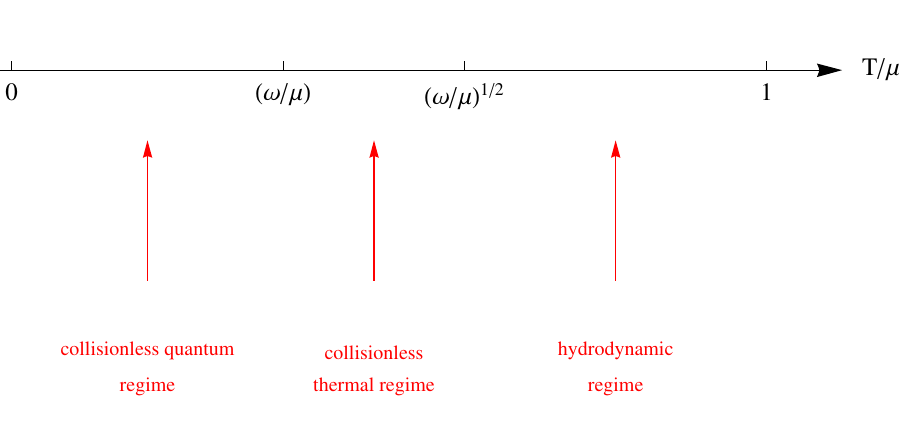}
  \caption{The three regimes of charge density transport in a LFL, with increasing $T/\mu$.}
  \label{fig:LFLregimes}
\end{figure}

In a LFL at $T=0$, the zero sound mode dispersion relation $k(\omega)$ at small $\omega/\mu$ has
\beq
\textrm{Re}\,k(\omega) \propto \omega, \qquad \textrm{Im}\,k(\omega) \propto \omega^2.
\eeq
If expressed instead as $\omega(k)$, then the dispersion relation at small $k/\mu$ has
\beq
\label{eq:lflzerosoundomega}
\textrm{Re}\,\omega(k) \propto k, \qquad \textrm{Im}\,\omega(k) \propto k^2.
\eeq
At non-zero $T$, the attenuation $\textrm{Im}\,k(\omega)$ of the zero sound mode changes as $T$ increases~\cite{pinesnozieres,zerosoundreview}. The attenuation $\textrm{Im}\,k(\omega)\propto\nu$, where $\nu$ is the quasiparticle collision frequency,
  \begin{equation}
	\label{eq:FLcollisionfrequency}
	\nu\propto\frac{\omega^2+\pi^2T^2}{\mu\left(1+e^{-\omega/T}\right)}.
  \end{equation}
On the right-hand-side of eq.~\eqref{eq:FLcollisionfrequency}, in the numerator the $\omega^2$ term is due to quantum interactions between the quasiparticles whereas the $T^2$ term is due to thermal scattering. At very low temperatures, $T\ll\omega$, the quantum term dominates and thus $\textrm{Im}\,k(\omega)\propto\omega^2/\mu$. This is called the ``collisionless quantum'' regime. As $T$ increases, the thermal term begins to dominate when $T\approx\omega$, at which point the system enters the ``collisionless thermal'' regime, where $\textrm{Im}\,k(\omega)\propto T^2/\mu$. At even higher temperatures, the collision frequency becomes significantly greater than the propagating frequency of the mode, $\nu\gg\omega$. In this regime, the mode is very short-lived and so does not contribute significantly to the low-energy properties of the system. However, in precisely this limit we expect hydrodynamic behaviour, such as hydrodynamic sound propagation and diffusion of charge, since the high frequency of thermal collisions brings about 
local thermal equilibrium. This collisionless/hydrodynamic crossover at $\omega\simeq T^2/\mu$ thus leads to a qualitative change in the response of the liquid to density perturbations. We summarize our discussion of the three regimes of a LFL with increasing $T$ in fig.~\ref{fig:LFLregimes}.

\subsection{Collective Modes in Holographic Quantum Liquids}
\label{collectiveholo}

We are interested in compressible states of the D3/D$p$ systems in the quantum liquid regime, that is, with a fixed, large density of matter present. We will thus normalise all dimensionful quantities by the appropriate power of the density, $d$, such that they are dimensionless. Specifically, we will define the dimensionless quantities
\beq
\label{eq:BarredVariables1}
\Tbar\equiv\frac{\pi T}{d^{1/n}}, \quad \wbar\equiv\frac{\omega}{d^{1/n}}, \quad \kbar\equiv\frac{k}{d^{1/n}}.
\eeq
We will always work with $\Tbar,\wbar,\kbar \ll1$. Similarly, $\bar{\chi}^{\mu \nu}(\omega,k)$ will denote the spectral function normalised to be dimensionless by an appropriate power of $d$. Notice that since $\Tbar\ll1$, normalising dimensionful quantities by powers of $d$ is equivalent, up to numerical factors, to normalising by appropriate powers of the chemical potential $\mu$, via eq.~\eqref{eq:muzeroT}.

Both of our holographic quantum liquids support a holographic zero sound excitation at $T=0$~\cite{Karch:2008fa}. For now, we will follow the convention of refs.~\cite{Karch:2008fa,Kulaxizi:2008kv,Nickel:2010pr,Ammon:2011hz,Davison:2011ek} and write the dispersion relation of the holographic zero sound in the form $\omega(k)$,
\beq
\label{eq:zeroTdispersionrelation}
\wbar=\pm \frac{1}{\sqrt{n}}\kbar-i\frac{1}{2n}\frac{d^{1/n}}{\mu_0}\,\kbar^2+O\left(\kbar^3\right).
\eeq
(Recall that $\mu_0$ was defined in eq.~\eqref{eq:mu0def}.) These modes are called holographic zero sound because this $T=0$ dispersion relation has $\textrm{Re}\,\omega(k) \propto k$ and $\textrm{Im}\,\omega(k) \propto k^2$, just like the zero sound excitation of a LFL, eq.~\eqref{eq:lflzerosoundomega}~\cite{Karch:2008fa}.

\begin{figure}[!htb]
\centering
\subfigure{\includegraphics[width=0.45\textwidth]{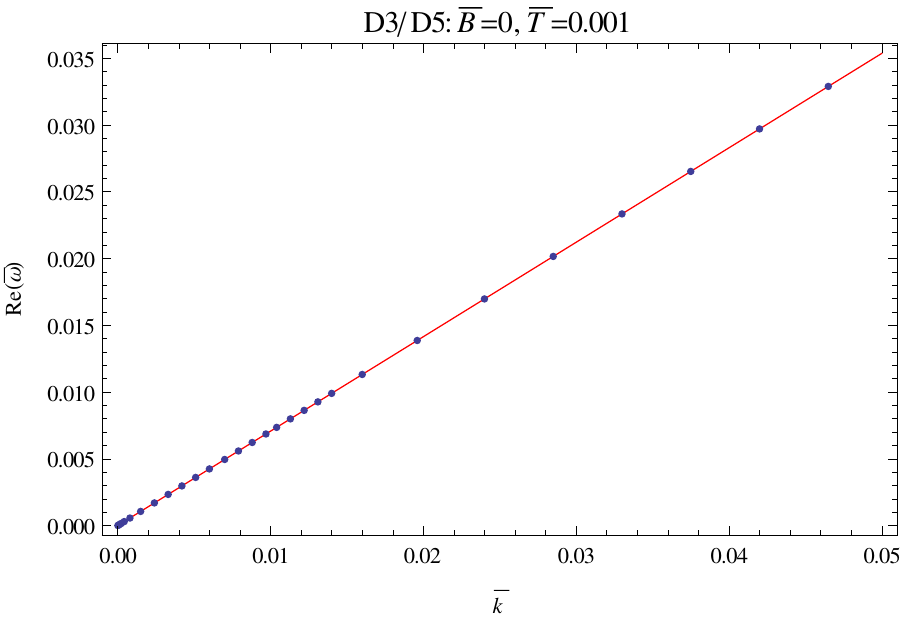}}
\subfigure{\includegraphics[width=0.45\textwidth]{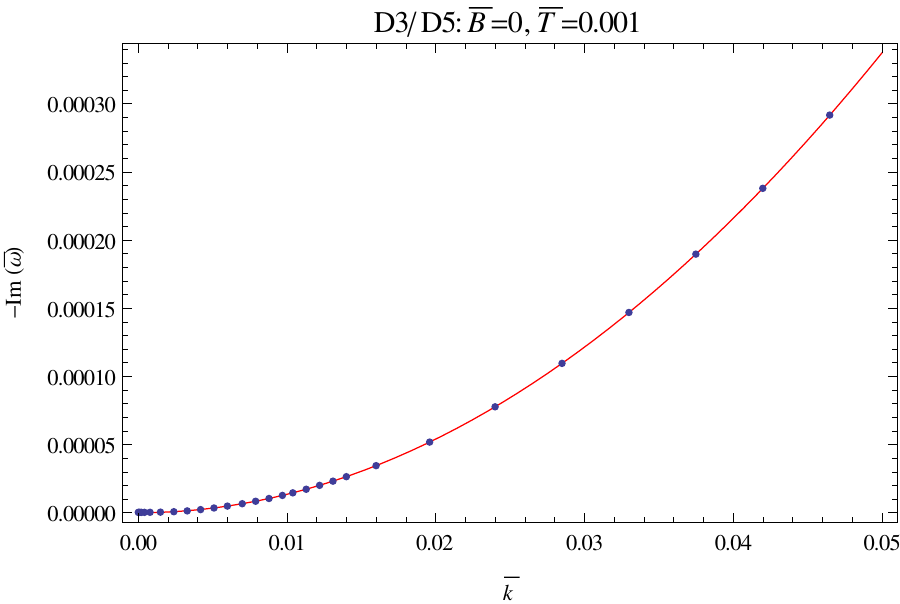}}
\caption{\textbf{Left:} $\textrm{Re}\,\wbar$ as a function of $\kbar$ for the holographic zero sound dispersion relation in the D3/D5 system at $B=0$ and $\Tbar = 0.001$, within the range $\Tbar^2\ll\Tbar\ll\kbar$. \textbf{Right:} $-\textrm{Im}\,\wbar$ versus $\kbar$ for the same mode at the same $\Tbar$. In both figures the dots are our numerical results while the red lines come from the $T=0$ dispersion relation in eq.~\eqref{eq:zeroTdispersionrelation}. Clearly the $T=0$ result in eq.~\eqref{eq:zeroTdispersionrelation} is a very good approximation to our numerical results for the low-temperature dispersion relation.}
  \label{fig:D3D5CQQNMs-ZeroB}
\end{figure}

The holographic zero sound mode persists at small but non-zero temperatures in the range $\Tbar^2\ll\Tbar\ll\kbar$, with a dispersion relation described to a very good approximation by the $T=0$ form, eq.~\eqref{eq:zeroTdispersionrelation}, as we show in fig.~\ref{fig:D3D5CQQNMs-ZeroB}. This regime is analogous to the collisionless quantum regime of a LFL, where quantum interactions of quasiparticles dominate thermal collisions and hence the decay rate of zero sound $-\textrm{Im}\,\omega(k)\propto k^2$.

\begin{figure}[!htb]
  \centering
  \includegraphics[width=0.45\textwidth]{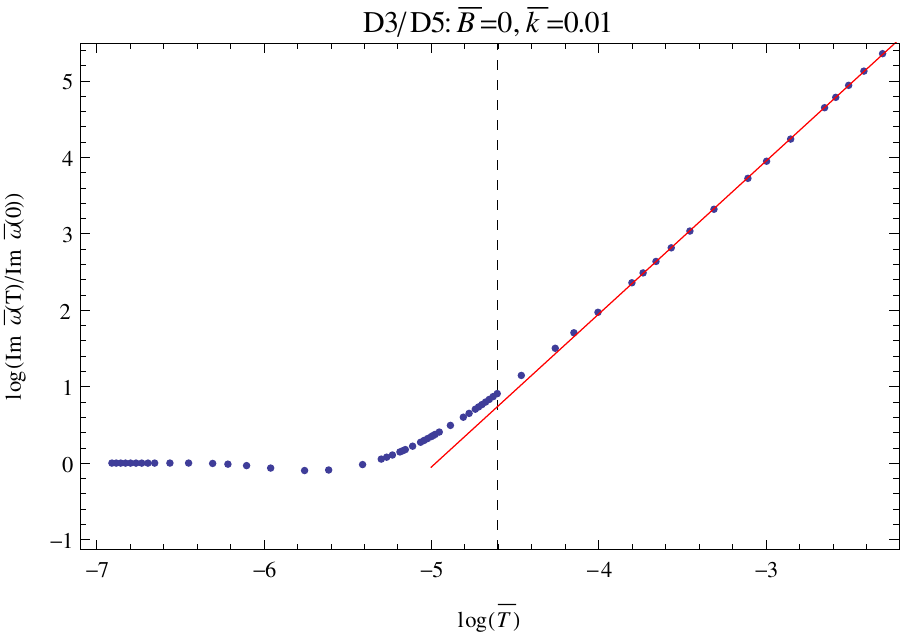}
  \caption{Logarithm of the decay rate of the holographic zero sound mode in the D3/D5 system at $B=0$, normalized to the $T=0$ value, $\log \left(\textrm{Im}\,\wbar(\Tbar)/\textrm{Im}\,\wbar(0)\right)$, as a function of $\log \Tbar$ for $\kbar=0.01$, within the regime $\Tbar^2\ll\kbar$. The dots are our numerical results, the vertical dashed black line denotes $\Tbar=\kbar$, and the solid red best-fit line has gradient 2, indicating that $\textrm{Im}\,\wbar(\kbar) \propto \Tbar^2$ for $\Tbar \gtrsim \kbar$.}
  \label{fig:D3D5CTQNMs-ZeroB-Tdep}
\end{figure}

As $\Tbar$ increases and enters the range $\Tbar^2\ll\kbar\ll\Tbar$, the decay rate of the holographic zero sound mode increases as $-\textrm{Im}\,\wbar(\kbar) \propto \Tbar^2$, as shown in fig.~\ref{fig:D3D5CTQNMs-ZeroB-Tdep}. This is analogous to the collisionless thermal regime of a LFL, where thermal collisions of quasiparticles dominate quantum interactions, and the decay rate of zero sound $-\textrm{Im}\,\omega(k)\propto T^2/\mu$.

\begin{figure}[!htb]
  \centering
\subfigure{
      \includegraphics[width=0.45\textwidth]{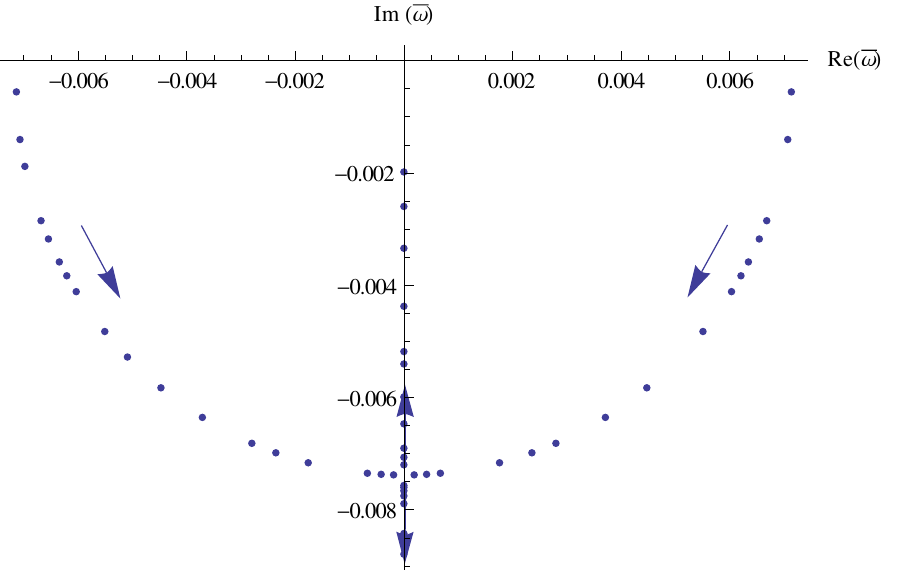}
}
\subfigure{
      \includegraphics[width=0.45\textwidth]{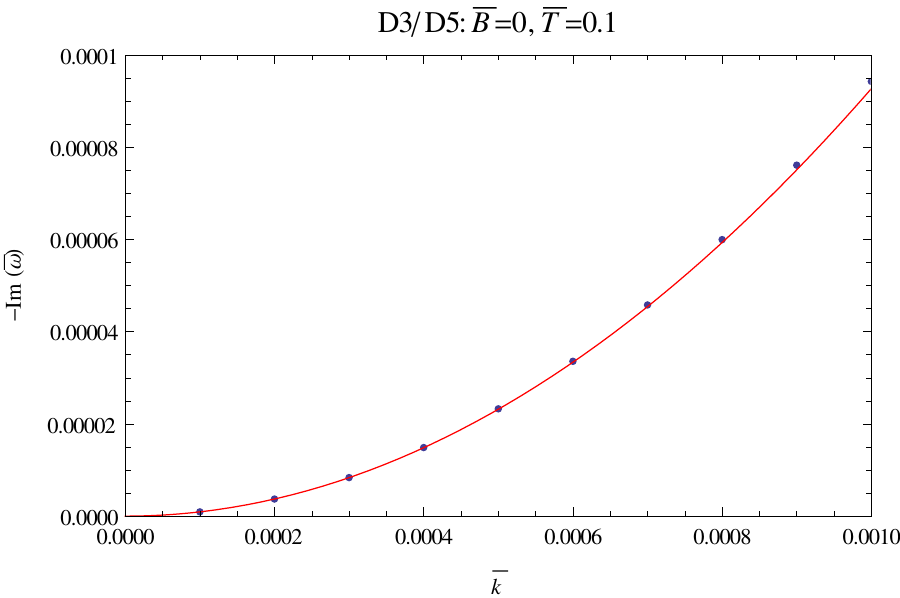}
      }
  \caption{\textbf{Left:} The complex $\wbar$ plane, where the dots are our numerical results for the dominant poles of $G^{tt}_R(\omega,k)$ in the D3/D5 system at $B=0$ and $\kbar=0.01$ as $\Tbar$ changes: the arrows point in the direction of increasing $\Tbar$. The two complex poles are due to holographic zero sound and the purely imaginary pole closest to the origin is due to charge diffusion. We define the location of the collisionless/hydrodynamic crossover as the value of $\Tbar$ where the holographic zero sound poles collide on the imaginary axis. \textbf{Right:} $-\textrm{Im}\,\wbar(\kbar)$ versus $\kbar$ for the charge diffusion mode at $\Tbar = 0.1$, in the regime $\kbar\ll\Tbar^2$. The dots are our numerical results and the red line is the dispersion relation in eq.~\eqref{eq:zeroBhydrodispersion}.}
  \label{fig:D3D5hydrocrossover-ZeroB}
\end{figure}

As $\Tbar$ increases further still, the two holographic zero sound modes, viewed as poles of $G^{tt}_R(\omega,k)$ in the complex $\wbar$ plane, move down and also closer to the imaginary axis, following an approximately semi-circular path, and eventually collide on the imaginary axis to form two purely imaginary modes, as shown in fig.~\ref{fig:D3D5hydrocrossover-ZeroB}. The purely imaginary mode farther from the origin is short-lived, and does not contribute significantly to the low-energy physics of the system. The purely imaginary mode closer to the origin is the hydrodynamic charge diffusion mode, with dispersion relation~\cite{Myers:2008me,Wapler:2009tr,Pal:2012gr}
\beq
\label{eq:zeroBhydrodispersion}
\wbar = - i \left( \frac{1}{2} \Tbar \sqrt{1+\Tbar^4} \, _2F_1\left[-\frac{3}{4},\frac{1}{2};\frac{1}{4};-\frac{1}{\Tbar^4}\right]-\frac{1}{2}\Tbar^3 \right) \kbar^2 + O\left(\kbar^3\right),
\eeq
as shown in fig.~\ref{fig:D3D5hydrocrossover-ZeroB}. The decay rate $-\textrm{Im}\,\wbar(\kbar)$ of this hydrodynamic mode decreases as $\Tbar$ increases. This is analogous to the hydrodynamic regime of a LFL.

\begin{figure}[!htb]
  \centering
  \subfigure{
      \includegraphics[width=0.45\textwidth]{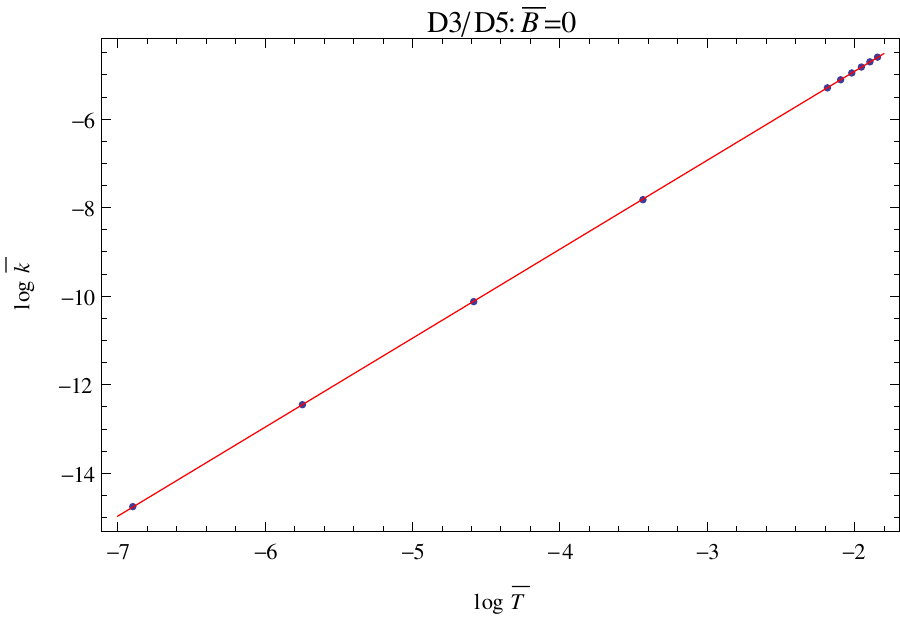}
}
\subfigure{
      \includegraphics[width=0.45\textwidth]{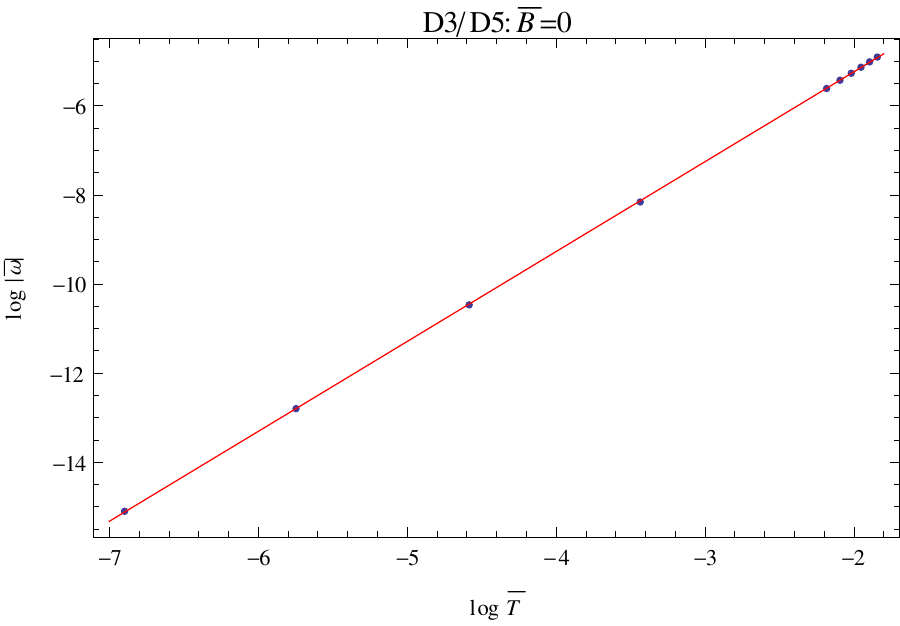}
      }
  \caption{\textbf{Left:} $\log \kbar$ versus $\log \Tbar$ and \textbf{Right:} $\log |\wbar|$ versus $\log \Tbar$ at the location of the collisionless/hydrodynamic crossover for the D3/D5 system at $B=0$. In both figures the dots are our numerical results and the solid red best-fit line has gradient 2, indicating that $\left|\wbar\right|\propto \Tbar^2$ and $\kbar\propto\Tbar^2$ at the crossover.}
  \label{fig:D3D5hydrocrossover3-ZeroB}
\end{figure}

These results for the D3/D5 system are qualitatively similar to those for the D3/D7 system~\cite{Davison:2011ek}. In both systems we define the location of the collisionless/hydrodynamic crossover as the value of $\Tbar$ where the holographic zero sound poles collide on the imaginary axis to form the purely imaginary poles. In both of the D3/D$p$ systems, we find that the crossover occurs when $\left|\wbar\right|\propto \Tbar^2$ and $\kbar\propto\Tbar^2$: for the D3/D5 system we show this in fig.~\ref{fig:D3D5hydrocrossover3-ZeroB}.
To find the exact proportionality constant in $\left|\wbar\right|\propto \Tbar^2$, we performed numerical fits to the data: we find that the crossover occurs when
\beq
\label{eq:ZeroBCrossoverLocation}
\left|\wbar\right| \approx \begin{cases} 0.30 \, \Tbar^2 & \text{for D3/D5,} \\ 0.45 \, \Tbar^2 & \text{for D3/D7.} \end{cases}
\eeq

\begin{figure}[!htb]
  \centering
  \subfigure{
      \includegraphics[width=0.45\textwidth]{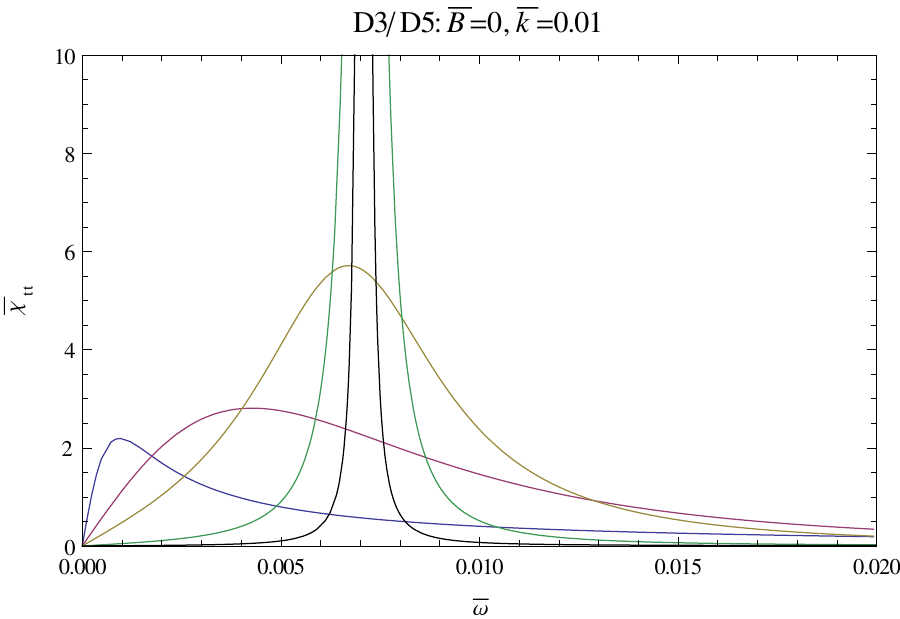}
}
\subfigure{
      \includegraphics[width=0.45\textwidth]{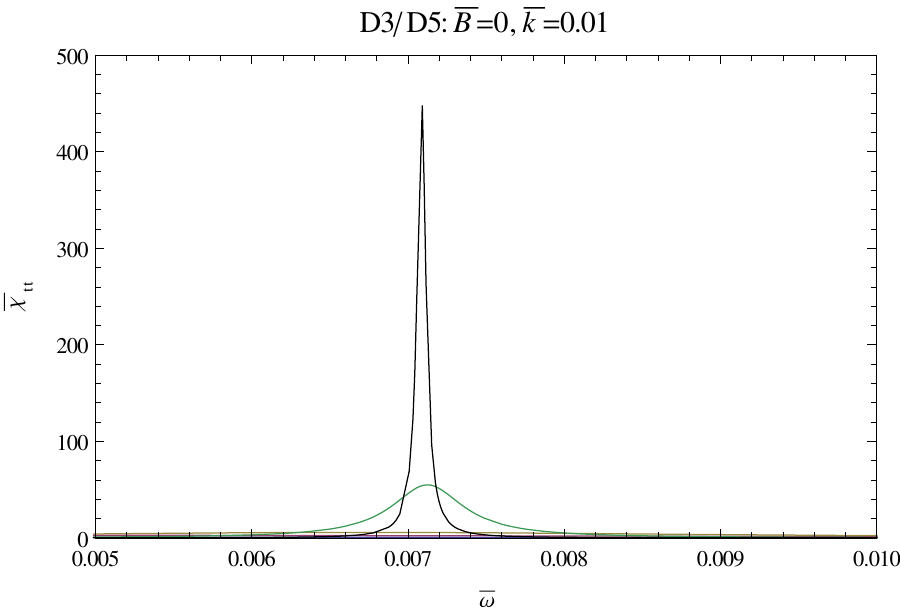}
      }
  \caption{The charge density spectral function $\bar{\chi}_{tt}\left(\wbar\right)$ in the D3/D5 system at $B=0$ as a function of $\wbar$ and at fixed $\kbar=0.01$. Moving from the tallest peak to the shortest peak corresponds to raising the temperature: $\Tbar=0.01$ (black), $\Tbar=0.032$ (green), $\Tbar=0.1$ (yellow), $\Tbar=0.16$ (magenta), $\Tbar=0.32$ (blue).  The right-hand plot shows the tallest peak in full.}
  \label{fig:D3D5chargespectral-ZeroB}
\end{figure}

The crossover is also visible in the charge density spectral function $\bar{\chi}_{tt}\left(\wbar,\kbar\right)$. In fig.~\ref{fig:D3D5chargespectral-ZeroB} we plot $\bar{\chi}_{tt}\left(\wbar\right)$ in the D3/D5 system at a fixed $\kbar$ as a function of $\wbar$, for various $\Tbar$. For small $\Tbar$, $\bar{\chi}_{tt}\left(\wbar\right)$ is dominated by a narrow peak due to the holographic zero sound mode. As $\Tbar$ increases, the peak due to the holographic zero sound mode becomes wider as its decay rate $-\textrm{Im}\,\wbar(\kbar)$ increases: in fig.~\ref{fig:D3D5hydrocrossover-ZeroB}, the holographic zero sound poles move down into the complex $\omega$ plane. Near the collisionless/hydrodynamic crossover, the peak moves towards the origin which reflects the location of the corresponding poles of $G^{tt}_R(\omega,k)$ in the complex $\wbar$ plane. As $\Tbar$ increases further and the system enters the hydrodynamic regime, the peak near the origin becomes narrower due to the decreasing decay rate $-\textrm{Im}\,\wbar(\kbar)$ of the hydrodynamic charge diffusion mode. Again, these results for $\bar{\chi}_{tt}\left(\wbar,\kbar\right)$ in the D3/D5 system are qualitatively similar to those of the D3/D7 system.

In summary, in the D3/D$p$ systems, the three regimes of behaviour in response to density perturbations are very similar to those of a LFL. In particular, the decay rate $-\textrm{Im}\,\wbar(\kbar)$ of the holographic zero sound mode has the same form as that of the LFL zero sound mode in both the collisionless quantum and collisionless thermal regimes. Additionally, the boundaries between the regimes are at $\kbar\simeq\Tbar$ and $\kbar\simeq\Tbar^2$, so since $|\wbar|\simeq\kbar$ for the holographic zero sound mode (the poles in the complex $\wbar$ plane in fig.~\ref{fig:D3D5hydrocrossover-ZeroB} follow an approximately semi-circular path), the boundaries are located at $\wbar\simeq\Tbar$ and $\wbar\simeq\Tbar^2$, as in a LFL. These similarities between the D3/D$p$ systems and LFLs are surprising because, as mentioned in sections~\ref{intro} and \ref{review}, the D3/D$p$ systems are \textit{not} LFLs. Indeed, no evidence has been found of a Fermi surface in the D3/D$p$ systems, and in fact the low-energy dynamics may be 
governed by a (0+1)-dimensional CFT~\cite{Jensen:2010ga,Nickel:2010pr,Ammon:2011hz,Goykhman:2012vy}.

\begin{figure}[!htb]
  \centering
  \subfigure{
      \includegraphics[width=0.45\textwidth]{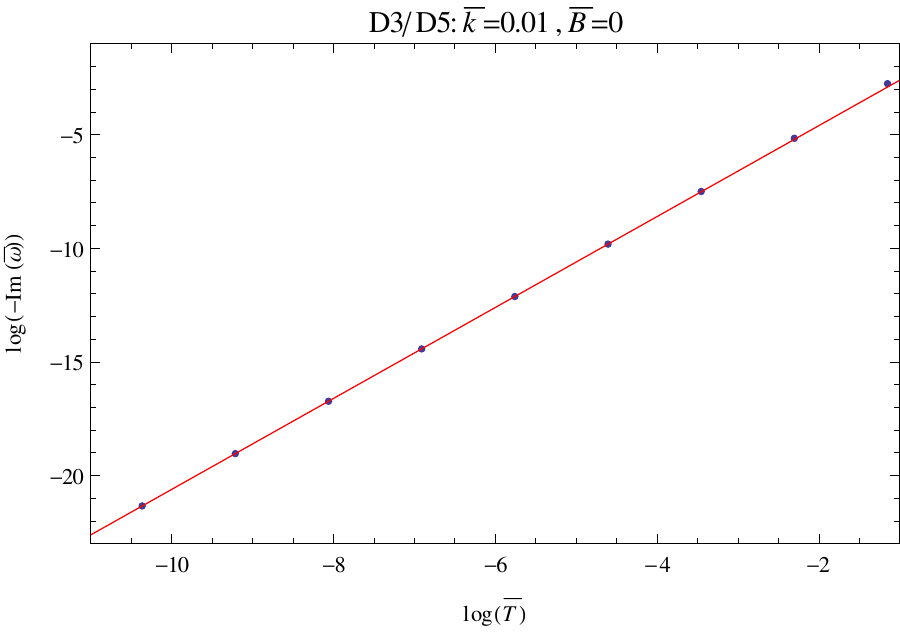}
}
\subfigure{
      \includegraphics[width=0.45\textwidth]{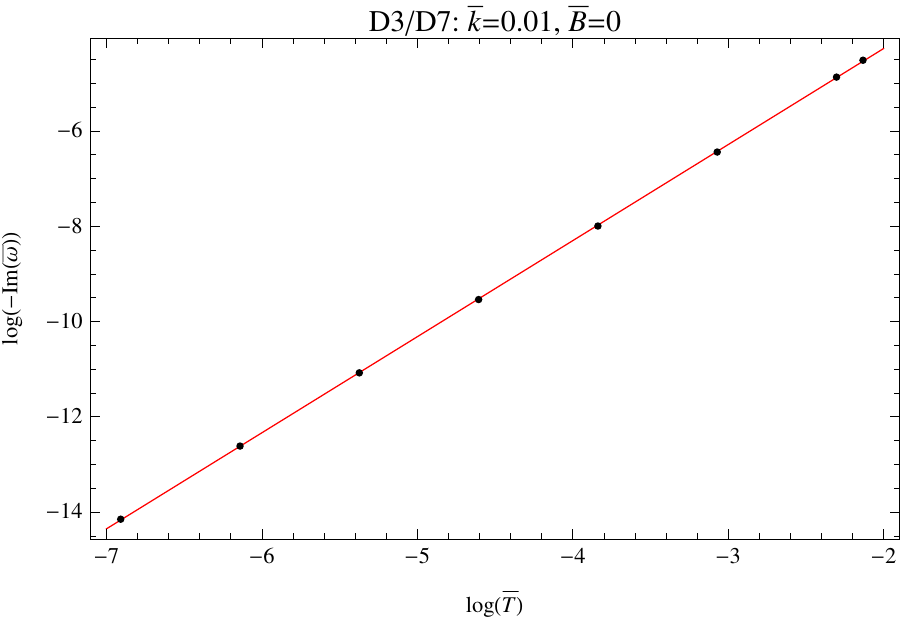}
      }
  \caption{\textbf{Left:} $\log \left(- \textrm{Im}\,\wbar(\kbar)\right)$ as a function of $\log \Tbar$ for the pole in $G^{yy}_R(\omega,k)$ closest to the origin of the complex $\wbar$ plane in the D3/D5 system at $B=0$, with $\kbar=0.01$, within the collisionless regime $\kbar\gg\Tbar^2$. \textbf{Right:} The same as in the left figure, but in the D3/D7 system. In both figures the dots are our numerical results and the red best-fit lines have gradient $2$, indicating that in both cases $\textrm{Im}\,\wbar(\kbar)\propto\Tbar^2$.}
  \label{fig:transverseQNMsBzero}
\end{figure}

In both of the D3/D$p$ systems, $J^y$'s retarded Green's function, $G^{yy}_R(\omega,k)$, also exhibits poles in the complex $\wbar$ plane. As discussed in section~\ref{eoms}, when $B$ is non-zero and parity is broken, these poles will mix with those of $G^{tt}_R(\omega,k)$, and hence will affect the crossover. As preparation for the non-zero-$B$ case, let us now discuss the poles in $G^{yy}_R(\omega,k)$ when $B=0$. When $B=0$, in both of the D3/D$p$ systems $G^{yy}_R(\omega,k)$ has no poles corresponding to long-lived modes, by which we mean poles satisfying $\wbar\rightarrow0$ as $\kbar\rightarrow0$. The pole in $G^{yy}_R(\omega,k)$ closest to the origin of the complex $\wbar$ plane is purely imaginary, but is not diffusive: for this mode $\wbar\left(\kbar\right)$ is approximately independent of $\kbar$. At low temperatures, for this mode $\textrm{Im}\,\wbar(\kbar)\propto\Tbar^2$, as shown in fig.~\ref{fig:transverseQNMsBzero}.

Following the convention of refs.~\cite{Karch:2008fa,Kulaxizi:2008kv,Nickel:2010pr,Ammon:2011hz,Davison:2011ek}, we described the poles in $G^{tt}_R(\omega,k)$ using complex frequency $\wbar(\kbar)$ with real momentum $\kbar$. We could then define the crossover by a collision of poles in the complex $\wbar$ plane, and could identify three distinct regimes of response to density perturbations, dependening upon the relative magnitudes of $\kbar$, $\Tbar$ and $\Tbar^2$. For the holographic zero sound mode $\left|\wbar\right|\propto\kbar$, so we could equivalently characterize the three regimes by the relative magnitudes of $\wbar$, $\Tbar$ and $\Tbar^2$. Crucially, when $B$ is non-zero such a simple translation from real $\kbar$ to real $\wbar$ will in general \textit{not} be possible, as we will see in the next section.

\section{Crossover at Non-zero Magnetic Field}
\label{finiteb}

In this section we study the collisionless/hydrodynamic crossover in both of the D3/D$p$ systems, with non-zero $B$. We will normalise $B$ by the appropriate power of $d$ to be dimensionless,
\beq
\label{eq:BarredVariables2}
\Bbar\equiv\frac{B}{d^{2/n}}.
\eeq
We will always work with a large, fixed density, such that $\Bbar,\Tbar,\wbar,\kbar\ll 1$. In particular, we will always work with values of $B$ below any known phase transitions. In general, the critical value of $B$ for the known transitions is a complicated function of $T$, but at any given $T$ the critical value of $B$ is bounded from below by the $T=0$ values given in section~\ref{review}. We will work with $\Tbar\geq10^{-3}$ and values of $\Bbar/\Tbar^2$ up to $10^4$ for the D3/D5 system and up to $10^5$ for the D3/D7 system, well below the known transitions.

In section~\ref{hzsnonzerob} we determine the dispersion relations of the holographic zero sound mode in our systems for $T=0$ and non-zero $B$ analytically (without numerics), following refs.~\cite{Goykhman:2012vy,Gorsky:2012gi}. In section~\ref{diffnonzerob} we determine the dispersion relation of the charge diffusion mode at non-zero $B$, following refs.~\cite{Wapler:2009tr,Pal:2012gr}. In section~\ref{sec:NumericalCollisionsSmallB} we repeat the analysis of section~\ref{zerob} but with non-zero $B$, that is, we study the poles of $G^{tt}_R(\omega,k)$ in the complex $\omega$ plane with non-zero $B$. For sufficiently small $B$ we will show that a collision of poles occurs as we increase $T$, similar to the $B=0$ case, but that such a collision does not occur for sufficiently high $B$. We thus turn to the spectral function $\bar{\chi}_{tt}\left(\wbar,\kbar\right)$ in section~\ref{spectralfunctions}, which will show a clear crossover from holographic zero sound to charge diffusion as $T$ increases with respect to $\omega$. The crossover is simplest to understand by studying the poles of $G^{tt}_R(\omega,k)$ in the complex \textit{momentum}\ plane, as we discuss in section~\ref{sec:NumericalResultsKPlane}. We summarize our results in section~\ref{sec:NumericalResultsSummarySection}.

\subsection{Holographic Zero Sound at Non-zero Magnetic Field}
\label{hzsnonzerob}

In this subsection we derive the dispersion relation of the holographic zero sound mode when $T=0$ and with $B$ non-zero. For the D3/D5 system we reproduce the results of ref.~\cite{Goykhman:2012vy}, while for the D3/D7 system our results are novel. We will also show, using our numerical solutions, that the $T=0$ dispersion relation is a very good approximation to the low-temperature dispersion relation, as was the case when $B=0$ (see fig.~\ref{fig:D3D5CQQNMs-ZeroB}).

To determine the dispersion relation without numerics, we follow the procedure of refs.~\cite{Karch:2008fa,Goykhman:2012vy}: we solve the equations of motion, eqs.~\eqref{eq:Eeom} and~\eqref{eq:ayeom}, in two different limits and then match the two solutions in a regime where the limits overlap. More specifically, we take a near-horizon limit of the equations of motion, meaning $r\to0$, solve the resulting equations, and then take a low-frequency limit of the solution. We then repeat these operations in opposite order: we take a low-frequency limit of the equations of motion, solve the resulting equations, and then take a near-horizon limit of the solution. 

We begin by setting $T=0$, or equivalently $r_H=0$, in eqs.~\eqref{eq:Eeom} and~\eqref{eq:ayeom}, in which case $|g_{tt}| = g_{rr}^{-1} = g_{xx} = r^2$. Our near-horizon $r\to0$ limit consists of taking $r/d^{1/n} \ll 1$ with $r/\sqrt{B} \approx O(1)$, so that
\beq
g_{xx}^{n-2} \left(g_{xx}^2 + B^2\right) + d^2 \to d^2, \qquad \omega^2 - u(r)^2 k^2 \to \omega^2.
\eeq
The resulting equations have solutions
\bea
 \label{eq:Bfinitenhdata}
\vec{V}(r,\omega,k) &=&  e^{i \omega/r}\left( \begin{array}{cc} r & \left( 1-\frac{i \omega}{r} \right) \frac{B}{\omega} \\ \frac{i}{\omega} \left( 1-\frac{i \omega}{r} \right) \frac{B}{\omega} & \frac{i}{\omega} \, r \end{array} \right) \vec{c},
\eea
with constant normalization vector $\vec{c}$, which is the $T=0$ analogue of the $\vec{V}_{\mathrm{nh}}$ defined for non-zero $T$ in eq.~\eqref{eq:vnhdef}. On the right-hand-side of eq.~\eqref{eq:Bfinitenhdata} the $e^{i \omega/r}$ factor describes an in-going wave as $r\to0$, as we discussed in section~\ref{eoms}.

We now want to take the low-frequency limit of the solution in eq.~\eqref{eq:Bfinitenhdata}, which means $\omega/r \ll 1$~\cite{Karch:2008fa}. As emphasized in ref.~\cite{Goykhman:2012vy}, however, Kohn's theorem suggests that the holographic zero sound dispersion may exhibit a gap when $B$ is non-zero, so if we wish to obtain a non-trivial result we must be careful not to take $\omega$ small with fixed $B$. Instead, we should take $\omega$ small with $B/\omega$ fixed, \textit{i.e.}\ we should scale the gap with $\omega$ as we make $\omega$ small. In this limit the solution of eq.~\eqref{eq:Bfinitenhdata} becomes
\bea
\label{Eq: NHExpansion}
\vec{V}(r,\omega,k) &=& \left( \begin{array}{cc} r + i \omega & \frac{B}{\omega} \\ \frac{i\,B}{\omega^2} & \frac{i}{\omega} \left(r+ i \omega\right) \end{array} \right)\vec{c}. 
\eea

We now need to perform the same operations in the opposite order. First, we take the low-frequency limit of eqs.~\eqref{eq:Eeom} and~\eqref{eq:ayeom}, meaning $\omega/r \ll 1$ and $k/r \ll 1$ keeping $\omega/ k$ fixed. As above, when we take $\omega$ to be small we also take $B$ to be small, in the sense that we keep $B/\omega$ fixed. In these limits, in eqs.~\eqref{eq:Eeom} and~\eqref{eq:ayeom} we drop all terms not involving derivatives of $E$ or $a_y$, as these are sub-leading, and moreover in the remaining terms we drop all factors of $B$. The resulting equations have solutions
\bea
E(r,\omega,k) & = & E^{(0)} + c_E \, r^{1-n} \left[ \frac{k^2}{n\sqrt{1+d^2/r^{2n}}} + \frac{k^2-n\omega^2}{n(n-1)} \left . _2F_1\left(\frac{1}{2},\frac{1}{2}-\frac{1}{2n},\frac{3}{2}-\frac{1}{2n},-\frac{d^2}{r^{2n}}\right)\right] \right ., \nonumber \\
& & \nonumber \\
a_y(r,\omega,k) & = & a_y^{(0)} + c_y \, \frac{r^{1-n}}{1-n} \left . _2F_1\left(\frac{1}{2},\frac{1}{2}-\frac{1}{2n},\frac{3}{2}-\frac{1}{2n},-\frac{d^2}{r^{2n}}\right)\right., \nonumber
\eea
where $E^{(0)}$, $c_E$, $a_y^{(0)}$, and $c_y$ are independent of $r$ but can depend on $\omega$ and $k$. In an expansion near the AdS boundary, $E^{(0)}$ and $a_y^{(0)}$ are the leading, non-normalizable terms, and hence act as sources for the dual operators $J^E$ and $J^y$. Now we take the near-horizon limit, $r/d^{1/n} \ll 1$,
\begin{subequations}
\label{eq:lowfreqnh}
\bea
E(r,\omega,k) & = & E^{(0)} + c_E \, \frac{\mu_0}{d}\left(\frac{1}{n}\,k^2 - \omega^2 \right) + c_E \, \frac{\omega^2}{d} \, r + O\left(r^{n+1}/d^{1+1/n}\right), \\
a_y(r,\omega,k) & = & a_y^{(0)} - c_y \, \frac{\mu_0}{d} + c_y \, d^{-1} r + O\left(r^{n+1}/d^{1+1/n}\right).
\eea
\end{subequations}

We can now match solutions in the regimes where the two limits overlap. Specifically, we match the terms constant in $r$ and the terms linear in $r$ in eqs.~\eqref{Eq: NHExpansion} and \eqref{eq:lowfreqnh}. That gives us four equations involving the two components of $\vec{c}$ as well as $E^{(0)}$, $c_E$, $a_y^{(0)}$, and $c_y$. We can use two of those equations to eliminate the two components of $\vec{c}$, leaving us with two equations that we can express in matrix form:
\beq
\label{eq:matrixeq}
\begin{pmatrix} E^{(0)} \\ a_y^{(0)} \end{pmatrix} = \begin{pmatrix} i \frac{\omega^3}{d} + \frac{\mu_0}{d}\left(\omega^2 - \frac{k^2}{n}\right) & -\frac{i\,B}{d} \\ \frac{i\,B}{d} & \left(\frac{i \, \omega}{d} + \frac{\mu_0}{d} \right)\end{pmatrix} \begin{pmatrix} c_E \\ c_y \end{pmatrix}.
\eeq
As discussed in section~\ref{eoms}, for a quasi-normal mode the sources $E^{(0)}=0$ and $a_y^{(0)}=0$. We can then obtain a non-trivial solution for $\vec{V}(r,\omega,k)$ only if the matrix on the right-hand-side of eq.~\eqref{eq:matrixeq} has vanishing determinant.

When $B=0$, the matrix on the right-hand-side of eq.~\eqref{eq:matrixeq} becomes diagonal, \textit{i.e.} $E$ and $a_y$ decouple. In that case the determinant of the matrix will vanish when either diagonal entry vanishes. Setting the upper left entry to zero gives us the holographic zero sound dispersion relation at $B=0$, eq.~\eqref{eq:zeroTdispersionrelation}. Setting the lower right entry to zero gives us a quasi-normal mode with dispersion $\omega = - i \mu_0$, dual to a pole in $G^{yy}_R(\omega,k)$. This purely-imaginary, $k$-independent mode is distinct from the purely-imaginary, $k$-independent mode in $G^{yy}_R(\omega,k)$ that we discussed near the end of section~\ref{collectiveholo}, which had $\textrm{Im}\,\wbar(\kbar) \propto \Tbar^2$ (recall fig.~\ref{fig:transverseQNMsBzero}), and hence $\textrm{Im}\,\wbar(\kbar) \to 0$ as $\Tbar \to 0$.

When $B$ is non-zero, $E$ and $a_y$ couple, and so the poles in $G^{EE}_R(\omega,k)$ and $G_R^{yy}(\omega,k)$ mix. Demanding that the determinant of the matrix on the right-hand-side of eq.~\eqref{eq:matrixeq} vanish when $B$ is non-zero gives us
\beq
\label{eq:polynomialeq}
\wbar^4 - 2i \frac{\mu_0}{d^{1/n}}\,\wbar^3 - \frac{\mu_0^2}{d^{2/n}}\,\wbar^2 + i \frac{\mu_0}{d^{1/n}} \frac{\kbar^2}{n} \, \wbar + \frac{\mu_0^2}{d^{2/n}} \frac{\kbar^2}{n} + \Bbar^2 = 0,
\eeq
which determines the dispersion relation of the holographic zero sound: expressed as $\wbar(\kbar)$, when both $\Bbar$ and $\kbar$ are non-zero and small, but of the same order, we find
\beq
\label{HZSanalytic}
\wbar(\kbar) = \pm \sqrt{\frac{1}{n} \, \kbar^2 + \frac{d^{2/n}}{\mu_0^2} \Bbar^2} - i \frac{d^{1/n}}{\mu_0} \left [ \frac{1}{2n} \kbar^2 + \frac{d^{2/n}}{\mu_0^2} \, \Bbar^2 \right] + O\left(\kbar^3\right).
\eeq
When $B=0$ we recover the result of ref.~\cite{Karch:2008fa}, namely eq.~\eqref{eq:zeroTdispersionrelation}, and when $B$ is non-zero, for $n=2$ we recover the result of ref.~\cite{Goykhman:2012vy}. In particular, we have found that the holographic zero sound dispersion relation is gapped: defining $\wbar_{\mathrm{gap}}\equiv \lim_{\kbar\to0}\textrm{Re}\,\wbar(\kbar)$, we have
\beq
\label{eq:kofwgap}
\wbar_{\mathrm{gap}} = \frac{d^{1/n}}{\mu_0} \, \Bbar.
\eeq
We can alternatively express the holographic zero sound dispersion relation as $\kbar(\wbar)$,
\beq
\label{eq:HZSanalyticKofW}
\kbar(\wbar)^2 = n \left( \wbar^2 - \frac{d^{2/n}}{\mu_0^2} \Bbar^2 \right) + i n \, \wbar \, \frac{d^{1/n}}{\mu_0} \left( \wbar^2 + \frac{d^{2/n}}{\mu_0^2} \Bbar^2 \right).
\eeq

\begin{figure}[!htb]
  \centering
\subfigure{
      \includegraphics[width=0.45\textwidth]{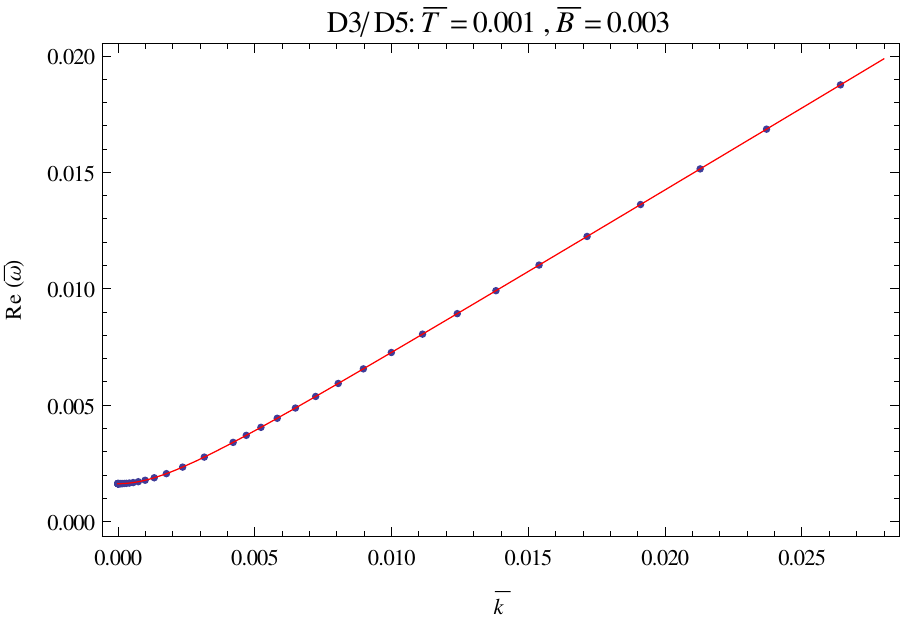}
}
\subfigure{
      \includegraphics[width=0.45\textwidth]{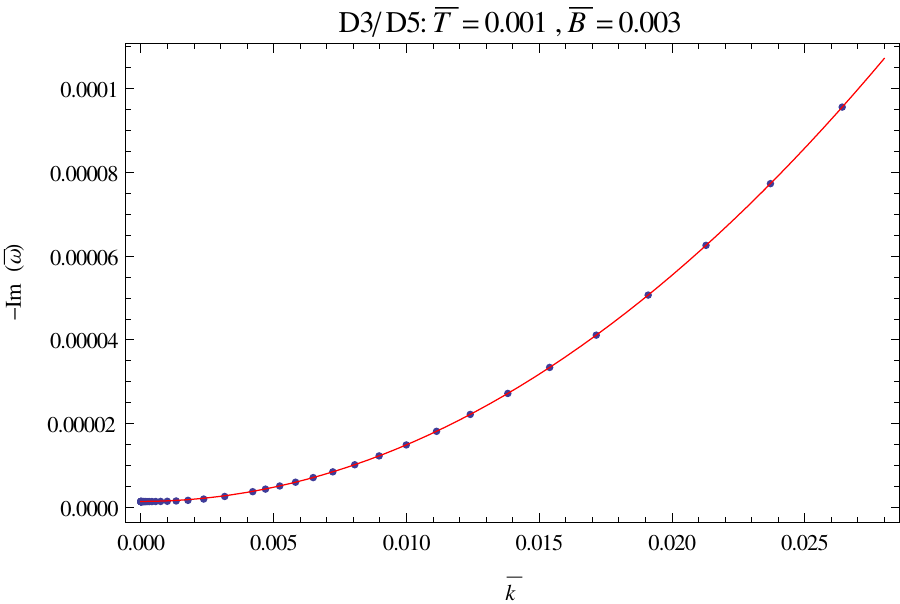}
      }
\\
\subfigure{
      \includegraphics[width=0.45\textwidth]{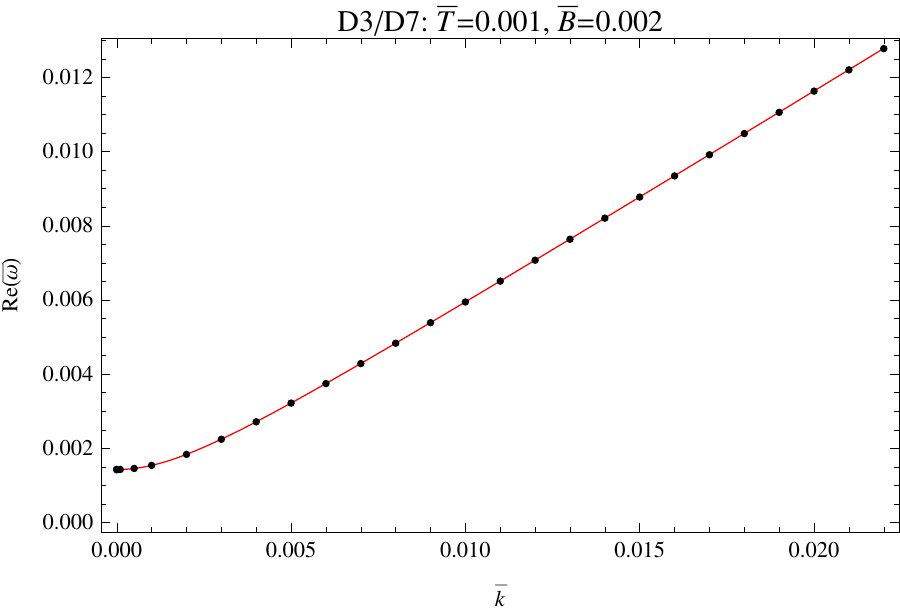}
}
\subfigure{
      \includegraphics[width=0.45\textwidth]{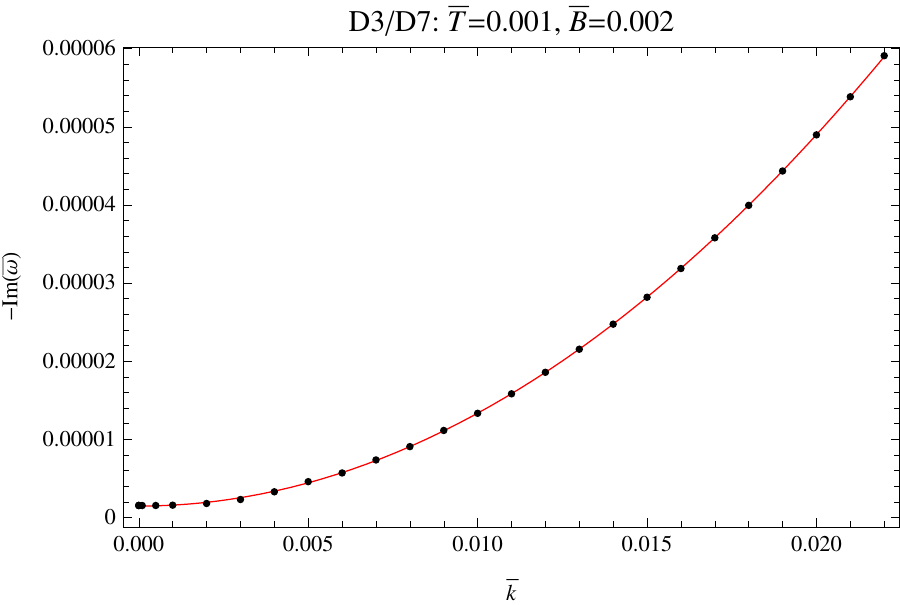}
      }
  \caption{\textbf{Top:} $\textrm{Re}\,\wbar(\kbar)$ (\textbf{Left}) and $-\textrm{Im}\,\wbar(\kbar)$  (\textbf{Right}) versus $\kbar$ for the holographic zero sound mode in the D3/D5 system with $\Bbar=0.003$ and $\Tbar=0.001$, in the range $\kbar\gg\Tbar$. \textbf{Bottom:} $\textrm{Re}\,\wbar(\kbar)$ (\textbf{Left}) and $-\textrm{Im}\,\wbar(\kbar)$ (\textbf{Right}) versus $\kbar$ for the holographic zero sound mode in the D3/D7 system with $\Bbar=0.002$ and $\Tbar=0.001$, in the range $\kbar\gg\Tbar$. In all four plots the dots are our numerical results and the solid red lines come from the $T=0$ dispersion relation, eq.~\eqref{HZSanalytic}. Clearly the $T=0$ result eq.~\eqref{HZSanalytic} is a very good approximation to the low-temperature dispersion relation, as was the case when $B=0$ (recall fig.~\ref{fig:D3D5CQQNMs-ZeroB}).}
  \label{fig:collisionlessQNMs-kdependence}
\end{figure}

In fig.~\ref{fig:collisionlessQNMs-kdependence} we present our numerical results for the dispersion relation of the holographic zero sound in both D3/D$p$ systems with non-zero $B$, at low temperatures. We find that the $T=0$ dispersion relation, eq.~\eqref{HZSanalytic}, is a very good approximation to the low-temperature dispersion relation, as occured when $B=0$ (recall fig.~\ref{fig:D3D5CQQNMs-ZeroB}).

\subsection{Charge Diffusion at Non-zero Magnetic Field}
\label{diffnonzerob}

As discussed in section~\ref{collectiveholo}, when $B=0$ a charge diffusion pole appears in $G^{tt}_R(\omega,k)$ in the hydrodynamic regime $\kbar,\wbar\ll \Tbar^2$. In this subsection, we will determine the dispersion relation of the longest-lived collective excitation in this hydrodynamic regime when $B$ is non-zero. We will find that the longest-lived mode is that of charge diffusion.

Notice that previous holographic studies of hydrodynamics with a non-zero magnetic field, such as those of refs.~\cite{Hartnoll:2007ih,Hartnoll:2007ip,Buchbinder:2008dc,Hansen:2008tq,Buchbinder:2008nf,Buchbinder:2009mk,Hansen:2009xe}, employed the typical ``hydrodynamic'' limit, in which $T$ is larger than all other scales. In contrast, as mentioned at the beginning of this section, we work in a limit where the charge density is the largest scale.

As in the previous subsection, to determine the dispersion relation of the charge diffusion mode without using numerics, we will solve the equations of motion, eqs.~\eqref{eq:Eeom} and~\eqref{eq:ayeom}, in two different limits, the near-horizon and low-frequency limits, and then match the two solutions in a regime where the limits overlap.

We begin with the near-horizon limit: we expand the equations of motion, eqs.~\eqref{eq:Eeom} and~\eqref{eq:ayeom}, in $(r-r_H)$ up to order $(r-r_H)$. We then solve the resulting equations, and take the limits of low-frequency and low momentum, $\omega/T$, $k/T\ll1$, assuming $|\omega| \propto |k^2|$. In these limits we discard $a_y$'s contribution to $E$'s equation of motion, being sub-leading. To find a quasi-normal mode, then, we only need the solution for $E$ in these limits:
\beq
\label{eq:deffenh}
E(r,\omega,k) =  \left( 1 +  i \, \frac{k^2}{\omega} \, \left[\lim_{r\to r_H}\frac{\left(u(r)^2\right)'}{\sqrt{|g_{tt}'(r)|\left(g_{rr}^{-1}(r)\right)'}} \right]\, \left(r-r_H\right)\right)\,E_{\mathrm{nh}},
\eeq
where $E_{\mathrm{nh}}$ is the upper component of $\vec{V}_{\mathrm{nh}}$. Notice that here, unlike the previous subsection, we do not keep $B/\omega$ fixed when we take the low-frequency limit.

We now perform the same operations in the opposite order. First, in the equations of motion, eqs.~\eqref{eq:Eeom} and~\eqref{eq:ayeom}, we take the low-frequency and low-momentum limits, $\omega/T$, $k/T\ll1$, assuming $|\omega|\propto |k^2|$. In those limits, in the equation of motion for $E$, eq.~\eqref{eq:Eeom}, we discard the terms without derivatives on $E$ or $a_y$, as these are sub-leading, and moreover we take
\beq
\omega^2 - u(r)^2 k^2 \to - u(r)^2 k^2,
\eeq
since we assume $|\omega^2| \propto |k^4|$, which is suppressed relative to $k^2$ in our limits. The solution for $E$ in these limits is then
\beq
\label{eq:diffelowfreq}
E(r,\omega,k) = E^{(0)} + C \int_r^{\infty} d\hat{r} \, \frac{u(\hat{r})^3 \left(g_{xx}^2(\hat{r}) + B^2 \right)}{g_{xx}^{(n+1)/2}(\hat{r}) |g_{tt}(\hat{r})| g_{rr}^{-1/2}(\hat{r})},
\eeq
where $E^{(0)}$ and $C$ are independent of $r$ but can depend on $\omega$ and $k$. Next we perform the near-horizon limit, expanding the solution in eq.~\eqref{eq:diffelowfreq} in $(r-r_H)$ up to order $(r-r_H)$. We then match the constant term and the term linear in $(r-r_H)$ to the corresponding terms in eq.~\eqref{eq:deffenh}. The matching gives us two equation for $E^{(0)}$ and $C$. Using one of those equations to eliminate $C$, and using the fact that each of $g_{tt}(r)$, $g_{rr}^{-1}(r)$, and $u(r)^2$ has a simple zero ar $r=r_H$, we find
\beq
 E^{(0)} = E_{\mathrm{nh}} \left( 1 + \frac{ i k^2}{\omega} \frac{g_{xx}^{(n+1)/2}(r_H)}{\left(g^2_{xx}(r_H) + B^2 \right)} \sqrt{ \frac{|g'_{tt}(r_H)|}{\left(u(r_H)^2\right)'} } \int_{r_H}^{\infty} dr \frac{u^3(r) \left(g_{xx}^2(r) + B^2 \right)}{g_{xx}^{(n+1)/2}(r) |g_{tt}(r)| g_{rr}^{-1/2}(r)} \right).
\eeq
Imposing the Dirichlet condition, $E^{(0)}=0$, we thus find the dispersion relation of a charge diffusion mode,
\beq
\label{eq:D3Dpdiffusionanalytic}
\omega(k) = - i \left[ \frac{g_{xx}^{(n+1)/2}(r_H)}{\left(g^{2}_{xx}(r_H) + B^2 \right)} \sqrt{ \frac{|g'_{tt}(r_H)|}{\left(u(r_H)^2\right)'} } \int_{r_H}^{\infty} dr \frac{u^3(r) \left(g_{xx}^2(r) + B^2 \right)}{g_{xx}^{(n+1)/2}(r) |g_{tt}(r)| g_{rr}^{-1/2}(r)} \right] k^2 + O\left(k^3\right).
\eeq
In the D3/D5 system we can perform the integral in eq.~\eqref{eq:D3Dpdiffusionanalytic}, giving us
\bea
\label{eq:D3D5diffusionanalytic}
\wbar(\kbar) &=& - i \Tbar^{2} \frac{\sqrt{1 + \Bbar^2 + \Tbar^{4}}}{\Tbar^4 + \Bbar^2} \left( - \frac{\Tbar}{2} \frac{2 \Bbar^4 + \Tbar^4 + 2 \Bbar^2 \left(1+\Tbar^4\right)} {\left(1+\Bbar^2\right)^2 \sqrt{1 + \Bbar^2 + \Tbar^{4}}} \right. \\ &\;& \left. \hphantom{\Tbar^{2} \frac{\sqrt{1 + \Bbar^2 + \Tbar^{4}}}{\Tbar^4 + \Bbar^2}} + \frac{\Tbar^3}{2} \frac{1+ 2 \Bbar^2}{\left(1+\Bbar^2\right)^2} \;_{2} F_{1} \left[ - \frac{3}{4}, \frac{1}{2}, \frac{1}{4}; - \left(\frac{1+\Bbar^2}{\Tbar^4}\right) \right] \right) \kbar^2 + O\left(\kbar^3\right). \nonumber
\eea
When $\Bbar=0$ we thus recover the result of eq.~\eqref{eq:zeroBhydrodispersion}, and when $\Bbar$ is non-zero we recover the result of refs.~\cite{Wapler:2009tr,Pal:2012gr}. In the D3/D7 system we have been unable to evaluate the integral in eq.~\eqref{eq:D3Dpdiffusionanalytic} in closed form:
\bea
\label{eq:D3D7diffusionanalytic}
\wbar &=& - i \Tbar^{2} \frac{\sqrt{1 + \Bbar^2 \Tbar^{2} + \Tbar^{6}}}{\Tbar^4 + \Bbar^2} \int^{\infty}_{\Tbar} d\rho \frac{\rho^2 \left( \rho^4 + \Bbar^2 \right)}{\left(1 + \Bbar^2 \rho^2 + \rho^6\right)^{\frac{3}{2}} } \kbar^2 + O\left(\kbar^3\right).
\eea
When $B=0$ we recover the result of refs.~\cite{Kim:2008bv,Mas:2008qs}, while to our knowledge our result with non-zero $\Bbar$ is novel.

While we have expressed the dispersion relations for the diffusive modes as $\omega(k)$, the expression for $k(\omega)$ is trivial to obtain by inverting eqs.~\eqref{eq:D3D5diffusionanalytic} and~\eqref{eq:D3D7diffusionanalytic}. Notice that, when expressed as $k(\omega)$, the dispersion relation of the charge diffusion mode will have a non-zero real part.

\begin{figure}[!htb]
\centering
\subfigure{\includegraphics[width=0.45\textwidth]{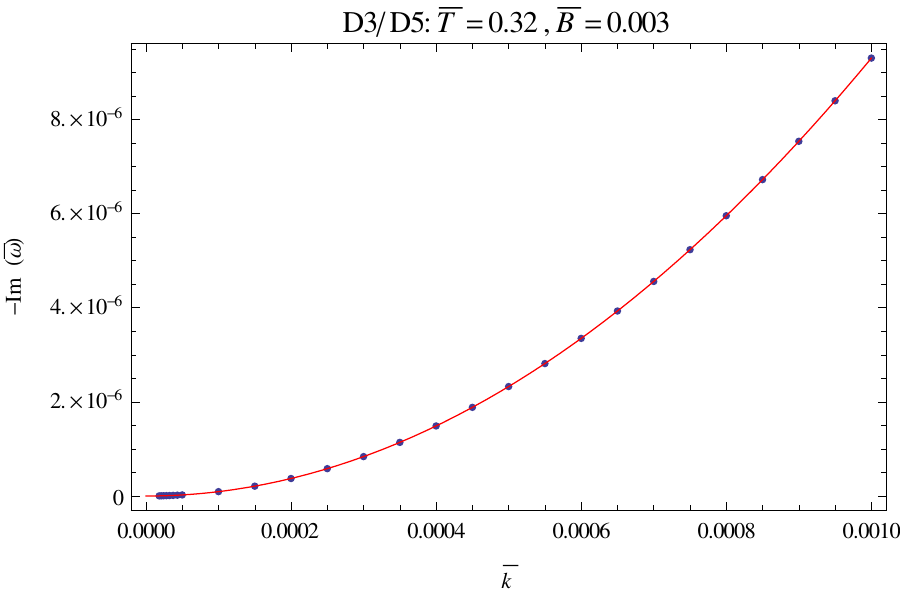}}
\subfigure{\includegraphics[width=0.45\textwidth]{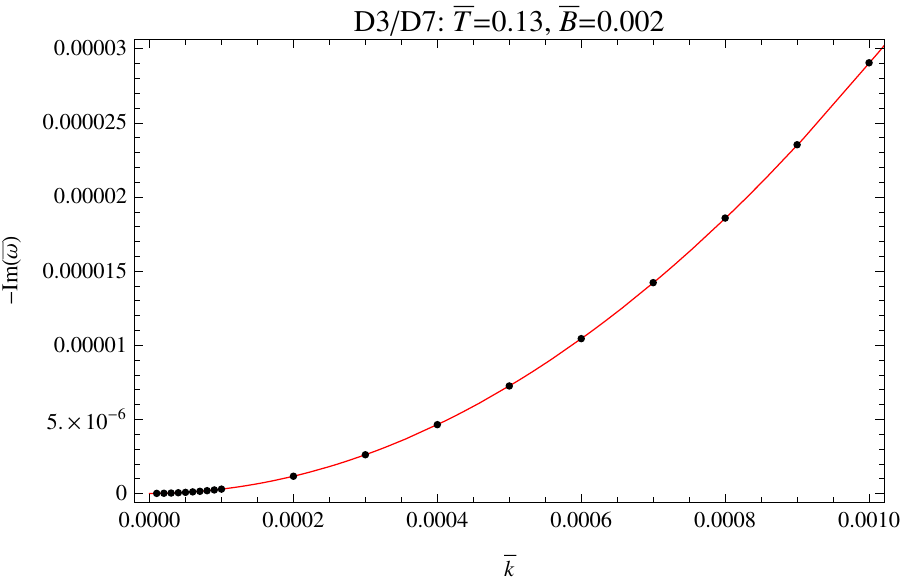}}
\caption{\textbf{Left:} $-\textrm{Im}\,\wbar$ versus $\kbar$ for the charge diffusion mode in the D3/D5 system with $\Bbar=0.003$ and $\Tbar=0.32$, within the range $\kbar\ll\Tbar^2$. \textbf{Right:} The same for the D3/D7 system, with $\Bbar=0.002$ and $\Tbar=0.13$. For both modes, $\Re\left(\wbar\right)=0$. In both plots the dots are our numerical results and the solid red lines are the dispersion relation in eq.~\eqref{eq:D3Dpdiffusionanalytic}.}
\label{fig:hydroQNMs-diffusion}
\end{figure}

In fig.~\ref{fig:hydroQNMs-diffusion} we present our numerical results for the dispersion relation of the charge diffusion mode in both D3/D$p$ systems with non-zero $B$ in the regime $\kbar\ll \Tbar^2$, which are extremely good approximations to the dispersion relations in eqs.~\eqref{eq:D3D5diffusionanalytic} and~\eqref{eq:D3D7diffusionanalytic}, as was the case when $B=0$ (recall fig.~\ref{fig:D3D5hydrocrossover-ZeroB}).

\subsection{Poles in the Complex Frequency Plane}
\label{sec:NumericalCollisionsSmallB}

Having found the dispersion relations of the long-lived excitations in both the $T=0$ limit and the hydrodynamic limit, we will now numerically investigate the collisionless/hydrodynamic crossover between these two limits. In this subsection we repeat the analysis of section~\ref{collectiveholo}, studying the motion of the poles in $G^{tt}_R(\omega,k)$ in the complex $\wbar$ plane as we increase $\Tbar$.

\begin{figure}[!htb]
\centering
  \subfigure{
      \includegraphics[width=0.45\textwidth]{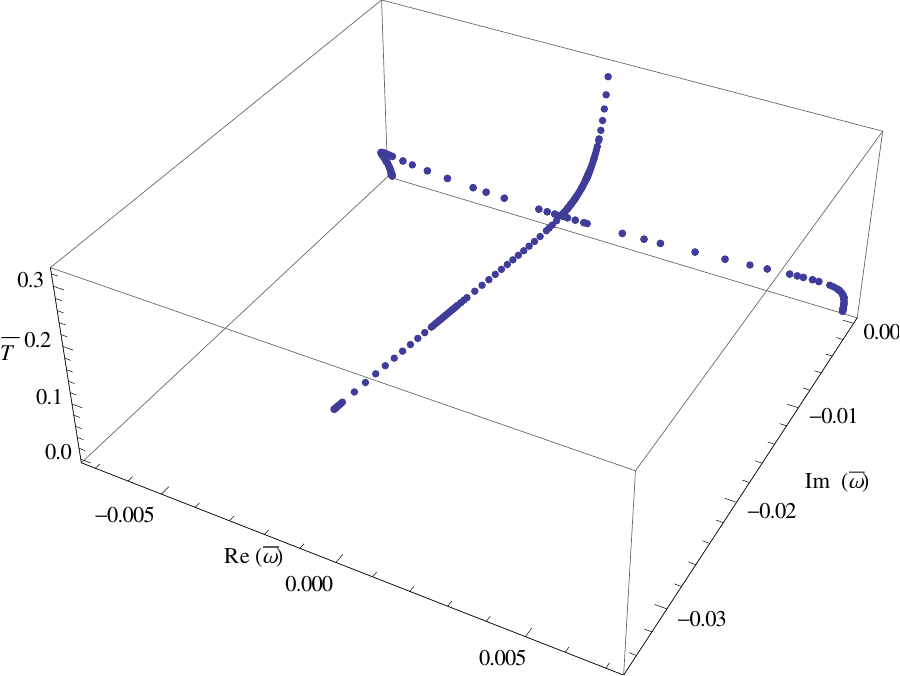}
}
\subfigure{
      \includegraphics[width=0.45\textwidth]{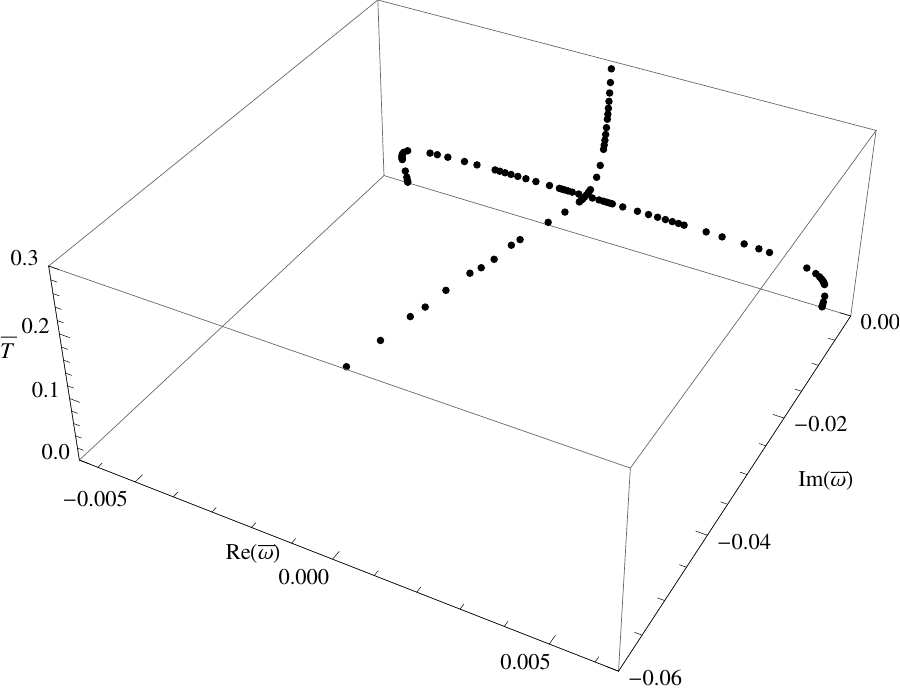}
      }
\begin{center}
\setlength{\unitlength}{0.1\columnwidth}
\begin{picture}(0.1,0.25)(0,0)
\put(0.8,3.75){\makebox(0,0){\textsf{sound modes}}}
\put(-0.1,3.7){\vector(-4,-1){2.1}}
\put(-0.1,3.7){\vector(-1,-1){0.85}}
\put(-4.65,4){\makebox(0,0){\textsf{diffusion mode}}}
\put(-3.75,3.95){\vector(4,-1){2.05}}
\put(0.45,0.75){\makebox(0,0){\textsf{short-lived mode}}}
\put(-0.55,0.75){\vector(-3,2){2.05}}
\end{picture}
\vskip-2em
\caption{\textbf{Left:} The space ($\Re\wbar,\Im\wbar,\Tbar$), with blue dots indicating our numerical results for the dominant poles in $G^{tt}_R(\omega,k)$ for the D3/D5 system with $\Bbar=0$ and $\kbar=0.01$. This plot is a three-dimensional representation of the left plot in fig.~\ref{fig:D3D5hydrocrossover-ZeroB}. \textbf{Right:} The same as the left plot, but for the D3/D7 system with $\Bbar=0$ and $\kbar=0.01$, and with black dots indicating our numerical results. In both cases we see that as we increase $\Tbar$ (moving vertically in the plot) the two holographic zero sound poles move along (approximate) semi-circles and eventually collide on the imaginary $\wbar$ axis, producing two purely imaginary poles. The purely imaginary pole closest to the origin is that of charge diffusion, while the other purely imaginary pole is short-lived.}
\label{fig:D3D5Fig4-ZeroB}
\end{center}
\vskip-1em
\end{figure}
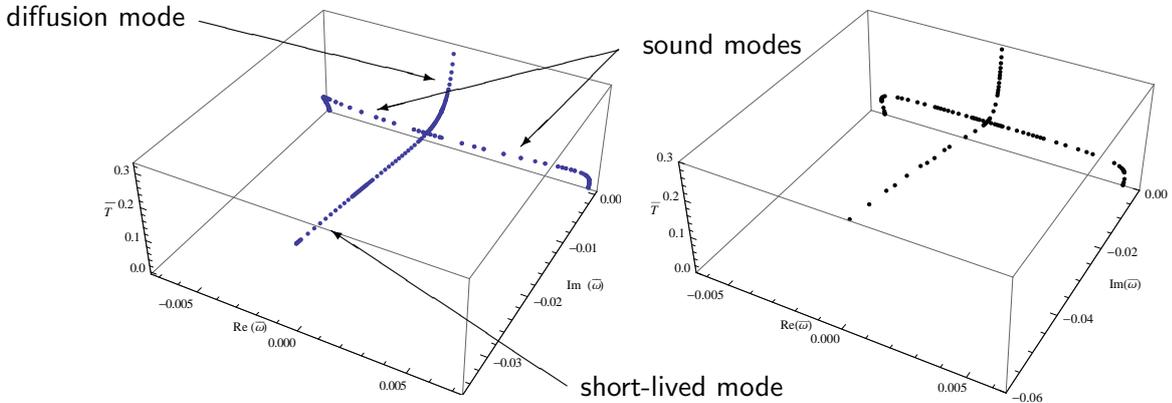

Recall from section~\ref{collectiveholo} that in both D3/D$p$ systems, when $B=0$ a collision of $G^{tt}_R(\omega,k)$'s poles occurs in the complex $\wbar$ plane: as we increase $\Tbar$ at fixed $\kbar$, the holographic zero sound poles move (approximately) along semi-circles before colliding on the imaginary $\wbar$ axis to form two purely imaginary poles. The purely imaginary pole closest to the origin is the charge diffusion mode. Indeed, as $\Tbar$ continues to increase, the charge diffusion pole approaches the origin, \textit{i.e.}\ becomes more stable, while the other purely imaginary pole moves away from the origin, becoming less stable. We summarize these $B=0$ results, for both D3/D$p$ systems, in fig.~\ref{fig:D3D5Fig4-ZeroB}.

\begin{figure}[!htb]
\centering
  \subfigure{
      \includegraphics[width=0.45\textwidth]{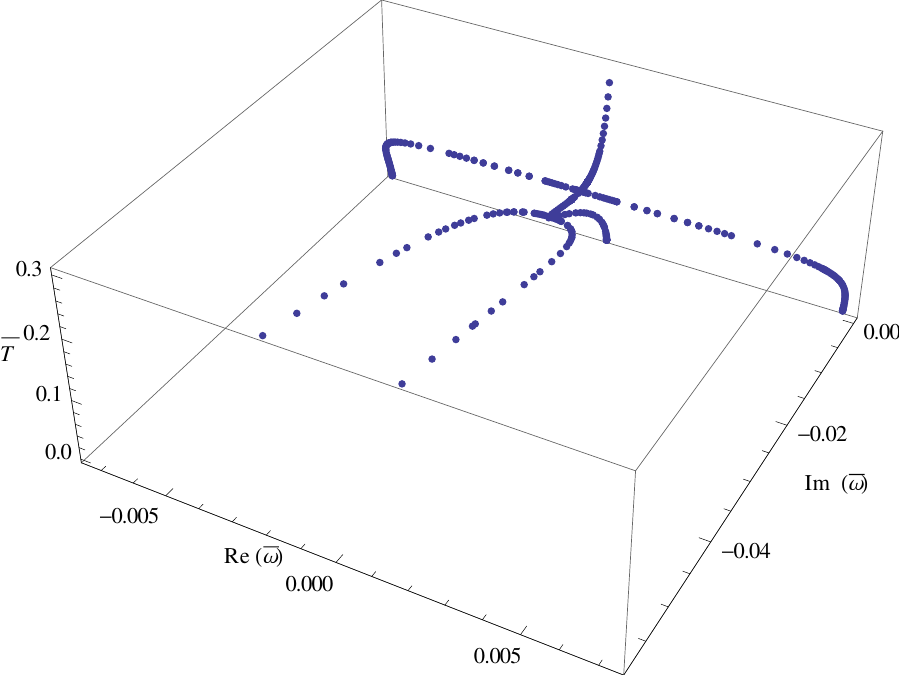}
}
\subfigure{
      \includegraphics[width=0.45\textwidth]{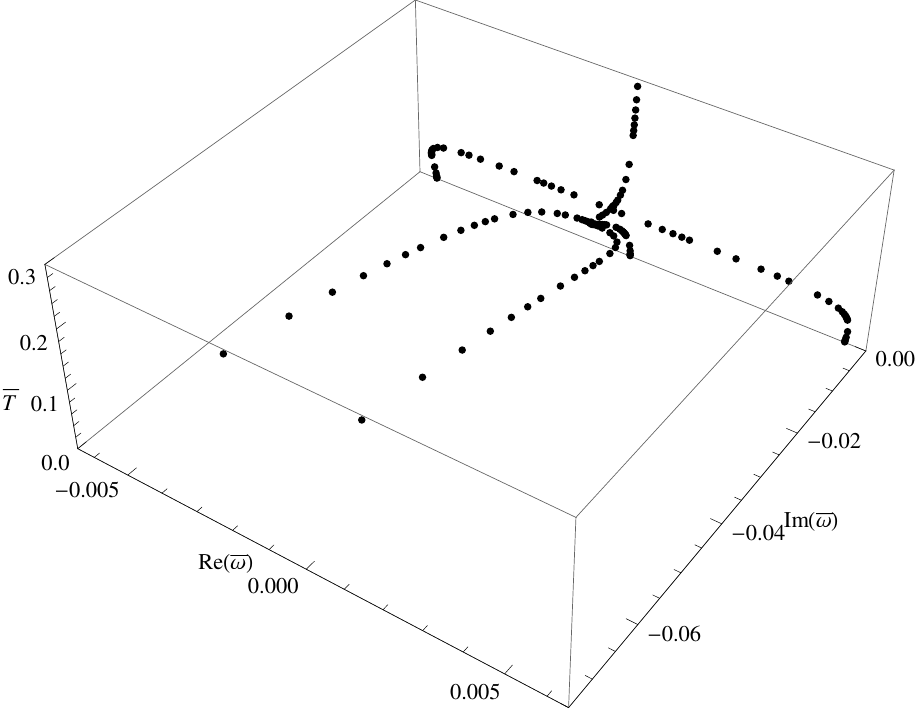}
      }
  \caption{\textbf{Left:} The space ($\Re\wbar,\Im\wbar,\Tbar$), with blue dots indicating our numerical results for the dominant poles in $G^{tt}_R(\omega,k)$ for the D3/D5 system with $\Bbar=0.003$ and $\kbar=0.01$. \textbf{Right:} The same as the left plot, but for the D3/D7 theory with $\Bbar=0.002$ and $\kbar=0.01$, and with black dots indicating our numerical results. In both plots, as we increase $\Tbar$ (moving vertically in the plot), the purely imaginary transverse pole moves down the $\textrm{Im}\,\wbar$ axis, while the holographic zero sound poles eventually collide on the $\textrm{Im}\,\wbar$ axis, above the transverse pole, producing two purely imaginary poles. One of these, corresponding to charge diffusion, moves up the $\textrm{Im}\,\wbar$ axis, while the other moves down, eventually intersecting the transverse pole and producing two propagating but short-lived modes (the ``horseshoe'' in the plot).}
  \label{fig:D3D5Fig4-NonZeroB}
\end{figure}

When $\Bbar$ is non-zero, the poles of $G_R^{yy}(\omega,k)$ mix with those of $G^{tt}_R(\omega,k)$, and thus the latter has an ``extra'' purely imaginary pole near the origin, which we henceforth call the ``transverse pole.'' For sufficiently small $\Bbar$ a collision of poles still occurs, however. At very low $\Tbar$, the three poles of $G^{tt}_R(\omega,k)$ closest to the origin are the two holographic zero sound poles and the transverse pole. As we increase $\Tbar$, the transverse pole moves down the $\textrm{Im}\,\wbar$ axis, whereas the holographic zero sound poles again collide on the $\textrm{Im}\,\wbar$ axis and produce two purely imaginary poles at a point closer to the origin than the transverse pole. One of these purely imaginary poles is the charge diffusion mode, which becomes more stable, moving up the $\textrm{Im}\,\wbar$ axis, as we increase $\Tbar$ further. The second purely imaginary mode becomes less stable as we increase $\Tbar$, and eventually collides with the transverse pole to form two propagating modes, \textit{i.e.}\ two poles with nonzero real parts. Increasing $\Tbar$ even more, these propagating modes move down into the complex $\wbar$ plane, becoming very short-lived, with decay rates $-\textrm{Im}\,\wbar(\kbar)$ increasing as $\Tbar$ increases. We present some small-$\Bbar$ results in fig.~\ref{fig:D3D5Fig4-NonZeroB}, where we depict the motion of poles in the complex $\wbar$ plane with $\Bbar=0.003$ for the D3/D5 system and $\Bbar=0.002$ for the D3/D7 system.

We emphasise that at a fixed value of $\Tbar$, \textit{i.e.}\ on a horizontal slice of figure \ref{fig:D3D5Fig4-NonZeroB}, two kinds of pole appear near the origin of the complex $\wbar$ plane. At low temperatures, these are the transverse pole and the poles of the holographic zero sound modes. At high temperatures, the three poles are the charge diffusion mode and the short-lived propagating modes.

\begin{figure}[!htb]
\centering
  \subfigure{
      \includegraphics[width=0.45\textwidth]{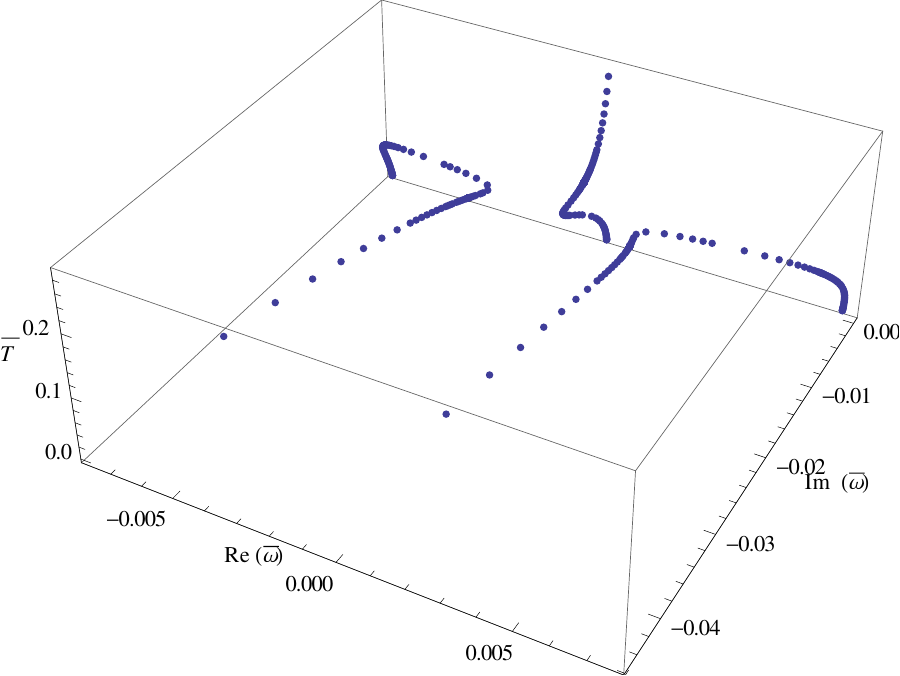}
}
\subfigure{
      \includegraphics[width=0.45\textwidth]{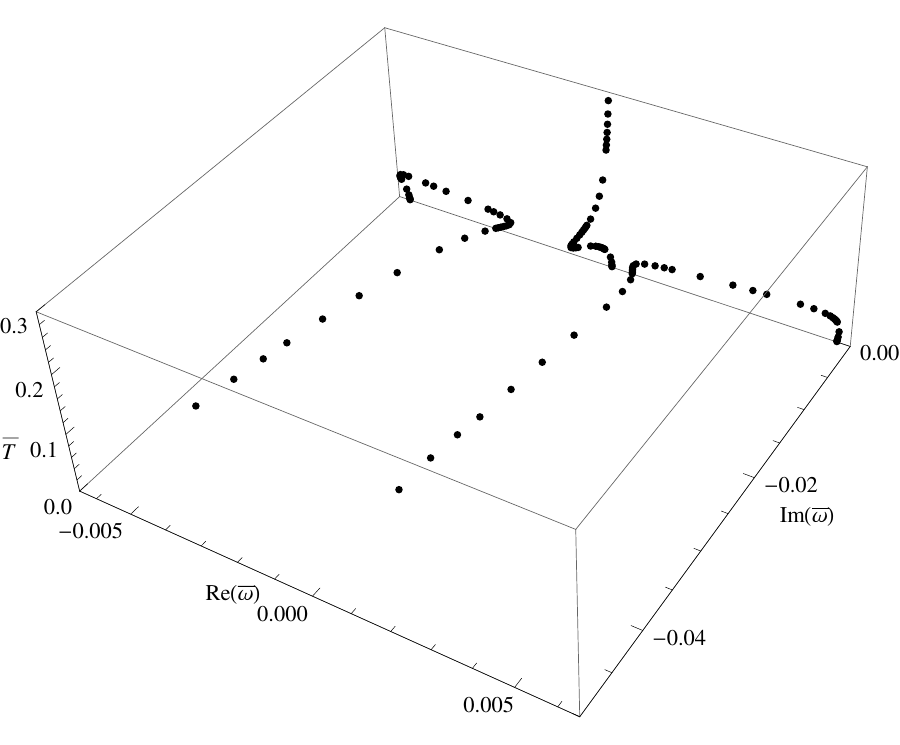}
      }
  \caption{\textbf{Left:} The space ($\Re\wbar,\Im\wbar,\Tbar$), with blue dots indicating our numerical results for the dominant poles in $G^{tt}_R(\omega,k)$ for the D3/D5 system with $\Bbar=0.005$ and $\kbar=0.01$. \textbf{Right:} The same as the left plot, but for the D3/D7 theory with $\Bbar=0.003$ and $\kbar=0.01$, and with black dots indicating our numerical results. In both plots, as we increase $\Tbar$ (moving vertically in the plot), the holographic zero sound poles move towards the $\textrm{Im}\,\wbar$ axis, reach a minimum value of $\textrm{Re}\,\wbar$, and then move away from the $\textrm{Im}\,\wbar$ axis, remaining as propagating modes at higher values of $\Tbar$. The purely imaginary transverse pole moves down the $\textrm{Im}\,\wbar$ axis, reaching a minimum value, and then moves back up the axis, eventually becoming the charge diffusion pole.}
  \label{fig:D3D5Fig4-NonZeroLargeB}
\end{figure}

For small $\Bbar$ we can define the location of the collisionless/hydrodynamic crossover by the collision of poles in the complex $\wbar$ plane, as in the $\Bbar=0$ case, however the story is not so simple for larger $\Bbar$. For sufficiently large $\Bbar$, as we increase $\Tbar$ the holographic zero sound poles begin to move towards the $\textrm{Im}\,\wbar$ axis before turning around and remaining as propagating modes at higher $\Tbar$. Most importantly, the holographic zero sound poles never intersect the $\textrm{Im}\,\wbar$ axis, \textit{i.e.}\ no collision of poles occurs. As we increase $\Tbar$, the transverse pole moves down the $\textrm{Im}\,\wbar$ axis, reaches a minimum, and then moves back up the axis, eventually becoming the charge diffusion pole. We present some large-$\Bbar$ results in fig.~\ref{fig:D3D5Fig4-NonZeroLargeB}, where we depict the motion of poles in the complex $\wbar$ plane with $\Bbar=0.005$ for the D3/D5 system and $\Bbar=0.003$ for the D3/D7 system.

At even higher values of $\Bbar$ than discussed above, the holographic zero sound poles do not appear to approach the $\textrm{Im}\,\wbar$ axis at all, but rather seem to move directly away from it as $\Tbar$ increases. The transverse pole's behavior is similar to that in fig.~\ref{fig:D3D5Fig4-NonZeroLargeB}, \textit{i.e.}\ as $\Tbar$ increases the transverse pole moves down the $\textrm{Im}\,\wbar$ axis, reaches a minimum, and then moves back up the axis, eventually becoming the charge diffusion mode.\footnote{We have presented numerical results for poles of $G^{tt}_R(\omega,k)$ in the complex $\wbar$ plane with fixed momentum $\kbar$ and increasing $\Tbar$. If we instead fix $\Tbar$ and increase $\kbar$, then we find results similar to those of the compressible states studied, using holography, in ref.~\cite{Jokela:2012vn}, namely $\N=4$ SYM coupled to (2+1)-dimensional flavor fermions alone, rather than a hypermultiplet, with non-zero $U(1)_b$ charge density. In that case, the holographic dual was a probe D-brane with non-zero worldvolume electric and magnetic fluxes, including magnetic fluxes along directions of the internal space wrapped by the D-brane.}

\begin{figure}[!htb]
  \centering
  \subfigure{
      \includegraphics[width=0.45\textwidth]{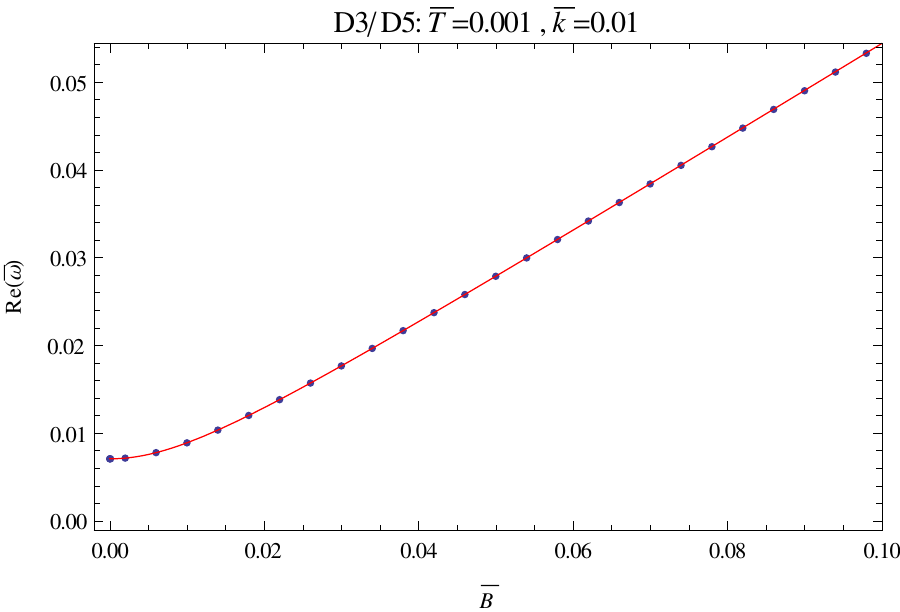}
}
\subfigure{
      \includegraphics[width=0.45\textwidth]{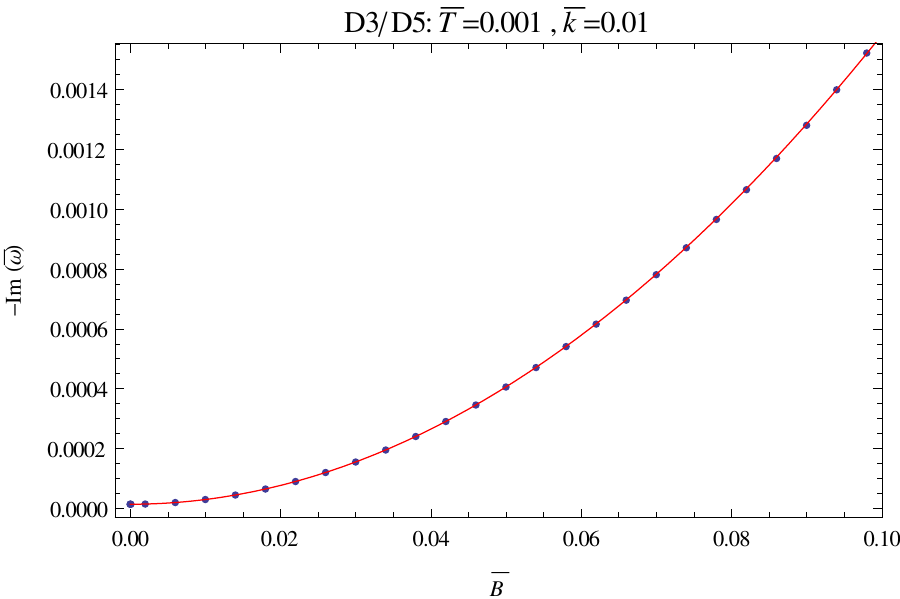}
      }
\\
\subfigure{
      \includegraphics[width=0.45\textwidth]{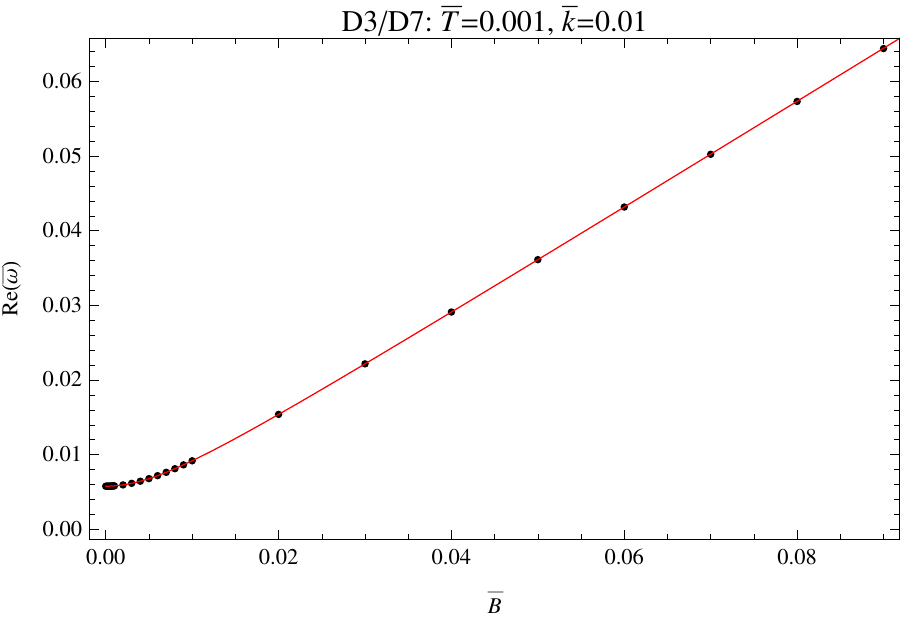}
}
\subfigure{
      \includegraphics[width=0.45\textwidth]{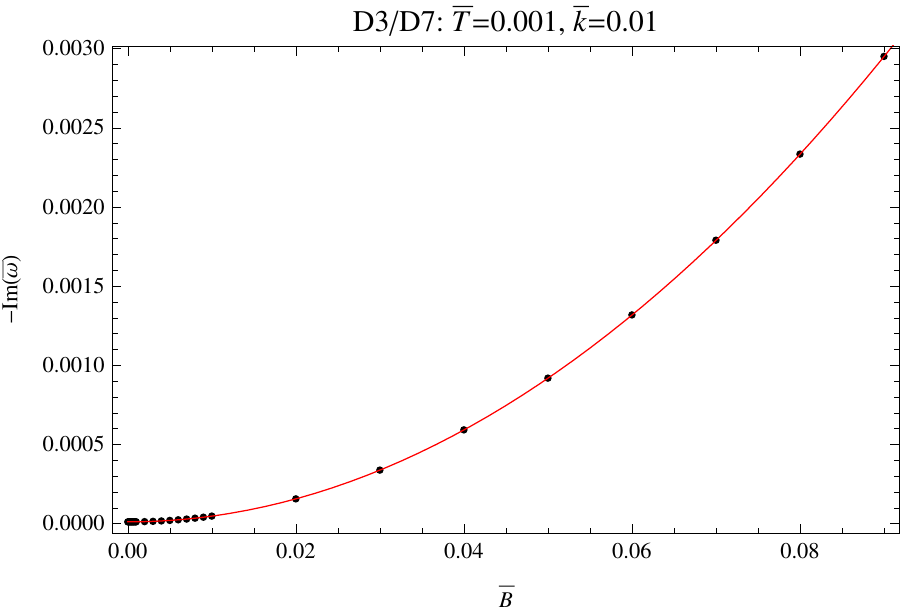}
      }
  \caption{\textbf{Top:} $\textrm{Re}\,\wbar$ (\textbf{Left}) and $-\textrm{Im}\,\wbar$ (\textbf{Right}) as functions of $\Bbar$ for the holographic zero sound poles of the D3/D5 system with $\kbar=0.01$ and $\Tbar=0.001$, within the range $\kbar\gg\Tbar$. \textbf{Bottom:} The same as the top plots, but for the D3/D7 system. In all plots the dots are our numerical results and the solid red lines come from the $\Tbar=0$ dispersion relation in eq.~\eqref{HZSanalytic}. Clearly eq.~\eqref{HZSanalytic} provides a very good approximation to the numerical results.}
  \label{fig:collisionlessQNMs-Bdependence}
\end{figure}

\begin{figure}[!htb]
  \centering
  \subfigure{
      \includegraphics[width=0.45\textwidth]{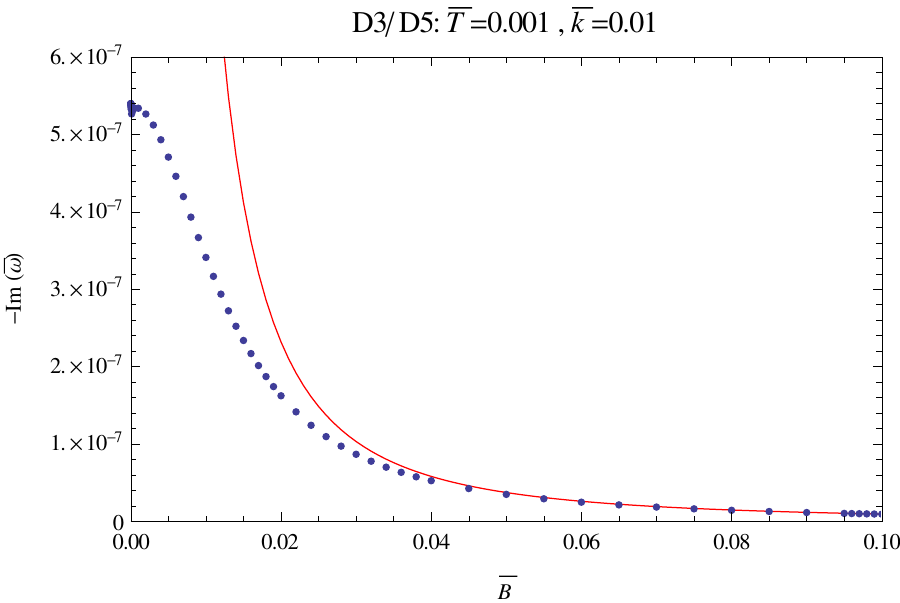}
}
\subfigure{
      \includegraphics[width=0.45\textwidth]{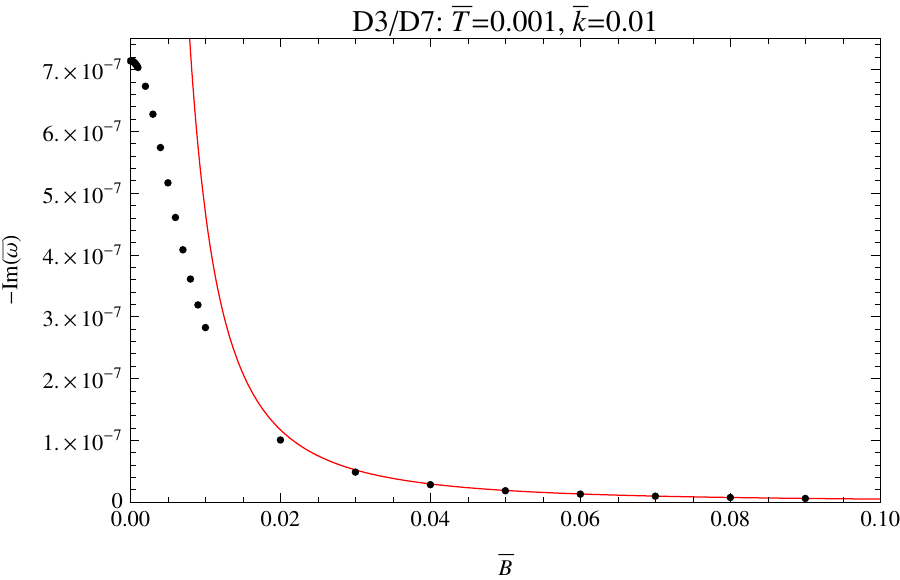}
      }
  \caption{\textbf{Left:} $-\textrm{Im}\,\wbar$ as a function of $\Bbar$ for the longest-lived purely imaginary pole of the D3/D5 system with $\kbar=0.01$ and $\Tbar=0.001$, within the range $\kbar\gg\Tbar$. \textbf{Right:} The same as the left plot, but for the D3/D7 system. In each plot the dots are our numerical results and the solid red line comes from the dispersion relations in eq.~\eqref{eq:D3D5diffusionanalytic} or~\eqref{eq:D3D7diffusionanalytic}. For sufficiently large $\Bbar$, we see that the purely imaginary pole becomes that of charge diffusion.}
  \label{fig:collisionlessQNMsPI}
\end{figure}

As at smaller $\Bbar$, at a fixed temperature $\Tbar$ in the large-$\Bbar$ regime, \textit{i.e.}\ on a horizontal slice of fig.~\ref{fig:D3D5Fig4-NonZeroLargeB}, two kinds of mode are always present, one purely imaginary and the other propagating, however, unlike the small-$\Bbar$ regime, here we can show that the purely imaginary mode is charge diffusion and the propagating modes are holographic zero sound. Consider for example the temperature range $\kbar\gg\Tbar$ where, when $\Bbar=0$, the holographic zero sound mode is present, with a dispersion relation that approximately obeys the $T=0$ dispersion relation eq.\eqref{eq:zeroTdispersionrelation}, as shown in fig.~\ref{fig:D3D5CQQNMs-ZeroB}. In fig.~\ref{fig:collisionlessQNMs-Bdependence} we present our numerical results for the dispersion relation of the holographic zero sound as a function of $\Bbar$ in both D3/D$p$ systems. We find that our numerical results are very well approximated by the $\Tbar=0$ result, eq.~\eqref{HZSanalytic}, for all $\Bbar$. In fig.~\ref{fig:collisionlessQNMsPI}, we present our numerical results for the dispersion relation of the longest-lived purely imaginary pole as a function of $\Bbar$ in the temperature range $\kbar\gg\Tbar$. We find that for large enough $\Bbar$, this purely imaginary pole is simply the charge diffusion mode. In short, we find that for a fixed $\kbar\gg\Tbar$, both the holographic zero sound and charge diffusion modes coexist at sufficiently high $\Bbar$.

In physical terms, we have learned that if we perturb these systems at a fixed momentum $\kbar$, then for sufficiently large $\Bbar$ we will see both holographic zero sound and charge diffusion, hence for large $\Bbar$ we cannot characterize the reponse of the system as either collisionless or hydrodynamic. We then seem to have no way of defining a collisionless/hydrodynamic crossover. The story will be very different if we perturb these systems with fixed real frequency $\wbar$, as we will now show.

\subsection{Charge Density Spectral Functions}
\label{spectralfunctions}

The poles of $G^{tt}_R(\omega,k)$ in the complex $\wbar$ plane are not directly observable. To gain perspective, we thus turn to a quantity that is directly observable (in principle), the charge density spectral function $\bar{\chi}_{tt}\left(\wbar,\kbar\right)$.

We have so far considered perturbations with fixed real momentum $\kbar$. In such cases, the charge density spectral function $\bar{\chi}_{tt}\left(\wbar,\kbar\right)$, as a function of $\wbar$, will exhibit peaks at frequencies $\wbar$ determined by the positions (and residues) of the poles in the complex $\wbar$ plane. For example, when $\Bbar=0$, as we increase $\Tbar$, the spectral function always exhibits only a single dominant peak that moves towards the origin, reflecting the movement of the corresponding pole in the complex $\wbar$ plane towards the $\textrm{Im}\,\wbar$ axis, as we saw in fig.~\ref{fig:D3D5chargespectral-ZeroB}.

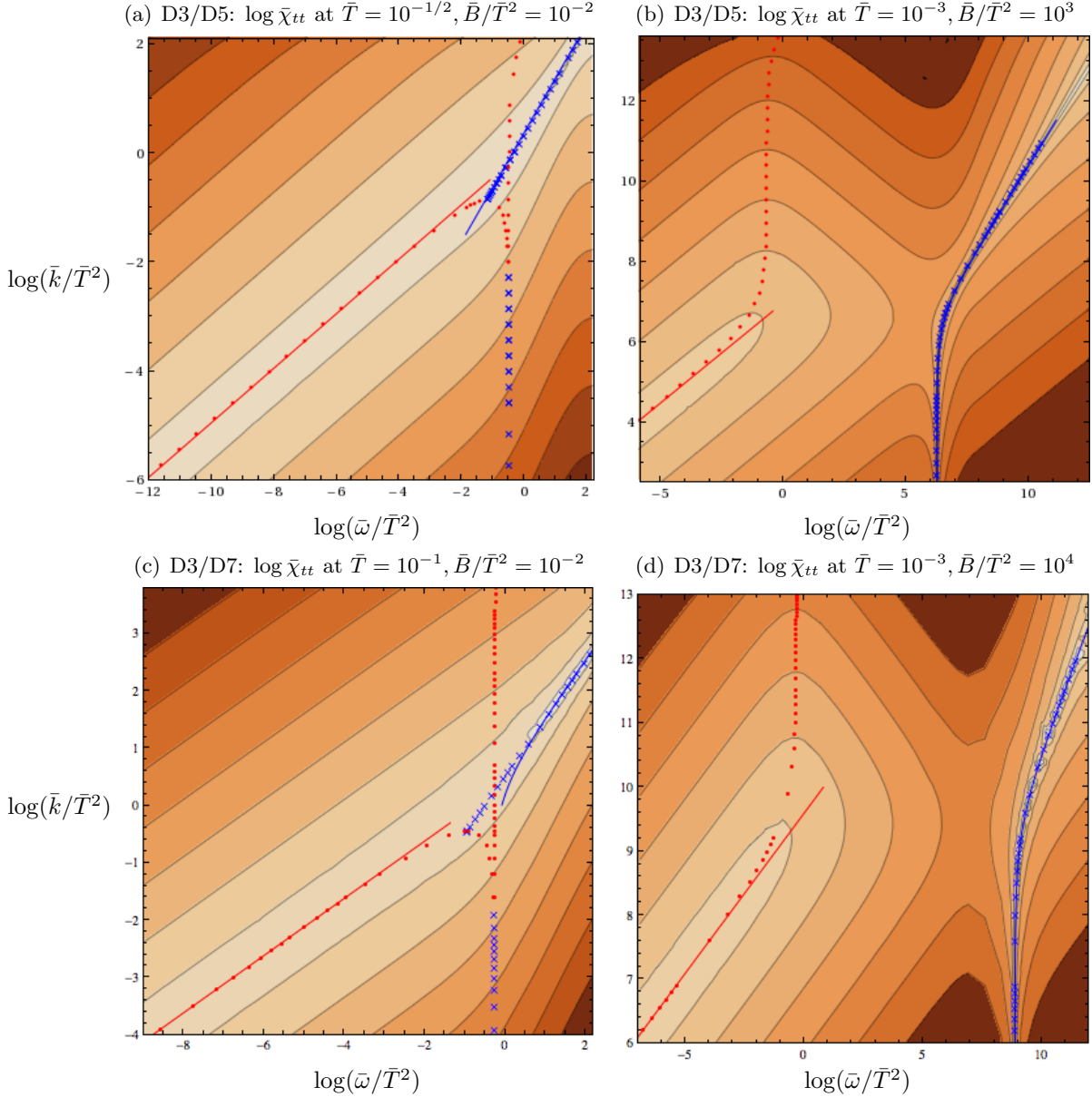
\begin{figure}[!htb]
\centering
\subfigure[D3/D5: $\log\bar{\chi}_{tt}$ at $\Tbar=10^{-1/2}, \Bbar/\Tbar^2=10^{-2}$]{      \includegraphics[width=0.45\textwidth]{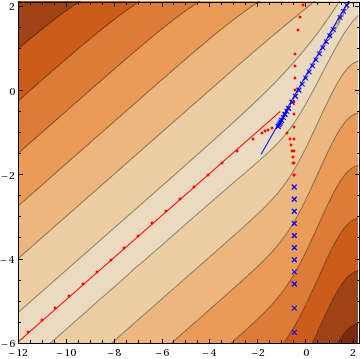}
\vspace{0.25em}
}
\subfigure[D3/D5: $\log\bar{\chi}_{tt}$ at $\Tbar=10^{-3}, \Bbar/\Tbar^2=10^{3}$]{
\includegraphics[width=0.45\textwidth]{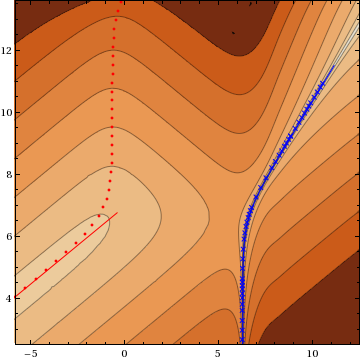}
}
\\ 
\vskip1em
\subfigure[D3/D7: $\log\bar{\chi}_{tt}$ at $\Tbar=10^{-1}, \Bbar/\Tbar^2=10^{-2}$]{      \includegraphics[width=0.45\textwidth]{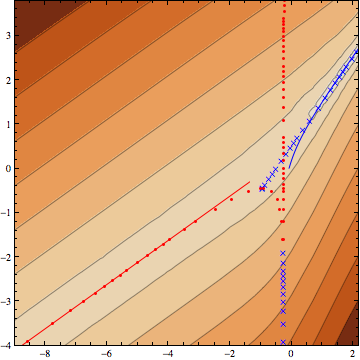}
}
\subfigure[D3/D7: $\log\bar{\chi}_{tt}$ at $\Tbar=10^{-3}, \Bbar/\Tbar^2=10^{4}$]{
\includegraphics[width=0.45\textwidth]{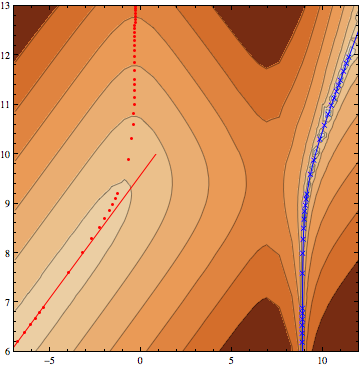}
}
\begin{center}
\setlength{\unitlength}{0.1\columnwidth}
\begin{picture}(0.1,0.25)(0,0)
\put(-2.25,5.75){\makebox(0,0){\small $\log(\wbar/\Tbar^2)$}}
\put(2.45,5.75){\makebox(0,0){\small $\log(\wbar/\Tbar^2)$}}
\put(-2.25,0.4){\makebox(0,0){\small $\log(\wbar/\Tbar^2)$}}
\put(2.45,0.4){\makebox(0,0){\small $\log(\wbar/\Tbar^2)$}}
\put(-5.2,8.15){\makebox(0,0){\small $\log(\kbar/\Tbar^2)$}}
\put(-5.2,3.05){\makebox(0,0){\small $\log(\kbar/\Tbar^2)$}}
\end{picture}
\vskip-1.5em
\caption{\textbf{(a)}  Contour plot of our numerical results for $\log \bar{\chi}_{tt}(\wbar,\kbar)$ over the $\left(\log\wbar/\Tbar^2,\log\kbar/\Tbar^2\right)$ plane in the D3/D5 system with $\Tbar=10^{-1/2}$ and $\Bbar/\Tbar^2=10^{-2}$. Lighter colours represent larger values of the spectral function, with $\log\bar{\chi}_{tt}=-15,-12.5,-10,\ldots,2.5$ from darkest to lightest. The red dots and blue crosses are our numerical results for the locations of purely imaginary and propagating poles in $G^{tt}_R(\omega,k)$, respectively. The solid red and blue lines are the values of $\left|\wbar(\kbar)\right|$ for the dispersion relations of the charge diffusion, eq.~\eqref{eq:D3Dpdiffusionanalytic}, and holographic zero sound, eq.~\eqref{HZSanalytic}, respectively. \textbf{(b)} The same as (a), but with $\Tbar=10^{-3}$ and $\Bbar/\Tbar^2=10^3$. \textbf{(c)} The same as (a) and (b), but for the D3/D7 system with $\Tbar=10^{-1}$ and $\Bbar/\Tbar^2=10^{-2}$ and with $\log\bar{\chi}_{tt}=-10,-8,-6,\ldots,2$ from darkest to lightest. \textbf{(d)} The same as (c), but with $\Tbar=10^{-3}$ and $\Bbar/\Tbar^2=10^4$.} 
\label{fig:3dSpectral}
\end{center}
\vskip-1.5em
\end{figure}

In fig.~\ref{fig:3dSpectral} we present our numerical results for $\log\bar{\chi}_{tt}\left(\wbar,\kbar\right)$ in the $\left(\log\wbar/\Tbar^2,\log\kbar/\Tbar^2\right)$ plane for both D3/D$p$ systems at two different values of $\Bbar/\Tbar^2$. Here we chose to normalize $\wbar$ and $\kbar$ by $\Tbar^2$ because in the $B=0$ case $\wbar/\Tbar^2$ and $\kbar/\Tbar^2$ are order one at the crossover. Given those normalizations and the fact that the gap in the $T=0$ holographic zero sound dispersion relation goes like $\wbar_{\mathrm{gap}} \propto \Bbar$, the natural normalization for $\Bbar$ was then $\Bbar/\Tbar^2$. To make the relationship between poles of $G^{tt}_R(\omega,k)$ and peaks in $\bar{\chi}_{tt}\left(\wbar,\kbar\right)$ explicit, in fig.~\ref{fig:3dSpectral} we also superimposed on the plots our numerical results for the locations of poles of $G^{tt}_R(\omega,k)$ in the complex $\wbar$ plane (the red dots and blue crosses) as well as the absolute values of the dispersion relations of the long-lived modes, \textit{i.e.}\ we plot the value of $\left|\wbar\left(\kbar\right)\right|$ for the holographic zero sound dispersion relation in eq.~\eqref{HZSanalytic} and the charge diffusion dispersion relations in eqs.~\eqref{eq:D3D5diffusionanalytic} and~\eqref{eq:D3D7diffusionanalytic}.\footnote{\label{footnote}In general, the precise location of the peak in $\bar{\chi}_{tt}\left(\wbar,\kbar\right)$ actually depends on the location and residue of the pole in $G^{tt}_R(\omega,k)$ in a complicated way. The results depicted in fig.~\ref{fig:3dSpectral} show that for our systems $\left|\wbar\left(\kbar\right)\right|$ gives a very good approximation to the location of the peak. Notice that for a long-lived, propagating mode, this agrees with the usual expectation that the peak is located at the propagating frequency.}

Consider first the plots of $\log\bar{\chi}_{tt}\left(\wbar,\kbar\right)$ with $\Bbar/\Tbar^2\ll1$, figs.~\ref{fig:3dSpectral} (a) and (c). For a fixed value of the momentum $\kbar$, \textit{i.e.}\ along a horizontal line across the contour plot, the spectral function exhibits only a single peak, which for $\kbar\gg\Tbar^2$, near the top of each plot, is due to the holographic zero sound pole, as indicated by the blue crosses and solid blue line, and for $\kbar\ll\Tbar^2$, near the bottom of each plot, is due to the charge diffusion mode, as indicated by the red dots and solid red line. When $\kbar\simeq\Tbar^2$, these two peaks join, reflecting the collision of the corresponding poles of $G^{tt}_R(\omega,k)$ in the complex $\wbar$ plane. The ``extra'' poles due to mixing with $G^{yy}_R(\omega,k)$ do not produce any significant peaks in the spectral function at these low values of $\Bbar/\Tbar^2$. These plots provide additional evidence that at low $\Bbar$ and with fixed $\kbar$, a crossover occurs similar to the crossover at $\Bbar=0$.

Now consider the plots of $\log\bar{\chi}_{tt}\left(\wbar,\kbar\right)$ with $\Bbar/\Tbar^2\gg1$, figs.~\ref{fig:3dSpectral} (b) and (d). These are qualitatively different from the $\Bbar/\Tbar^2\ll1$ plots: now, at a fixed value of $\kbar$ (a horizontal line across fig.~\ref{fig:3dSpectral} (b) or (d)), the spectral function exhibits \textit{two} peaks, not one peak. At very high $\kbar/\Tbar^2$, near the top of each plot, the holographic zero sound peak is dominant, and at very low $\kbar/\Tbar^2$, near the bottom of each plot, the charge diffusion peak is dominant. In the intermediate region, however, the holographic zero sound and charge diffusion peaks coexist with roughly equal spectral weight. These results agree with what we found by studying poles in the complex frequency plane in the last subsection. In particular, in figs.~\ref{fig:3dSpectral} (b) and (d) the holographic zero sound and charge diffusion peaks are no longer joined: as we saw in the previous subsection, the corresponding poles in the $\wbar$ plane do not collide.

We have so far merely confirmed the analysis of the previous subsection. Now, however, let us consider $\bar{\chi}_{tt}(\wbar,\kbar)$ as a function of $\kbar$ at a fixed frequency $\wbar$, \textit{i.e.}\ a \textit{vertical} line in any of the plots of fig.~\ref{fig:3dSpectral}. With fixed $\wbar$, we find that for any $\Bbar$ and any $\wbar$ the spectral function exhibits \textit{only one peak}, which at low $\wbar/\Tbar^2$, near the left in each plot, is due to charge diffusion, and at high $\wbar/\Tbar^2$, near the right in each plot, is due to holographic zero sound. We also find that for larger $\Bbar$, figs.~\ref{fig:3dSpectral} (b) and (d), at intermediate frequencies the spectral function is suppressed by orders of magnitude relative to the holographic zero sound and charge diffusion peaks. That suppression is due to the gap in the holographic zero sound dispersion, as we discuss in more detail in the next subsection. At smaller $\Bbar$, such as figs.~\ref{fig:3dSpectral} (a) and (c), our numerical results indicate that if some suppression of spectral weight occurs in some non-zero range of frequencies at small $\Bbar$, then either the suppression is very small or the range of frequencies is very narrow.

By studying the spectral function $\bar{\chi}_{tt}(\omega,k)$, we have learned that the response of the system is simplest to understand using perturbations with fixed frequency $\wbar$ rather than fixed $\kbar$: with fixed $\wbar$ the spectral function always exhibits only a single peak for any $\Bbar$, in which case we should be able to characterize the response of the system as either collisionless or hydrodynamic. We are thus able to define a collisionless/hydrodynamic crossover, as we show in the next subsection.

\subsection{Poles in the Complex Momentum Plane}
\label{sec:NumericalResultsKPlane}

We saw in the previous subsection that the spectral function $\bar{\chi}_{tt}(\omega,k)$, as a function of momentum $\kbar$ at a fixed frequency $\wbar$, exhibits only a single peak, which in physical terms means the response of the system to an external perturbation with frequency $\wbar$ is dominated by a single excitation. To study that excitation, we will determine the poles of $G^{tt}_R(\omega,k)$ in the complex $\kbar$ plane. We will look for poles with $\Re(\kbar)>0$ and $\Im(\kbar)>0$, and will focus on the poles closest to the origin of the complex $\kbar$ plane, corresponding to the modes with the smallest propagating momentum $\Re(\kbar)$ and attenuation $\Im(\kbar)$.

\begin{figure}[!htb]
  \centering
\subfigure{
      \includegraphics[width=0.45\textwidth]{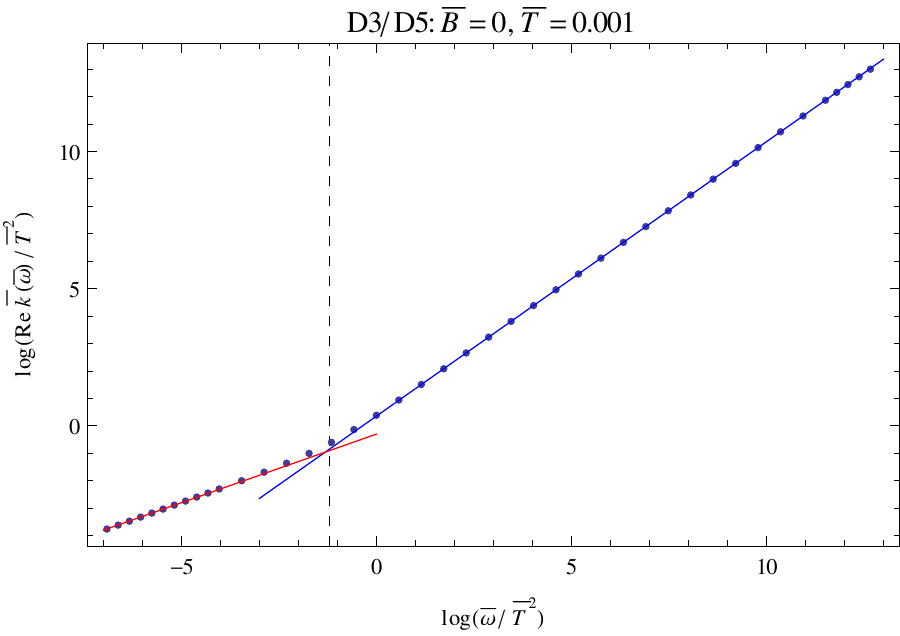}
}
\subfigure{
      \includegraphics[width=0.45\textwidth]{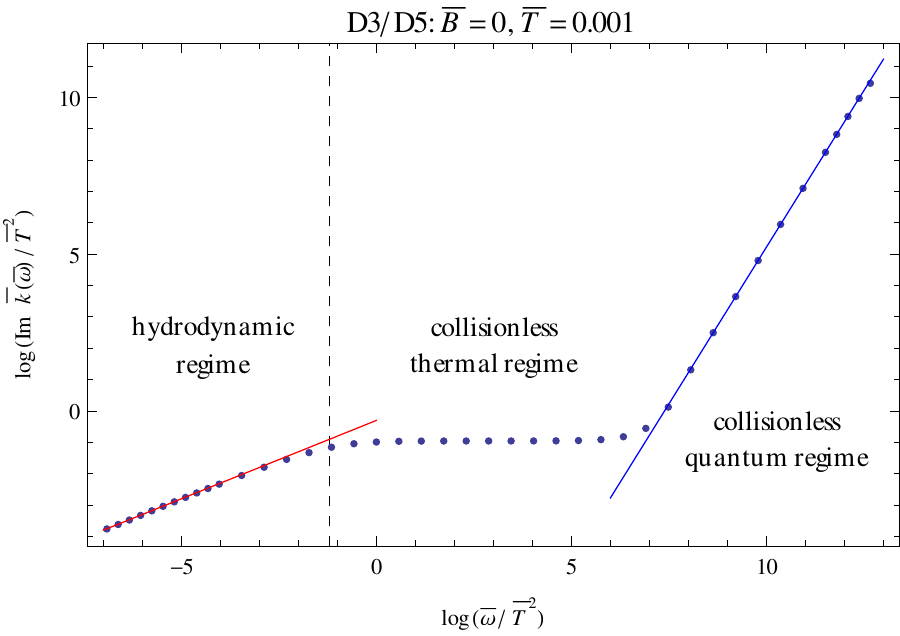}
      }
\subfigure{
      \includegraphics[width=0.45\textwidth]{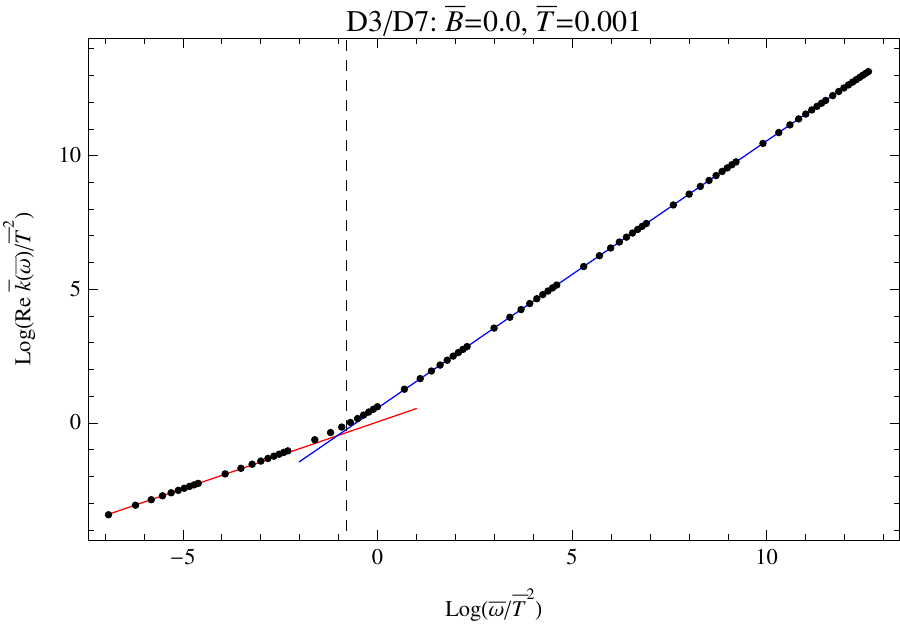}
}
\subfigure{
      \includegraphics[width=0.45\textwidth]{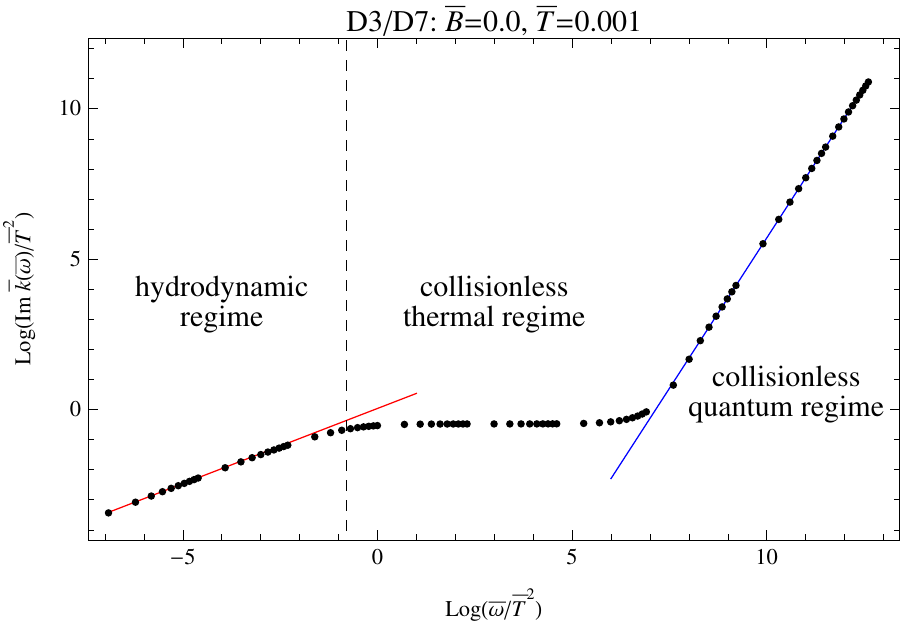}
      }
  \caption{\textbf{Top:} $\log \left(\textrm{Re}\,\kbar\left(\wbar\right)/\Tbar^2\right)$ (\textbf{Left}) and $\log \left(\textrm{Im}\,\kbar\left(\wbar\right)/\Tbar^2\right)$ (\textbf{Right}) as functions of $\log \left(\wbar/\Tbar^2\right)$ for the pole in $G^{tt}_R(\omega,k)$ closest to the origin of the complex $\kbar$ plane in the D3/D5 system with $\Bbar=0$ and $\Tbar=0.001$. The dots are our numerical results, the solid red lines, beginning on the left of each plot, come from the dispersion relation of the charge diffusion mode obtained by inverting eq.~\eqref{eq:D3Dpdiffusionanalytic}, the solid blue lines, beginning on the right of each plot, come from the holographic zero sound dispersion relation at $\Tbar=0$, eq.~\eqref{eq:HZSanalyticKofW}, and the vertical dashed line is the location of the collisionless/hydrodynamic crossover, as defined by a collision of poles in the complex $\wbar$ plane, eq.~\eqref{eq:ZeroBCrossoverLocation}. \textbf{Bottom:} The same as the top plots, but for the D3/D7 system. On the right plots we also indicate the hydrodynamic, collisionless thermal, and collisionless quantum regimes, as defined in section~\ref{collectiveholo}.}
  \label{fig:kofwB0}
\end{figure}

First, let us revisit the known collisionless/hydrodynamic crossover at $B=0$, now studying the behavior of poles in the complex $\kbar$ plane. In fig.~\ref{fig:kofwB0} we plot $\log \left(\textrm{Re}\,\kbar\left(\wbar\right)/\Tbar^2\right)$ and $\log \left(\textrm{Im}\,\kbar\left(\wbar\right)/\Tbar^2\right)$ of the pole closest to the origin as functions of $\log \left(\wbar/\Tbar^2\right)$, for both D3/D$p$ systems with $\Bbar=0$ and $\Tbar=0.001$. In fig.~\ref{fig:kofwB0} we also plot the dispersion relations of the holographic zero sound mode at $\Tbar=0$, eq~\eqref{eq:HZSanalyticKofW}, and of the charge diffusion poles in the complex $\kbar$ plane, obtained by inverting eqs.~\eqref{eq:D3D5diffusionanalytic} and~\eqref{eq:D3D7diffusionanalytic}, as well as the location of the collisionless/hydrodynamic crossover extracted from the collision of poles in the complex $\wbar$ plane, eq.~\eqref{eq:ZeroBCrossoverLocation}. We see that at the crossover, no collision of poles occurs in the complex $\kbar$ plane. Instead, we see a single pole whose dispersion relation $\kbar(\wbar)$ changes at the crossover. For example, we can see from the left plots of fig.~\ref{fig:kofwB0} that $\textrm{Re}\,\kbar(\wbar)$ exhibits a ``kink'' at the crossover in both D3/D$p$ systems.

\begin{figure}[!htb]
  \centering
\subfigure{
      \includegraphics[width=0.45\textwidth]{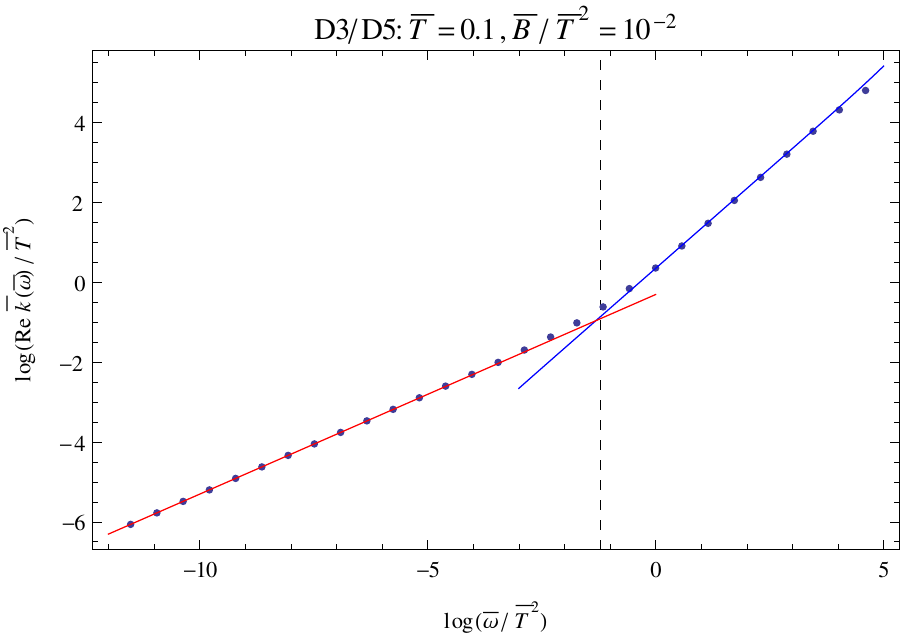}
}
\subfigure{
      \includegraphics[width=0.45\textwidth]{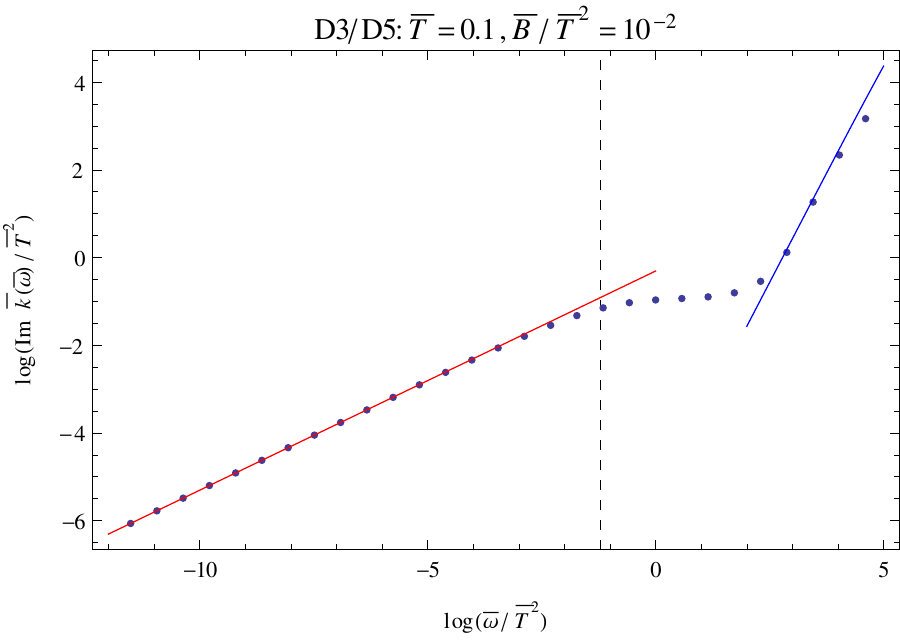}
      }
\subfigure{
      \includegraphics[width=0.45\textwidth]{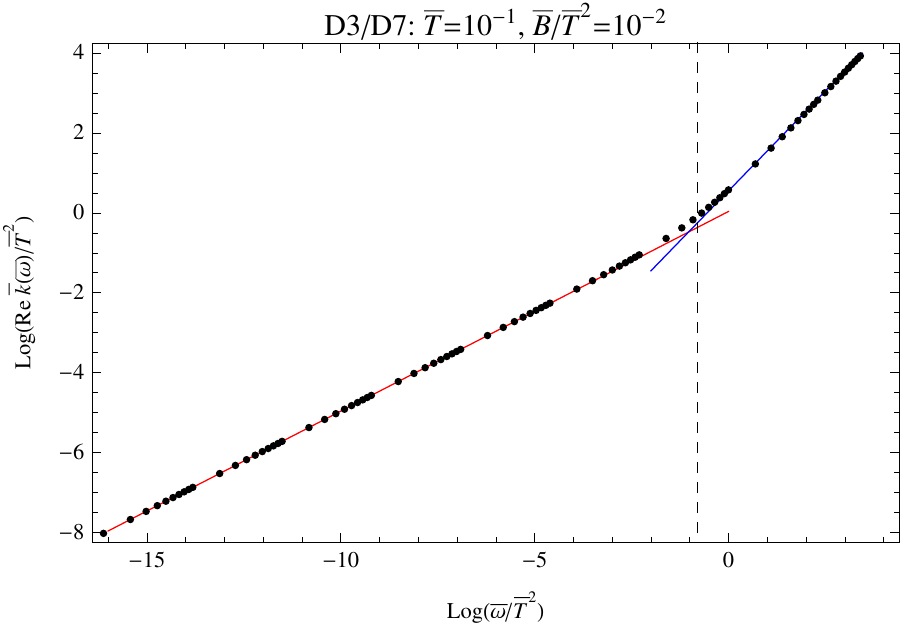}
}
\subfigure{
      \includegraphics[width=0.45\textwidth]{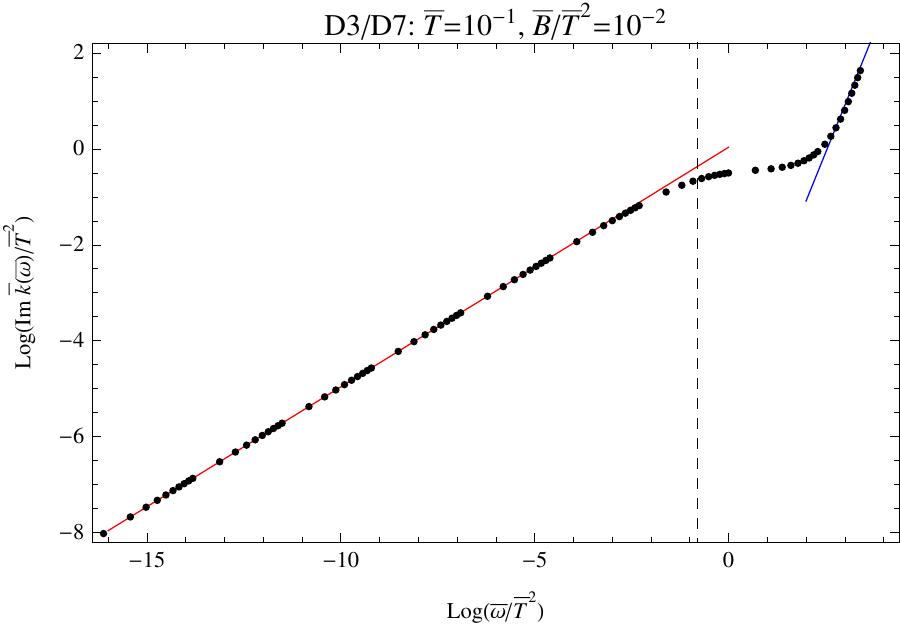}
      }
  \caption{\textbf{Top:} $\log \left(\textrm{Re}\,\kbar\left(\wbar\right)/\Tbar^2\right)$ (\textbf{Left}) and $\log \left(\textrm{Im}\,\kbar\left(\wbar\right)/\Tbar^2\right)$ (\textbf{Right}) as functions of $\log \left(\wbar/\Tbar^2\right)$ for the pole in $G^{tt}_R(\omega,k)$ closest to the origin of the complex $\kbar$ plane in the D3/D5 system with $\Tbar=0.1$ and $\Bbar/\Tbar^2=10^{-2}$, within the regime $\Bbar/\Tbar^2 \ll 1$. The dots are our numerical results, the solid red lines, beginning on the left of each plot, come from the dispersion relation of the charge diffusion mode obtained by inverting eq.~\eqref{eq:D3Dpdiffusionanalytic}, the solid blue lines, beginning on the right of each plot, come from the holographic zero sound dispersion relation at $\Tbar=0$, eq.~\eqref{eq:HZSanalyticKofW}, and the vertical dashed line is the location of the collisionless/hydrodynamic crossover, as defined by a collision of poles in the complex $\wbar$ plane at $\Bbar=0$, eq.~\eqref{eq:ZeroBCrossoverLocation}. \textbf{Bottom:} The same as the top plots, but for the D3/D7 system. All four plots, in this small $\Bbar$ regime, are qualitatively similar to the corresponding plots at $\Bbar=0$ in fig.~\ref{fig:kofwB0}.}
  \label{fig:AndyB10m2normalised}
\end{figure}

When $\Bbar$ is non-zero, we again find no collision of poles in the complex $\kbar$ plane. In fact, when $\Bbar\ll\Tbar^2$, the behavior of the poles is very similar to the $\Bbar=0$ case, as we show in fig.~\ref{fig:AndyB10m2normalised}, which is the same as fig.~\ref{fig:kofwB0}, but now with $\Bbar/\Tbar^2=10^{-2}$. In particular, we still see a kink in $\textrm{Re}\,\kbar(\wbar)$ at approximately the same value of $\wbar/\Tbar^2$ as at $\Bbar=0$.

\begin{figure}[!htb]
  \centering
\subfigure{
      \includegraphics[width=0.45\textwidth]{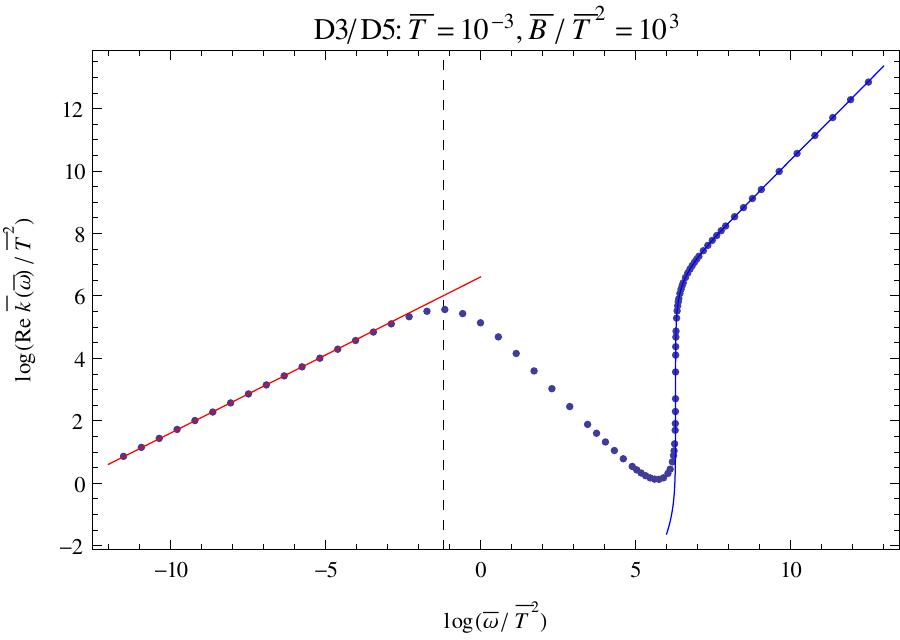}
}
\subfigure{
      \includegraphics[width=0.45\textwidth]{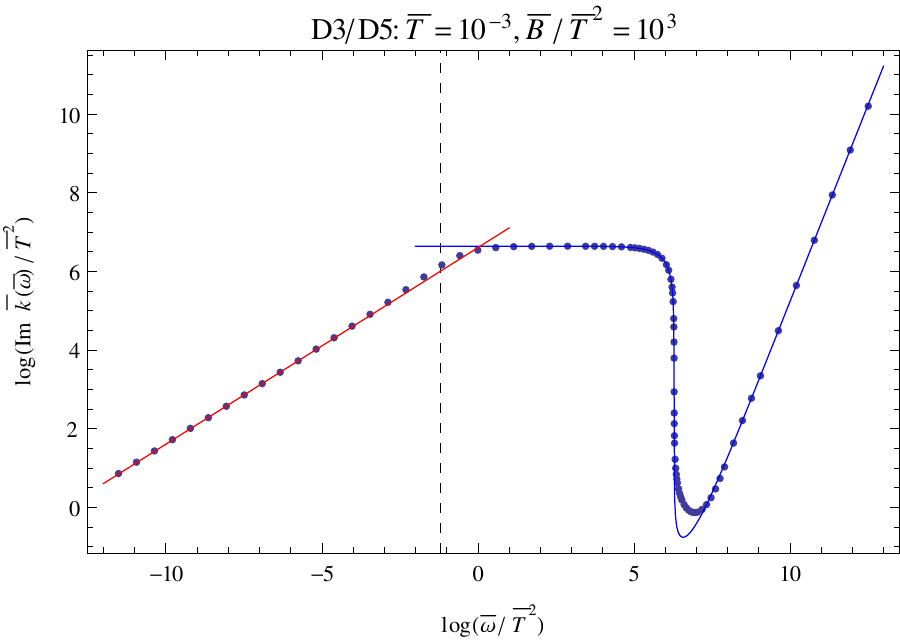}
      }
\subfigure{
      \includegraphics[width=0.45\textwidth]{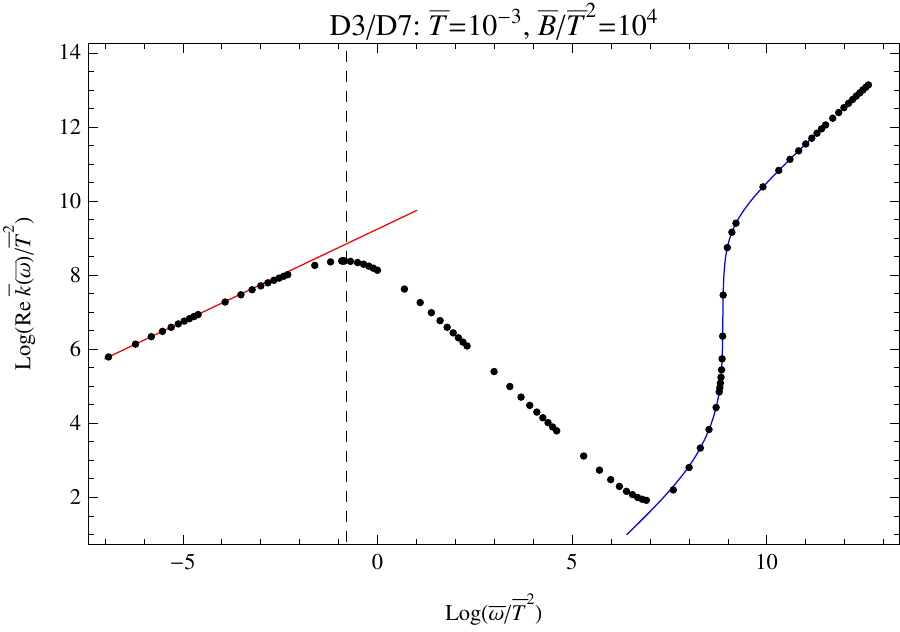}
}
\subfigure{
      \includegraphics[width=0.45\textwidth]{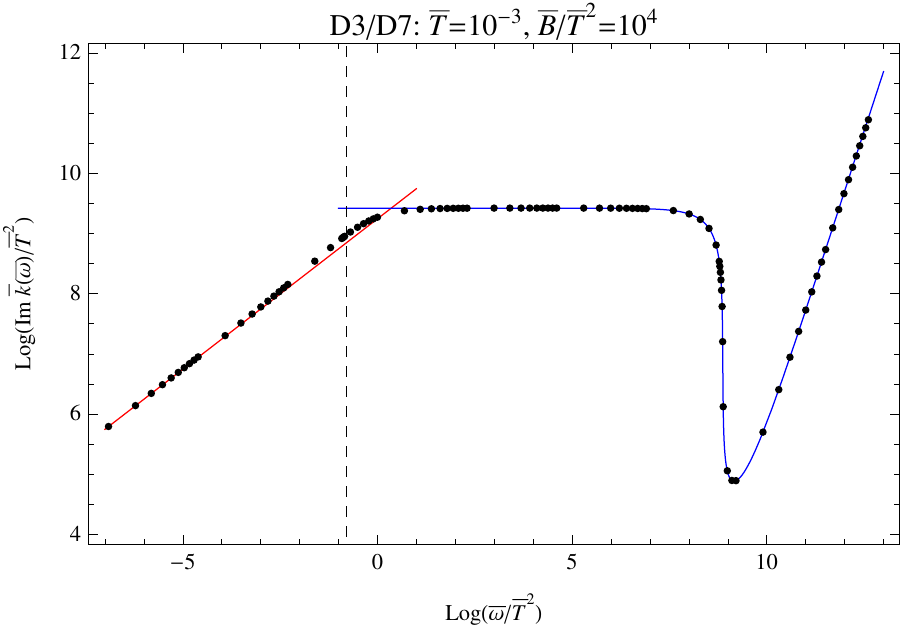}
      }
  \caption{\textbf{Top:} $\log \left(\textrm{Re}\,\kbar\left(\wbar\right)/\Tbar^2\right)$ (\textbf{Left}) and $\log \left(\textrm{Im}\,\kbar\left(\wbar\right)/\Tbar^2\right)$ (\textbf{Right}) as functions of $\log \left(\wbar/\Tbar^2\right)$ for the pole in $G^{tt}_R(\omega,k)$ closest to the origin of the complex $\kbar$ plane in the D3/D5 system with $\Tbar=10^{-3}$ and $\Bbar/\Tbar^2=10^3$, within the regime $\Bbar/\Tbar^2 \gg 1$. The dots are our numerical results, the solid red lines, beginning on the left of each plot, come from the dispersion relation of the charge diffusion mode obtained by inverting eq.~\eqref{eq:D3Dpdiffusionanalytic}, the solid blue lines, beginning on the right of each plot, come from the holographic zero sound dispersion relation at $\Tbar=0$, eq.~\eqref{eq:HZSanalyticKofW}, and the vertical dashed line is the location of the collisionless/hydrodynamic crossover, as defined by a collision of poles in the complex $\wbar$ plane at $\Bbar=0$, eq.~\eqref{eq:ZeroBCrossoverLocation}. \textbf{Bottom:} The same as the top plots, but for the D3/D7 system with $\Bbar/\Tbar^2 = 10^4$.}
  \label{fig:AndyB103normalised}
\end{figure}

When $\Bbar$ is non-zero and large, $\Bbar\gg\Tbar^2$, we find a qualitative difference from the $\Bbar\ll \Tbar^2$ case: at low $\wbar/\Tbar^2$ the mode is that of charge diffusion until a local \textit{maximum} in $\textrm{Re}\,\kbar(\wbar)$ at some $\wbar$. For larger $\wbar$, the mode is no longer that of charge diffusion.  We show this in fig.~\ref{fig:AndyB103normalised}, which is the same as figs.~\ref{fig:kofwB0} and~\ref{fig:AndyB10m2normalised}, but now with $\Bbar/\Tbar^2 = 10^3$ for the D3/D5 system and $\Bbar/\Tbar^2 = 10^4$ for the D3/D7 system. Remarkably, the local maximum in $\textrm{Re}\,\kbar(\wbar)$ occurs at approximately the same value of $\wbar/\Tbar^2$ as the kink in $\textrm{Re}\,\kbar(\wbar)$ at $\Bbar=0$ (the vertical dashed line in the left plots of fig.~\ref{fig:AndyB103normalised}).

In the intermediate regime $\Bbar \simeq \Tbar^2$, at low $\wbar/\Tbar^2$ the mode is still that of charge diffusion while at high $\wbar/\Tbar^2$ the mode is still that of holographic zero sound, and the transition between the two still occurs at approximately the same value of $\wbar/\Tbar^2$ as the kink in $\textrm{Re}\,\kbar(\wbar)$ at $\Bbar=0$, however at the transition neither a sharp kink nor a clear maximum in $\textrm{Re}\,\kbar(\wbar)$ were visible to us. The spectral function, at fixed $\wbar$, as a function of $\kbar$, still exhibits only a single peak, which is due to charge diffusion for small $\wbar/\Tbar^2$ and to holographic zero sound for large $\wbar/\Tbar^2$.

We have found that as we increase $\wbar$ the pole in $G^{tt}_R(\omega,k)$ in the complex $\kbar$ plane ceases to be that of charge diffusion approximately at the location of the crossover in the $\Bbar=0$ case, eq.~\eqref{eq:ZeroBCrossoverLocation}, \textit{for any value of $\Bbar$}. In other words, for any value of $\Bbar$, the response of the system to a density perturbation with fixed $\wbar$ below this value is dominated by charge diffusion, while the response to a perturbation with larger $\wbar$ is not. We have thus identified a clean boundary to the regime of hydrodynamic reponse that is approximately $\Bbar$-independent. We therefore define the location of the crossover from the hydrodynamic to the collisionless regime as occuring at this value of $\wbar/\Tbar^2$, eq.~\eqref{eq:ZeroBCrossoverLocation}: for the D3/D5 system $\wbar \approx 0.30 \Tbar^2$ and for the D3/D7 system $\wbar \approx 0.45 \Tbar^2$.

\begin{figure}[!htb]
\centering
\subfigure[D3/D5: $\bar{\chi}_{tt}$ at $\Tbar=10^{-1/2}, \Bbar/\Tbar^2=10^{-2}$]{      \includegraphics[width=0.45\textwidth]{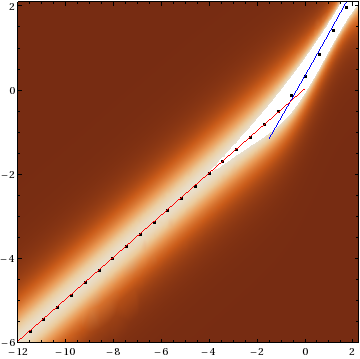}
}
\subfigure[D3/D5: $\bar{\chi}_{tt}$ at $\Tbar=10^{-3}, \Bbar/\Tbar^2=10^{3}$]{
\includegraphics[width=0.45\textwidth]{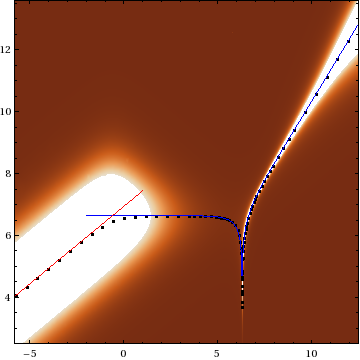}
}
\\ 
\vskip1em
\subfigure[D3/D7: $\bar{\chi}_{tt}$ at $\Tbar=10^{-1}, \Bbar/\Tbar^2=10^{-2}$]{      \includegraphics[width=0.45\textwidth]{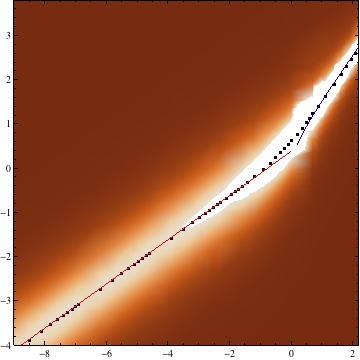}
}
\subfigure[D3/D7: $\bar{\chi}_{tt}$ at $\Tbar=10^{-3}, \Bbar/\Tbar^2=10^{4}$]{
\includegraphics[width=0.45\textwidth]{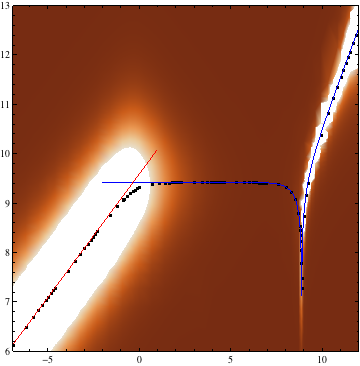}
}
\begin{center}
\setlength{\unitlength}{0.1\columnwidth}
\begin{picture}(0.1,0.25)(0,0)
\put(-2.25,5.75){\makebox(0,0){\small $\log(\wbar/\Tbar^2)$}}
\put(2.45,5.75){\makebox(0,0){\small $\log(\wbar/\Tbar^2)$}}
\put(-2.25,0.4){\makebox(0,0){\small $\log(\wbar/\Tbar^2)$}}
\put(2.45,0.4){\makebox(0,0){\small $\log(\wbar/\Tbar^2)$}}
\put(-5.2,8.15){\makebox(0,0){\small $\log(\kbar/\Tbar^2)$}}
\put(-5.2,3.05){\makebox(0,0){\small $\log(\kbar/\Tbar^2)$}}
\end{picture}
\vskip-1.5em
  \caption{\textbf{(a)} Density plot of our numerical results for $\bar{\chi}_{tt}(\wbar,\kbar)$ over the $\left(\log\wbar/\Tbar^2,\log\kbar/\Tbar^2\right)$ plane in the D3/D5 system with $\Tbar=10^{-1/2}$ and $\Bbar/\Tbar^2=10^{-2}$. (This plot uses the same data as that used to make fig.~\ref{fig:3dSpectral}.) Lighter colours represent larger values of the spectral function, as in fig.~\ref{fig:3dSpectral}. The black dots are our numerical results for the location of the pole in $G^{tt}_R(\omega,k)$ closest to the origin in the complex $\kbar$ plane. The solid red line starting at the bottom left of the plot is the value of $\left|\kbar\left(\wbar\right)\right|$ for the charge diffusion mode obtained by inverting the dispersion relation in eq.~\eqref{eq:D3Dpdiffusionanalytic}, and the solid blue line starting at the top right of the plot is the value of $\left|\kbar\left(\wbar\right)\right|$ for the holographic zero sound mode, eq.~\eqref{eq:HZSanalyticKofW}. \textbf{(b)} The same as (a), but with $\Tbar=10^{-3}$ and $\Bbar/\Tbar^2=10^3$. \textbf{(c)} The same as (a) and (b), but for the D3/D7 system with $\Tbar=10^{-1}$ and $\Bbar/\Tbar^2=10^{-2}$. \textbf{(d)} The same as (c), but with $\Tbar=10^{-3}$ and $\Bbar/\Tbar^2=10^4$.}
  \label{fig:3dSpectralwithkofw}
  \end{center}
  \vskip-1.5em
\end{figure}
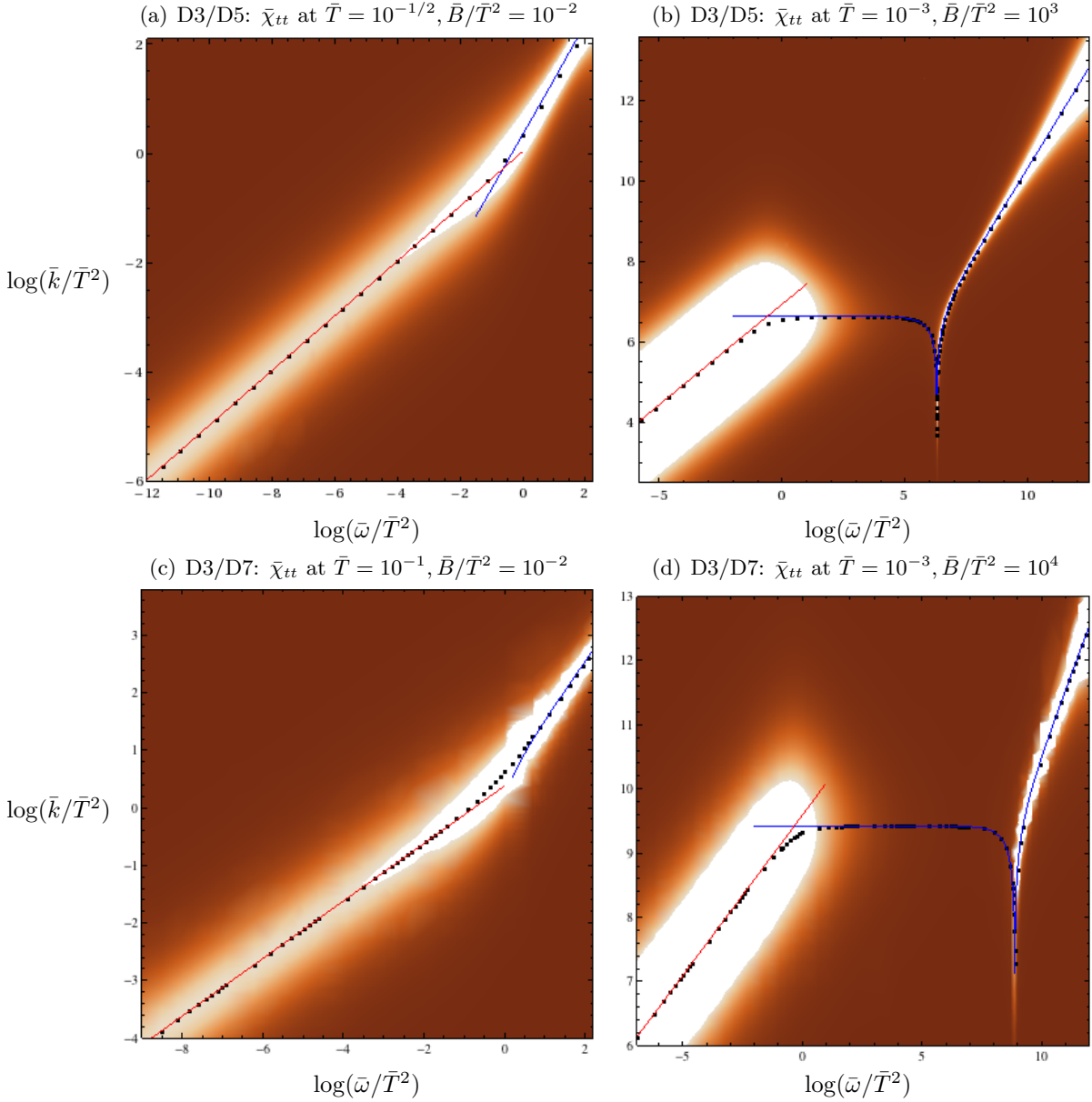

Let us now discuss in detail the region of suppressed spectral weight that we mentioned at the end of section~\ref{spectralfunctions}, in the regime of large magnetic field, $\Bbar \gg \Tbar^2$. In fig.~\ref{fig:3dSpectralwithkofw} we reproduce fig.~\ref{fig:3dSpectral}, the contour plots of $\log \bar{\chi}_{tt}(\wbar,\kbar)$ in the plane of $\left(\log\wbar/\Tbar^2,\log\kbar/\Tbar^2\right)$, but now with two differences: first, instead of a contour plot of $\log \bar{\chi}_{tt}(\wbar,\kbar)$ we present a density plot of $\bar{\chi}_{tt}(\wbar,\kbar)$, and second we superimpose on the density plots our numerical results for the locations of the pole of $G^{tt}_R(\omega,k)$ closest to the origin in the complex $\kbar$ plane as well as the values of $|\kbar(\wbar)|$ for the $T=0$ holographic zero sound dispersion relation, eq.~\eqref{eq:HZSanalyticKofW}, and for the charge diffusion, obtained by inverting eq.~\eqref{eq:D3Dpdiffusionanalytic}. In fig.~\ref{fig:3dSpectralwithkofw} we see clearly that at a fixed frequency $\wbar$ the peak in the spectral function $\bar{\chi}_{tt}\left(\wbar,\kbar\right)$ is given to good approximation by these values of $\left|\kbar\left(\wbar\right)\right|$.

\begin{figure}[!htb]
\centering
\subfigure{
\includegraphics[width=0.45\textwidth]{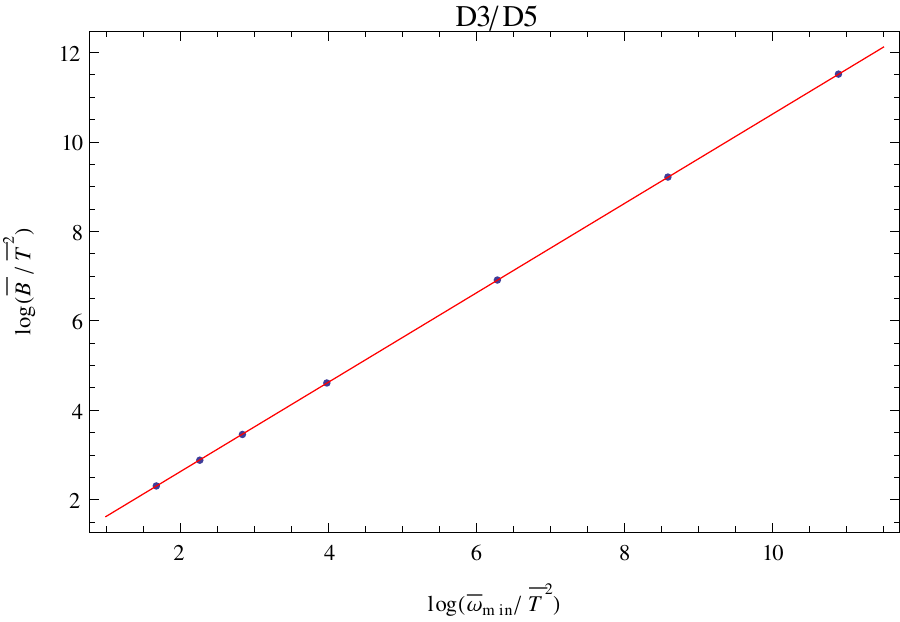}
}
\subfigure{
\includegraphics[width=0.45\textwidth]{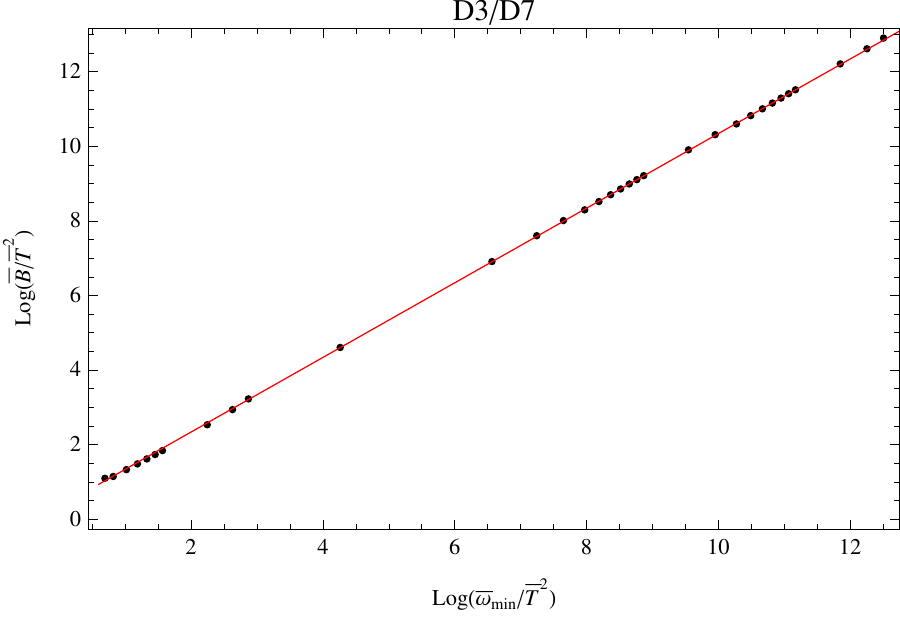}
}
  \caption{\textbf{Left:} $\log\left(\Bbar/\Tbar^2\right)$ versus $\log\left(\omega_{\mathrm{min}}/\Tbar^2\right)$ for the D3/D5 system, where $\wbar_{\mathrm{min}}$ is the high-frequency end of the region of suppressed spectral weight, \textit{i.e.}\ the cusp in the solid blue lines in figs.~\ref{fig:3dSpectralwithkofw} (b) and (d). The dots are our numerical results and the solid red line is derived from the gap in the $\Tbar=0$ holographic zero sound dispersion relation, the $\wbar_{\mathrm{gap}}$ in eq.~\eqref{eq:kofwgap}. Clearly eq.~\eqref{eq:kofwgap} is a very good approximation to the numerical result for $\wbar_{{\mathrm{min}}}$. \textbf{Right:} The same as the left figure, but for the D3/D7 system.}
  \label{fig:Andywedge}
\end{figure}

In the regime $\Bbar/\Tbar^2\gg1$, as shown in figs.~\ref{fig:3dSpectralwithkofw} (b) and (d), suppose we start with small $\wbar/\Tbar^2$, that is, we consider a vertical line near the left of each plot, where the spectral function exhibits a single peak due to charge diffusion. As we increase $\wbar/\Tbar^2$, moving to the right in each plot, we encounter the collisionless/hydrodynamic crossover, where the peak is no longer due to charge diffusion. For larger $\wbar/\Tbar^2$, a peak is still present in the spectral function, but is smaller by orders of magnitude compared to the charge diffusion peak. Only at sufficiently large $\wbar/\Tbar^2$ does the peak grow again by orders of magnitude, now being the peak due to holographic zero sound. In other words, for an intermediate range of frequencies $\wbar/\Tbar^2$, the spectral weight is suppressed. The high-frequency end of that intermediate region (on the right in the plots), is marked by a minimum in the value of $\left|\kbar\left(\wbar\right)\right|$ of the corresponding pole: in each of figs.~\ref{fig:3dSpectralwithkofw} (b) and (d), the solid blue line on the right has a cusp with a distinct minimum. The value of $\wbar$ at that minimum, $\wbar_{\mathrm{min}}$, is fixed by the gap in the holographic zero sound dispersion relation, $\wbar_{\mathrm{gap}}$ in eq.~\eqref{eq:kofwgap}: in fig.~\ref{fig:Andywedge} we plot $\log \left(\Bbar/\Tbar^2\right)$ versus the value of $\log\left(\wbar_{\mathrm{min}}/\Tbar^2\right)$, and find very good agreement with the value of $\wbar_{\mathrm{gap}}$. Notice that such a minimum does not occur for $\Bbar/\Tbar^2 \ll 1$, \textit{i.e.}\ the solid blue line in figs.~\ref{fig:3dSpectralwithkofw} (a) and (c) has no minimum like that in figs.~\ref{fig:3dSpectralwithkofw} (b) and (d).

\subsection{Overview: The Collisionless/Hydrodynamic Crossover}
\label{sec:NumericalResultsSummarySection}

\begin{figure}[!htb]
 \centering
\subfigure{
\includegraphics[width=0.45\textwidth]{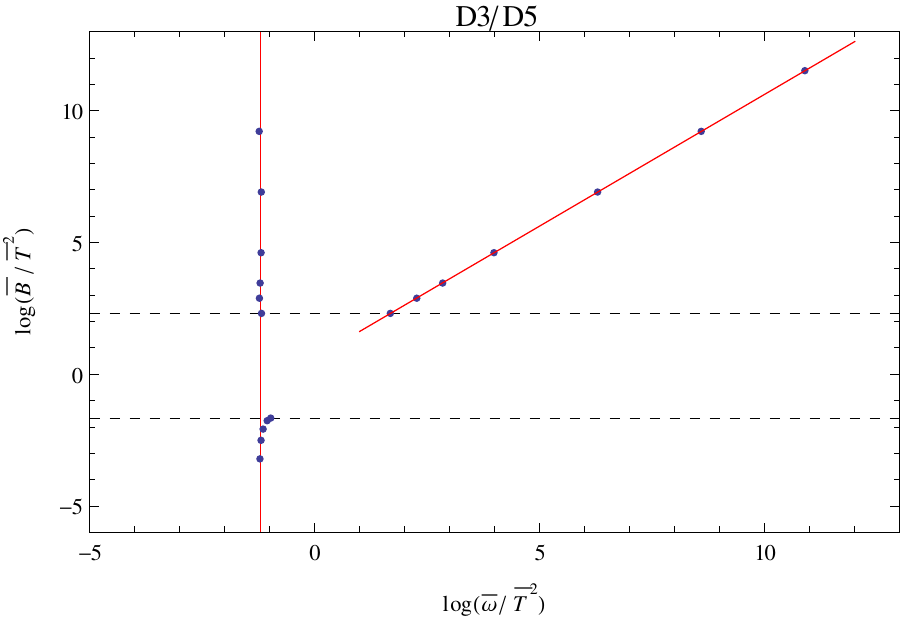}
}
\subfigure{
\includegraphics[width=0.45\textwidth]{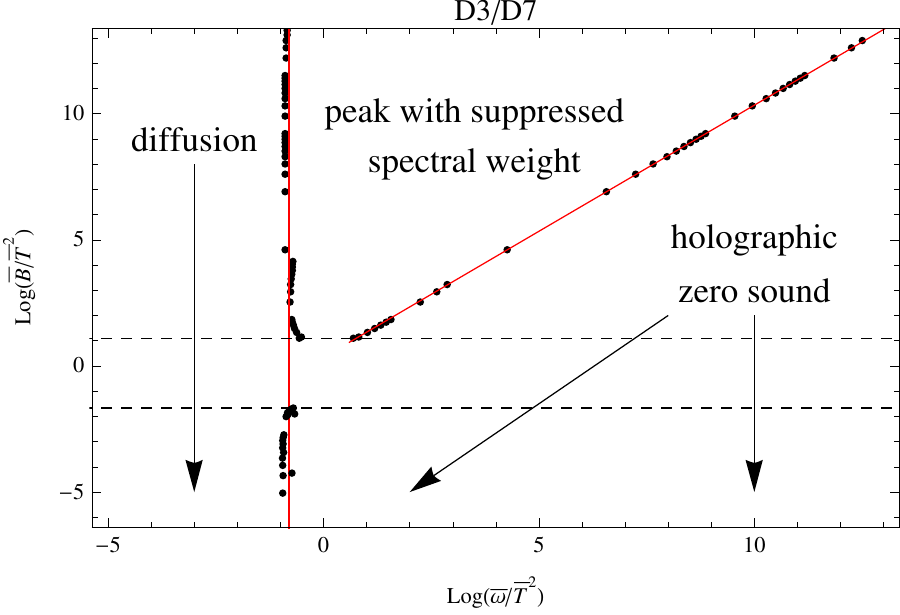}
}
\caption{An overview of our main results (see also fig.~\ref{fig:cartoon2}): $\log\left(\Bbar/\Tbar^2\right)$ versus $\log\left(\omega/\Tbar^2\right)$ for the D3/D5 system (\textbf{Left}) and the D3/D7 system (\textbf{Right}). The lines and dots have the same meanings in the two plots. The solid red vertical line denotes the collisionless/hydrodynamic crossover, defined for increasing $\Tbar$ with fixed $\wbar$ as the value of $\Tbar$ where the pole of $G^{tt}_R(\omega,k)$ closest of the origin of the complex $\kbar$ plane becomes that of charge diffusion. That value of $\Tbar$ is approximately the same as the $\Bbar=0$ value in eq.~\eqref{eq:ZeroBCrossoverLocation}. The dots on the solid red vertical line above the dashed lines are extracted from poles in the complex $\kbar$ plane, as discussed in section~\ref{sec:NumericalResultsKPlane}, while those below the dashed lines are extracted from a collision of poles in the complex $\wbar$ plane, as discussed in sections~\ref{collectiveholo} and ~\ref{sec:NumericalCollisionsSmallB}. In the region between the horizontal dashed lines we had difficulty extracting a precise location of the crossover from our numerics. The solid red diagonal line is the value of the gap in the $T=0$ dispersion relation for holographic zero sound, eq.~\eqref{eq:kofwgap}. The dots on the solid red diagonal line are our numerical results for $\wbar_{\mathrm{min}}$, reproduced from fig.~\ref{fig:Andywedge}. On the right figure we have indicated the nature of the dominant peak in the spectral function with fixed $\wbar$ as a function of $\kbar$.}  
\label{fig:Andywedge2}
\end{figure}

We can now give an overview of the collisionless/hydrodynamic crossover in our systems at non-zero $\Bbar$. When $\Bbar \ll \Tbar^2$, if we fix a real $\kbar$ and study poles in $G^{tt}_R(\omega,k)$ in the complex $\wbar$ plane, then we find that as we increase $\Tbar$ a collision of poles occurs, allowing us to define the crossover in a fashion similar to the $\Bbar=0$ case. When $\Bbar \gg \Tbar^2$, no such collision of poles occurs in the complex $\wbar$ plane. Indeed, in this regime of large $\Bbar$, the spectral function, for fixed $\kbar$ as a function of $\wbar$, generically exhibits two peaks, one for holographic zero sound and one for charge diffusion, preventing us from identifying this regime as either collisionless or hydrodynamic. We found that for any $\Bbar$, if we instead fix real $\wbar$ and study poles in $G^{tt}_R(\omega,k)$ in the complex $\kbar$ plane, then we find a single pole, and in the spectral function we find a single peak. As we increase $\Tbar$, we could then define the crossover as the value of $\Tbar$ where the single pole becomes that of charge diffusion, and hence the system enters the hydrodynamic regime. Remarkably, we found that the value of $\Tbar$ where that crossover occurs is approximately $\Bbar$-independent, being given by the $\Bbar = 0$ value, eq.~\eqref{eq:ZeroBCrossoverLocation}: $|\wbar| \approx 0.30 \, \Tbar^2$ for the D3/D5 system and $|\wbar| \approx 0.45 \, \Tbar^2$ for the D3/D7 system. In the language of LFL theory, this would indicate that the collision frequency of the quasiparticles is independent of $\Bbar$. We summarize our results in fig.~\ref{fig:Andywedge2}, which we represented schematically in fig.~\ref{fig:cartoon2}.

We can actually provide a quantitative estimate of the $O(1)$ value of $\Bbar/\Tbar^2$ at which, as $\Bbar$ increases in the collisionless regime, the gap in the holographic zero sound dispersion relation produces suppression of spectral weight. In fig.~\ref{fig:Andywedge2}, the diagonal red line represents the value of the gap $\wbar_{\mathrm{gap}}$ in the $\Tbar=0$ dispersion relation of the holographic zero sound, eq.~\eqref{eq:kofwgap}, while the vertical red line represents the location of the crossover, \textit{i.e.}\ the value of $\wbar$ in eq.~\eqref{eq:ZeroBCrossoverLocation}. By extending the diagonal red line down to the intersection with the vertical red line, we can estimate the $O(1)$ value of $\Bbar/\Tbar^2$ that we want: we simply equate the $\wbar_{\mathrm{gap}}\propto \Bbar$ in eq.~\eqref{eq:kofwgap} with the $\wbar\propto \Tbar^2$ in eq.~\eqref{eq:ZeroBCrossoverLocation} and solve for $\Bbar/\Tbar^2$,\footnote{For a holographic quantum liquid in which the $U(1)_b$ density is produced by fermions alone, a result of ref.~\cite{Jokela:2012vn} is that the holographic zero sound's dispersion relation becomes gapped when $\Bbar/\Tbar^2 \gtrsim 0.18$. The relationship between that value of $\Bbar/\Tbar^2$ and the value above which the spectral weight of the holographic zero sound is suppressed would be interesting to study.}
\beq
\label{eq:criticalb}
\Bbar/\Tbar^2 \approx  \begin{cases} 0.56 & \text{for D3/D5}, \\ 0.63 & \text{for D3/D7.} \end{cases}
\eeq

\section{Summary and Outlook}
\label{conclusions}

We used holography to study compressible states of the D3/D5 and D3/D7 theories that were not solids, superfluids, LFLs, or NFLs. In particular, we studied the effect of a non-zero magnetic field $B$ on the collisionless/hydrodynamic crossover in these systems as we increased the temperature $T$, for values of $B$ below any known phase transitions. We found that the crossover was simplest to understand by studying the spectrum of collective excitations of these systems with fixed frequency $\omega$, where we could define the crossover for any value of $B$ as the value of $T$ where the dominant pole in $G^{tt}_R(\omega,k)$ becomes that of charge diffusion. Remarkably, that value of $T$ is, to a very good approximation, independent of $B$. In the high-frequency, collisionless regime, for values of $B$ above a value $B/T^2 \simeq O(1)$ estimated in eq.~\eqref{eq:criticalb}, we also saw a suppression of spectral weight due to the magnetically-induced gap in the holographic zero sound dispersion relation. Our main results are summarized in figs.~\ref{fig:cartoon2} and~\ref{fig:Andywedge2}. We believe that, given our minimal ingredients, our fig.~\ref{fig:cartoon2} may be generic to compressible states described holographically by probe DBI actions with non-zero electric and magnetic flux. In other words, we believe that fig.~\ref{fig:cartoon2} may be characteristic of such compressible states, and hence may help in classifying them.

Various open questions remain about these, and other similar, holographic quantum liquids. Here we will list only a few such questions.

In the D3/D7 system we only studied excitations with momenta perpendicular to the magnetic field, but we have another option: the momentum can also have a non-zero component parallel to the magnetic field. A natural question is how the non-zero magnetic field affects modes with momentum parallel to the magnetic field, for example, how does non-zero $B$ affect the holographic zero sound mode that propagates parallel to the magnetic field? What happens to modes parallel to the magnetic field during the collisionless/hydrodynamic crossover?

We only considered values of $B$ below any known phase transitions. What happens at higher values of $B$, in the symmetry-broken phases? Clearly the spectrum of excitations will then include the Goldstone bosons associated with the symmetry breaking. How do those affect the collisionless/hydrodynamic crossover?

More generally, an open question is whether these compressible states of the D3/D$p$ theories are the true ground states, that is, whether these states are \textit{global} minima of the free energy. Indeed, as we mentioned in section~\ref{review}, for the D3/D7 system, at one point in the phase diagram in the grand canonical ensemble with non-zero $B$, these states are known to exhibit an instability towards an inhomogenous state. We hasten to add that even if the states we studied are not the true ground states, they are a useful theoretical laboratory that may reveal some guiding principle(s) to classify compressible states.

Another perspective on the collective excitations of these systems in the collisionless regime, which we have not explored here, is given by the semi-holographic description~\cite{Jensen:2010ga,Nickel:2010pr,Ammon:2011hz,Goykhman:2012vy}, an effective description in which some critical degrees of freedom, described by a $(0+1)$-dimensional CFT dual holographically to degrees of freedom in a near-horizon $AdS_2$, produce the dissipation of the holographic zero sound mode. Clearly in our systems the bulk geometry has no near-horizon $AdS_2$ (unlike extremal Reissner-Nordstr\"om-AdS), nevertheless quite generically fluctuations of the D$p$-brane worldvolume fields effectively ``see'' $AdS_2$ deep in the bulk, due to the worldvolume electric flux~\cite{Jensen:2010ga,Nickel:2010pr,Ammon:2011hz,Goykhman:2012vy}. An open question is what aspects of the collisionless/hydrodynamic crossover, at either zero or non-zero $B$, can be understood using semi-holographic techniques.

The behavior of collective excitations with increasing $T$ for the compressible states dual to Reissner-Nordstr\"om-AdS is qualitatively different from that of the D3/D$p$ systems~\cite{Edalati:2010pn,Davison:2011uk}. Although those states do exhibit a sound mode in the collisionless regime~\cite{Edalati:2010pn}, as $T$ increases no collision of poles in the complex $\omega$ plane occurs~\cite{Davison:2011uk}. Does a non-zero magnetic field produce a gap in the sound dispersion relation in those states? If so, how does that affect the behavior of the collective modes as $T$ increases?

\acknowledgments{We would like to thank A.~Cherman, K.~Jensen, N.~Jokela, M.~Lippert, A.~Parnachev, M.~Rangamani, K.~Schalm, D.T.~Son, B.~Withers, and J.~Zaanen for helpful discussions and correspondence. We especially thank A.~Starinets for reading and commenting on a preliminary draft of this paper. D.B., R.D., and S.G. are supported by STFC studentships. The research leading to these results has received funding from the European Research Council under the European Community's Seventh Framework Programme
(FP7/2007-2013) / ERC grant agreement no. 247252.}

\bibliographystyle{JHEP}
\bibliography{magfieldcrossover}

\end{document}